\documentclass[12pt]{article}
\usepackage{amscd,amssymb,amsmath,latexsym,enumerate}
\usepackage[mathscr]{euscript}
\usepackage{epsfig}
\usepackage{textcomp}
\usepackage{amsmath}
\usepackage{amssymb}
\usepackage{verbatim}

\title{Scattering theory for lattice operators in dimension $d\geq 3$}
\vspace{.5cm}

\author{Jean Bellissard$^{1}$, Hermann Schulz-Baldes$^{2}$\\
{\small $^1$ Georgia Institute of Technology, School of Mathematics, Atlanta GA 30332-0160, USA}\\
{\small $^2$ Department Mathematik, Universit\"at Erlangen-N\"urnberg, D-91054
Erlangen, Germany} \vspace{.2cm}
}

\date{ }

\newtheorem{theo}{Theorem}
\newtheorem{defini}{Definition}
\newtheorem{proposi}{Proposition}
\newtheorem{lemma}{Lemma}
\newtheorem{coro}{Corollary}
\newtheorem{rem}{Remark}
\newtheorem{exam}{Example}

\newcommand{\Bb}{{\mathcal B}}

\newcommand{\Dd}{{\mathcal D}}
\newcommand{\Ee}{{\mathcal E}}
\newcommand{\Ff}{{\mathcal F}} 

\newcommand{\Hh}{{\mathcal H}}

\newcommand{\Jj}{{\mathcal J}}
\newcommand{\Kk}{{\mathcal K}}

\newcommand{\Oo}{{\mathcal O}}

\newcommand{\Rr}{{\mathcal R}}
\newcommand{\Ss}{{\mathcal S}}
\newcommand{\Tt}{{\mathcal T}}
\newcommand{\Uu}{{\mathcal U}}
\newcommand{\Vv}{{\mathcal V}}
\newcommand{\Ww}{{\mathcal W}}

\newcommand{\RR}{{\bf R}}

\newcommand{\id}{{\mathbf 1}}

\newcommand{\CM}{{\mathbb C}}

\newcommand{\NM}{{\mathbb N}}
\newcommand{\QM}{{\mathbb Q}}

\newcommand{\RM}{{\mathbb R}}

\newcommand{\SM}{{\mathbb S}}
\newcommand{\TM}{{\mathbb T}}

\newcommand{\ZM}{{\mathbb Z}}

\newcommand{\hA}{{\widehat{A}}}

\newcommand{\hH}{{\widehat{H}}}

\newcommand{\hX}{{\widehat{X}}}

\newcommand{\as}{{\mathscr A}}

\newcommand{\es}{{\mathscr E}}

\newcommand{\hs}{{\mathscr H}}

\newcommand{\js}{{\mathscr J}}

\newcommand{\sss}{{\mathscr S}}

\newcommand{\vs}{{\mathscr V}}

\newcommand{\cA}{\overset{\;\,\circ}{A}}
\newcommand{\cS}{\overset{\;\circ}{S}}
\newcommand{\cT}{\overset{\;\circ}{T}}
\newcommand{\cH}{\overset{\;\circ}{H}}

\newcommand{\cO}{\overset{\;\circ}{O}}

\newcommand{\TR}{{\rm Tr}}                       
\newcommand{\Cs}{$C^{\ast}$-algebra }              
\newcommand{\CsS}{$C^{\ast}$-algebras}             
\newcommand{\supp}{\mbox{\rm supp}}                
\newcommand{\dist}{\mbox{\rm dist}}                
\newcommand{\rank}{\mbox{\rm Rank}}               
\newcommand{\Ran}{\mbox{\rm Ran}}                 
\newcommand{\Ker}{\mbox{\rm Ker}}                 
\newcommand{\SL}{\mbox{\rm SL}}                    
\newcommand{\Conv}{\mbox{\rm Conv}}                
\newcommand{\dive}{\mbox{\rm div}}                 

\newcommand{\cri}{{\sss}^*}                        
\newcommand{\criM}{{\sss}}                   
\newcommand{\one}{{\bf 1}}                         
\newcommand{\piso}{\Pi}                            

\def\Xint#1{\mathchoice
   {\XXint\displaystyle\textstyle{#1}}%
   {\XXint\textstyle\scriptstyle{#1}}%
   {\XXint\scriptstyle\scriptscriptstyle{#1}}%
   {\XXint\scriptscriptstyle\scriptscriptstyle{#1}}%
   \!\int}
\def\XXint#1#2#3{{\setbox0=\hbox{$#1{#2#3}{\int}$}
     \vcenter{\hbox{$#2#3$}}\kern-.5\wd0}}

\def\dashint{\Xint-}

\begin{document}

\maketitle

\begin{abstract} This paper analyzes the scattering theory for periodic tight-binding
Hamiltonians perturbed by a finite range impurity. The classical energy gradient flow is used to construct a conjugate (or dilation) operator to the unperturbed Hamiltonian. For dimension $d\geq 3$ the wave operator is given by an explicit formula in terms of this dilation operator, the free resolvent and the perturbation. From this formula the scattering and time delay operators can be read off. Using the index theorem approach, a Levinson theorem  is proved which also holds in presence of embedded eigenvalues and threshold singularities. 
\end{abstract}




\section{Introduction}
\label{scatt08.sect-intro}

The purpose of this work is to present the scattering theory for a quantum particle described by a tight-binding Hamiltonian $H=H_0+V$ acting on the Hilbert space $\ell^2(\ZM^d)$ where $H_0$ is a periodic operator with a single band and $V$ is a finite rank perturbation. Most of the present work is focusing on the case of dimension $d\geq 3$.  All along this paper, it will be assumed that the Fourier transform of $H_0$ acts on $L^2(\TM^d)$ as a multiplication operator by a real analytic Morse function $\Ee(k)$ having only one maximum and one minimum. Operators of this type appear in solid state physics as effective one-band Hamiltonians for electrons or holes in periodic media. The analyticity reflects the exponential decay of the hopping amplitude of the particle and the Morse condition is generic.

 \subsection{Main Results}
 \label{scatt08.ssect-results}

Scattering theory for a Schr\"odinger operator with a periodic potential has already been considered \cite{New2,BY,GN}. The above scattering problem has also been addressed in the physics literature, for example in \cite{Eco}. The present work is going further. Initially, it was motivated by the remark by Kellendonk and Richard \cite{KR06} that Levinson's theorem \cite{Lev} relating the number of bound states to the total scattering phase can be interpreted as a special case of the Atyiah-Singer index theorem. As it turns out, this nice basic idea requires a substantial amount of technicalities when it comes to mathematical justification \cite{KR2,KR3,KR4}. For indeed, the global character of Levinson's theorem requires several technical steps. First, a complete control on the nature of the singularities of the Green function of $H_0$ is needed, a task that is easy on the continuum, but more involved in the present case. In addition, it requires a conjugate operator to $H_0$ in order to shift the energy, replacing the dilation operator used for the continuum situation. Moreover, the potential term $V$ may create embedded eigenvalues and threshold singularities that must be analyzed thoroughly since they contribute to Levinson's theorem. At last, for the index theorem to apply, it is necessary to prove that both the wave operator and the scattering matrix can be expressed as suitable continuous function of the energy and dilation operators. This was achieved in \cite{KR3} through an explicit calculation in dimension $d=1$ using the techniques of \cite{Jen}. In higher dimension \cite{KR4}, the explicit calculation turns out to be harder, but it is possible to prove sufficient regularity of the wave operators and the scattering matrix. Unfortunately, the work \cite{BY} is insufficient to implement this program completely for the class of models considered here. The present paper is supplementing these points.

\vspace{.2cm}

In the light of the previous introduction, the main results of the paper can be summarized as follows:

\begin{itemize}
  \item A lattice analog of the dilation operator is constructed (see Theorem~\ref{theo-dilation}). It is a self-adjoint unbounded operator $A$ such that $\imath[H_0,A]=F(H_0)$ where $F$ is a positive function on the spectrum of $H_0$ vanishing only at the band edges. 

  \item Explicit expressions for the wave operators, the scattering matrix and the time delay operators are obtained in terms of $H_0$, $V$ and $A$ (see Theorem~\ref{scatt08.theo-waveop3d}, Theorem~\ref{scatt08.theo-Sin3d} and Theorem~\ref{scatt08.theo-Tin3d}).

 \item A series of results concerning the existence of embedded eigenvalues and threshold singularities.

  \item A Levinson type theorem is derived which is now briefly described (see Theorem~\ref {scatt08.theo-Levinson3d} for details). The essential spectrum of $H$ is the spectrum $[E_-,E_+]$ of $H_0$ and $E_\pm$ are called the {\em band edges} or {\em thresholds}. Let $P_{\mbox{\rm\tiny pp}}$ be the eigenprojection on the eigenvalues $N$ be the total number of eigenvalues, including the embedded and the threshold eigenvalues (the latter have to be distinguished from threshold resonances). Further let $S$ be the scattering operator and let $T=-\imath S^{-1}[A,S]$ be the {\em time delay operator} seen as acting on $\ell^2(\ZM^d)$. Finally let $m_\pm\in\{0,1\}$ be the degeneracies of the {\em threshold resonances} at $E_\pm$ also called {\em half-bound states} singularities (higher degeneracies are possible, but not dealt with here). Then for $d=3$ and isotropic extrema of $\Ee$, 

\begin{equation}
\label{scatt08.eq-LevinsonTh}
\frac{1}{2\imath\pi}\; \TR\left(S^{-1}[A,S]\right) \;+\; 
  \TR(P_{\mbox{\rm\tiny pp}})
  \;=\;
-\,\frac{m_+}{2}\,-\,\frac{m_-}{2}
\;.
\end{equation}
For $d\geq 5$ it is proved that always $m_\pm=0$ and that \eqref{scatt08.eq-LevinsonTh} holds. For $d=4$, \eqref{scatt08.eq-LevinsonTh} is proved under the hypothesis that $m_\pm=0$.

\end{itemize}

\noindent As already pointed out, this paper is restricted to dimension $d\geq 3$. Dimensions $d=1$ and $d=2$ require a detailed asymptotic expansions of the free Green function near the band edges \cite{New,BGDW,KR3}. The one-dimensional case has been treated in \cite{CK,HKS}. The case $d=2$ will be addressed in a future publication.

\vspace{.2cm}

While most of these results are technically new, similar results  have been already obtained in the past. The scattering problem in $\RM^d$ with standard Laplacian perturbed by a decaying potential together with a proof of Levinson's theorem can be found in standard textbooks such as \cite{New,RS}. In this situation, also threshold resonances have been analyzed in details (see \cite{Bol} for a review). The scattering of an electron in a periodic potential by an impurity has been considered by physicists for a long time, in connection with the transport properties of semiconductiors. This theory is based on the KKR equations (for Korriga, Kohn and Rostoker) and in this context Levinson's theorem given in equation~(\ref{scatt08.eq-LevinsonTh}) above is also known under the name of {\em Friedel sum rule}. This was investigated by Newton \cite{New2} for $d=3$ with an impurity potential that could lead to threshold resonances, but not embedded eigenvalues. Levinson's theorem for periodic potentials in dimension $d=1$ was proved by Firsova \cite{Fir}. The mathematical aspects of scattering theory in a periodic potential has been considered by Birman and Yafaev \cite{BY} and bears many similarities with the present approach. However, the latter work does not deal with the critical points of the band functions, and hence does not lead to a proof of Levinson's theorem.  The scattering of a lattice electron by a localized impurity is also addressed in the book by Economou \cite{Eco}, however, the explicit formulas obtained in the present work for the scattering matrix and the wave operators are missing. The problem of embedded eigenvalues has long been considered as  an irrelevant curiosity. Indeed, even though they are non-generic within the class of finite rank perturbations considered here (as follows from the arguments in Section~\ref{scatt08.ssect-spec}), they may occur in practical devices, in particular, when a compact part of the lattice is inaccessible to a particle coming from the outside (see Example~\ref{scatt08.exam-int} below). The eigenvalues of the Hamiltonian inside have an influence on the scattering outside in various ways, as can be seen in equation~(\ref{scatt08.eq-LevinsonTh}). An analogous effect occurs for microwaves reflected by a cavity, as was shown, for instance, in \cite{DSF}.

 \subsection{Strategy of proofs}
 \label{scatt08.ssect-discu}

As suggested by the formula $\imath[H_0,A]=F(H_0)$, the conjugate operator $A$ is the generator of an energy shift. For its construction the classical energy gradient flow is slowed down near the band edges which are also called thresholds. This flow can be implemented as a strongly continuous one-parameter group of unitary operators in the Hilbert space and $A$ is then simply the generator of this group. The unitary implementation of a vector field has already been carried out in [HS,ABG], however, these constructions excluded energy surfaces with critical points. Removing this constraint is crucial for the proof of Levinson's theorem and this is probably the main conceptual contribution of this paper to the scattering theory in periodic lattices. The proof also covers dimension $d=2$.

\vspace{.2cm}

The introduction of the conjugate operator $A$ is closely linked to an important tool of calculation used in this paper, namely the {\em rescaled energy and Fermi surface} (REF) representation giving an adequate spectral representation of both $H_0$ and $A$. It shows that the Hilbert space $\ell^2(\ZM^d)$ is isomorphic to $L^2(\RM)\otimes L^2(\Sigma,\nu)$ where $\RM$ is a rescaled energy variable and $\Sigma$ is a reference Fermi surface given by some level set of $\Ee$ furnished with a Riemannian volume $\nu$.  The influence of the critical points on the dynamics defined by $A$ lies on a set of zero Lebesgue measure explaining why the unitary group $e^{\imath tA}$ is globally defined. While Morse's theory shows that the topology of the level sets changes at the passage through a critical value, the previous result, on the opposite, shows that the topology of the Fermi surface does not play any role for the Hilbert space isomorphism. Changing $H_0$ into $B=f(H_0)$, for a suitable function $f$ such that $\imath[B,A]=\id$, leads to a representation where $B$ is the multiplication by a variable $b\in\RM$, called the {\em rescaled energy} while $A$ becomes the derivative $-\imath\partial_b$ and acts as an infinitesimal rescaled energy shift.

\vspace{.2cm}

Given the REF representation, it is possible to compute all standard objects of scattering theory explicitly. In order to limit the technical difficulties, this work will be restricted to dimension $d\geq 3$ and to a compactly supported perturbation. The wave operator, minus the identity, is then an explicit continuous function in $A$ and $B$ with values in the algebra of compact operators on $L^2(\Sigma,\nu)$. From this formula and the invariance principle, an expression for the on-shell scattering matrix is then readily deduced. Up to an explicit partial isometry, it is a finite dimensional unitary matrix expressed in terms of the perturbation $V$ and the Green function of $H_0$. The spectral property of the time delay operator linking it to the resolvent also follows form this analysis. This allows to give a first short proof of Levinson's theorem by a contour integration argument when there are no embedded eigenvalues and no threshold singularities. 

\vspace{.2cm}

However, both threshold singularities and embedded eigenvalues may occur for adequate choices of $V$, sometimes with physical meaning. This situation is covered by the second proof of Levinson's theorem which follows closely the $K$-theoretic arguments of Kellendonk and Richard \cite{KR06,KR2}. Following  these authors, a \Cs $\es$ is generated by continuous functions of $A$ and of $B$ with values in the compact operators on $L^2(\Sigma,\nu)$ and having well-defined limits at $\pm\infty$ which coincide in the four corners $A=\pm\infty$ and $B=\pm\infty$. It contains the ideal $\Jj$ of those functions vanishing at $\infty$ and the extension is precisely by the \Cs $\as$ of operators fibered over $A$ or $B$, again coinciding in the four corners. A large amount of effort is then dedicated to proving that the wave operator belongs to the Toeplitz extension $\es$, even when embedded eigenvalues and threshold singularities are present. It follows that the wave operator is a lift of the scattering operator which combined with contributions stemming from the thresholds is an element of $\as$. This leads to a $K$-theoretic version of the proof of Levinson's theorem.

\vspace{.2cm}

\noindent {\bf Notations:} As usual, $|A|=(A^*A)^{\frac{1}{2}}$, $\Re e\,A=\frac{1}{2}(A+A^*)$ and  $\Im m\,A=\frac{1}{2\imath}(A-A^*)$ for any operator $A$. Furthermore, throughout there is a rescaled energy variable $b=f(E)$ associated with the bijection $f$ from the spectrum of the unperturbed operator to $\RM$ which is defined in \eqref{scatt08.eq-energychange} below. For objects depending on energy both $E$ and $b$ will be used as indices, for example $P_b=P_E$, $\Pi_b=\Pi_E$, $C_b=C_E$ and so on.

\vspace{.2cm}

\noindent {\bf Acknowledgments:} We thank A. Knauf, S. Golenia, S. Richard and J. Kellendonk for numerous comments. The work of J.~B. was support in part by NSF Grant No. 0600956 and 0901514, that of H. S.-B. in part by the DFG. After the first version of this paper was submitted, two papers on related matters appeared on the archives. The article \cite{BSPL11} provides a variational principle giving an upper bound on the number of eigenvalues (including embedded ones) and \cite{KKN} analyzes the Friedel sum rule via the spectral shift function.

\vspace{.5cm}

\section{Analysis of the unperturbed lattice Hamiltonian}
\label{scatt08.sect-freeop}

 \subsection{Unperturbed Hamiltonian and its energy band}
 \label{sec-normal}
 
The tight-binding Hamiltonians considered in this work act on the Hilbert space
$\ell^2(\ZM^d)$ of square summable sequences of complex numbers indexed by the
$d$-dimensional lattice $\ZM^d$. The {\em free Hamiltonian} $H_0$ on
$\ell^2(\ZM^d)$ is supposed to be of the form

\begin{equation}
\label{scatt08.eq-freeHam} \langle n\,|\,H_0\,\phi\rangle \;=\;
   \sum_{m\in\ZM^d}
    \Ee_{n-m}\;\langle m|\phi\rangle\,,
\qquad
  \phi \in \ell^2(\ZM^d)\,,
\end{equation}

\noindent where the $\Ee_n$'s are the Fourier coefficients of a real-analytic real-valued function
$\Ee(k)=\sum_{n\in\ZM^d} e^{\imath k n} \Ee_n$  on the $d$-dimensional torus $\TM^d=\RM^d/ (2\pi \ZM^d)$. Hence we restrict ourselves to a free operator with a single band. As $H_0$ is translation invariant, it is diagonalized by the discrete Fourier transform $\Ff: \ell^2(\ZM^d)\to L^2(\TM^d)$, where $L^2(\TM^d)$ is the Hilbert space of square integrable functions on $\TM^d$. It is densely defined by
$$
(\Ff\phi)(k)\;=\;
\frac{1}{(2\pi)^{\frac{d}{2}}}\;
\sum_{n\in\ZM^d} e^{\imath k n}\,\langle n|\phi\rangle\;.
$$
and is unitary. Now $\hH_0=\Ff H_0 \Ff^*$ is a multiplication operator on $L^2(\TM^d)$ by the function $\Ee$.  The main hypothesis on $H_0$ are expressed in terms of this function $\Ee$. The set of critical points $\cri\subset\TM^d$ at which the gradient $\nabla \Ee$ w.r.t. the euclidean metric vanishes is finite due to the analyticity of $\Ee$ and each critical point is supposed to be non-degenerate, namely for any $k^*\in\cri$ the Hessian $\Ee''(k^*)$ is a real symmetric invertible $d\times d$ matrix. In other words, $\Ee$ is supposed to be a so-called Morse function \cite{Nic}.  Recall that the index of a critical point $k^*$ is the number of negative eigenvalues of $\Ee''(k^*)$.  Then the Morse inequalities state that the number of critical points with index $p$ is larger than or equal to the Betti number $\beta_p$ of the torus $\TM^d$, which is equal to the binomial coefficient $d$ over $p$.  In particular, there must exist critical points of $\Ee$ with signature $p$ for every $p=0,\ldots,d$. For the discrete Laplacian, the energy band $\Ee(k)=2\sum_{j=1}^d\cos(k_j)$ is a Morse function for which the Morse inequalities become equalities. We also assume that there are only two critical points $k^*_-$ and $k_+^*$ of definite signature corresponding to the minimal and maximal values $E_-=\Ee(k^*_-)$ and $E_+=\Ee(k^*_+)$ of $\Ee$. Hence all other critical points $k^*\in\cri$ are supposed to have critical values $\Ee(k^*)$ in $(E_-,E_+)$ and to be of indefinite signature. Note that $\Ee(\cri)$ is the set of all critical values. A spectral interval is called non-critical if it does not contain any critical value.

 \subsection{The classical energy flow}
 \label{scatt08.sec-flow}
 
Let $F:[E_-,E_+]\to\RM_{\geq 0}$ be a real analytic function vanishing only at the band edges $E_-$ and $E_+$ and satisfying $F(E-E_-)\leq C|E-E_-|$ and $F(E_+-E)\leq C|E_+-E|$ for some constant $C$. Below we will choose 

\begin{equation}
\label{scatt08.eq-Fchoice}
F(E)\;=\;2\;
\frac{(E-E_-)(E_+-E)}{E_+-E_-}\,,
\end{equation}

\noindent but this particular choice will only become relevant for the calculation of the wave operators in Section~\ref{scatt08.ssect-waveREF}. Then let $\hX$ be the vector field on $\TM^d$ defined by

\begin{equation}
\label{scatt08.eq-Xvec} \hX(k) \;=\;
F\bigl(\Ee(k)\bigr)\;
   \frac{\nabla \Ee(k)}{|\nabla\Ee(k)|^2}\;,\qquad k\in\TM^d\;.
\end{equation}

\noindent Apart from the factor $F\circ \Ee$, the vector field $\hX$ is precisely the one used in the standard argument of Morse theory \cite{Nic} as well as in the proof of the coarea formula \cite{Sak}. As $\Ee$ and $F$ are smooth, this vector field is smooth away from the set $\cri$ of critical points. At the critical points $k^*_\pm$ with extremal energy $\Ee(k^*_\pm)=E_\pm$,  the function $k\mapsto F(\Ee(k^*_\pm+k))$ vanishes linearly by the assumption on $F$ and hence the vector field has a source or a sink there. At all other critical points with critical values lying inside the band $[E_-,E_+]$, the vector field $\hX$ has a singularity which has to be dealt with below. Let $\theta_b:\TM^d\setminus\cri\to\TM^d$ be the flow of $\hX$, that is, $\partial_b\theta_b=\hX\circ\theta_b$ and $\theta_0=\mbox{\rm id}$. The somewhat unconventional choice of $b$ as notation for the time parameter is due to its interpretation as rescaled energy variable below, which is dual to the spectral parameter $a$ of the dilation operator $A$. The flow $\theta_b$ is not complete because an orbit can reach one of the critical points with indefinite signature in a finite time. Choosing orbits which stay away from these critical points or times which are sufficiently small, one can calculate the flow of energy along the orbits. By the definition of the vector field $\hX$,
$$
\partial_b \,\Ee(\theta_b(k))\;=\;F(\Ee(\theta_b(k)))
\;.
$$ 
This equation shows that the flow $\theta_b$ maps constant energy surfaces to constant energy surfaces. Moreover, the energy flow is governed by a simple ordinary differential equation of first order which can be integrated. Choosing some reference energy $E_r\in(E_-,E_+)$, it leads to the following invertible function

\begin{equation}
\label{scatt08.eq-energychange}
f(E)\;=\;
 \int^E_{E_r}\frac{de}{F(e)}\,.
\end{equation}

\noindent Then $b=f(\Ee(\theta_b(k)))-f(\Ee(k))$ and

\begin{equation}
\label{scatt08.eq-Hevolv}
\Ee(\theta_b(k))\;=\;
 f^{-1}\bigl(b+f(\Ee(k))\bigr)\,.
\end{equation}

\noindent If $F$ is given by equation~\eqref{scatt08.eq-Fchoice} and if $E_r=(E_++E_-)/2$, it gives

\begin{equation}
\label{scatt08.eq-Fchoiceconclusion}
f(E)\;=\;
  \frac{1}{2}\,
   \ln\left(\frac{E-E_-}{E_+-E}\right)\,,
\hspace{1cm}
     f^{-1}(b)\;=\;E_r+\Delta\,\tanh(b)\;,
  \hspace{1cm}
     F(f^{-1}(b))\;=\;\frac{\Delta}{\cosh^{2}(b)}\,,
\end{equation}

\noindent where $\Delta=(E_+-E_-)/2$. By restricting $\theta_b$ to an adequate subset of $\TM^d$, a complete flow can be constructed. Let $\criM$ be the union of $\cri$ and of the set of points reaching one of the critical points $k^*\in\cri$ in finite time (either positive or negative). It is important to remark that, under this flow, almost all points reach the maximum and the minimum eventually, but it takes an infinite time to do so. Therefore the finite time condition is a strong constraint. In fact, $\criM$ is the union of $\cri$ and the stable and unstable manifolds of all critical points of indefinite signature. 

\begin{proposi}
\label{scatt08.prop-complete} 
The set $\criM$ is compact and has zero Lebesgue measure. The flow $\theta_b:\TM^d\setminus\criM\to\TM^d\setminus\criM$ is defined for all $b\in\RM$, that is, $\hX$ is complete on  $\TM^d\setminus\criM$. In addition, $\lim_{b\to \pm\infty}\,\theta_b(k)=k^*_\pm$ for all $k\in\TM^d\setminus\criM$. Furthermore, for any open neighborhood $U$ of $\criM$ there exists an open subset $V\subset U$ which contains $\criM\setminus\{k^*_-,k^*_+\}$ and is invariant under the flow $\theta$.
\end{proposi}

\noindent {\bf Sketch of a proof.}  The vector field $\hX$ is gradient-like in the terminology of \cite{Nic} (it is actually a gradient vector field). Hence \cite[Section~2.4]{Nic} shows that $\lim_{b\to \pm\infty}\,\theta_b(k)\in\cri$ and that the stable and unstable manifolds of all critical points of indefinite signature are locally smooth submanifolds of $\TM^d$. For each critical point, the sum of the dimensions of the stable and unstable manifolds is equal to $d$. Along the flow on these submanifolds the energy increases with a finite speed, except in neighborhoods of $k_\pm^*$. Hence either the submanifolds reach another critical point in a finite time (non-generic) or the points $k_\pm^*$ in infinite time. Consequently, the points $k_\pm^*$ compactify the stable and unstable manifolds. As the number of critical points is finite, the set $\criM$ is compact with zero Lebesgue measure. To prove the last statement of the proposition, let $k^*$ be a critical point of indefinite signature. Then let $V(k^*)$ be an open neighborhood of $k^*$ contained in $U$. Then $V=\bigcup_{k^*}\bigcup_{b\in\RM}\theta_b(V(k^*))$ is an open set that is invariant  by the flow. A compactness argument can be used to show that $V\subset U$ by choosing $V(k^*)$ sufficiently small.
\hfill $\Box$

\vspace{.2cm}

The level set of $\Ee$ corresponding to an energy $E\in (E_-,E_+)$ is defined by

$$
\Sigma_E\;=\; 
 \left\{
  k\in\TM^d\setminus\criM\;\Big|\;\Ee(k)\;=\;E
 \right\}\,.
$$

\noindent These level sets will be called the {\em quasi-Fermi surfaces}. This terminology is introduced to stress that $\Sigma_E$ is a strict subset of Fermi surface $\Ee^{-1}(E)$ because the points on the stable and unstable manifolds of all critical points with indefinite signature are excluded. However, the difference is only of measure zero. A reference quasi-Fermi surface will be taken at energy $E_r$ and denoted by $\Sigma=\Sigma_{E_r}$. Because the singularities are excluded, the sets $\Sigma_E$ are smooth open submanifolds of $\TM^d$ of codimension $1$ which, for $d\geq 2$, have several connected components. Now the flow $\theta_b$ maps these connected components diffeomorphically into each other. By the above arguments, for each energy $E$, there is a time $b=f(E)$ such that the flow $\theta_b$ maps the reference quasi-Fermi surface $\Sigma$ diffeomorphically into $\Sigma_E$. Consequently we have:

\begin{proposi}
\label{scatt08.prop-Morsediffeo} For $E\in(E_-,E_+)$, the map
$\theta_{f(E)}:\Sigma\to\Sigma_E$ is a diffeomorphism.
\end{proposi}

For our purposes below, we will also need properties of the divergence of $\hX$. A straightforward calculation gives
$$\mbox{\rm div}(\hX)(k) \;=\;
   F'(\Ee(k)) + F(\Ee(k))\;
   \left(
     \frac{\Delta \Ee(k)}{|\nabla\Ee(k)|^2}
      -2\,
       \frac{\langle\nabla \Ee(k)|\Ee''(k)|\nabla \Ee(k)\rangle}
            {|\nabla\Ee(k)|^4}
   \right)\,,
$$
\noindent where Dirac notation is also used for vectors in $\RM^d$. Near a critical point $k^*$, one has $\nabla \Ee(k^*+k)=\Ee''(k^*)\,k+\Oo(k^2)$,  leading to
$$
\mbox{\rm div}(\hX)(k^*+k) \;=\;
 F'(\Ee(k^*+k))+
  F(\Ee(k^*+k))
   \left(
     \frac{\mbox{\rm Tr}(\Ee''(k^*))}{\langle k|\Ee''(k^*)^2|k\rangle}-
     2\,
     \frac{\langle k|\Ee''(k^*)^3|k\rangle}
      {\langle k|\Ee''(k^*)^2|k\rangle^2}
      +\Oo(|k|^{-1})
   \right)\,.
$$
\noindent Both $F$ and $F'$ are regular, hence $|\mbox{\rm div}(\hX)(k^*+k)| \leq C/|k|^2$ and thus $\mbox{\rm div}(\hX)$ is an integrable function for dimension $d\geq 3$. Furthermore, near the extrema $k^*_\pm$, namely at the band edges, $\Ee(k^*_\pm+k)=\Ee(k^*_\pm)+\frac{1}{2}\langle k|\Ee''(k^*_\pm)|k\rangle+\Oo(|k|^3)$, $F'(\Ee(k^*_\pm+k))=\mp 2+\Oo(|k|^2)$ and $F(\Ee(k^*_\pm+k))=\mp\langle k|\Ee''(k^*_\pm)|k\rangle+\Oo(|k|^3)$. Therefore, setting

$$
g_\pm(k)\;=\;
 \frac{
   \mbox{\rm Tr}(\Ee''(k_\pm^*))\,
    \langle k|\Ee''(k_\pm^*)|k\rangle\,
     \langle k|\Ee''(k_\pm^*)^2|k\rangle
      -2\,\langle k|\Ee''(k_\pm^*)|k\rangle\,
       \langle k|\Ee''(k_\pm^*)^3|k\rangle}
       {\langle k|\Ee''(k_\pm^*)^2|k\rangle^2}\,,
$$

\noindent leads to

\begin{equation}
\label{scatt08.eq-divXvicinity2}
\mbox{\rm div}(\hX)(k^*_\pm+k) \;=\; \mp\,
  \bigl(
    2\,+\,g_\pm(k)+\Oo(|k|)
  \bigr)\,.
\end{equation}

\noindent The functions $g_\pm$ are homogeneous of degree $0$ and can thus be seen as functions on the sphere $\SM^{d-1}$. In dimension $d=1$, one has $g_\pm(k)=-1$. In higher dimension, $g_\pm(k)=d-2$ whenever $\Ee''(k^*_\pm)$ is a multiple of the identity (isotropy of the extrema). Otherwise $g_\pm$ are non-trivial.

 \subsection{Construction of the dilation operator}
 \label{scatt08.ssect-conjOp}

\noindent The aim of this section is the construction of an unbounded conjugate (or dilation) operator $A$ such that $\imath [A,H_0]=F(H_0)$ where $F$ is as above. The basic idea is to implement the flow $\theta_b$ of $\hX$ in $L^2(\TM^d)$ as a strongly continuous group of unitaries. Let $\Dd$ denote the set of smooth functions on $\TM^d$ vanishing in some neighborhood of $\criM$. Since $\criM$ has zero Lebesgue measure and is compact, $\Dd$ is dense in $L^2(\TM^d)$. Furthermore, Proposition~\ref{scatt08.prop-complete} implies that every function in $\Dd$ vanishes on a flow invariant open subset containing $\criM\setminus\{k^*_-,k^*_+\}$. Hence for $\phi\in\Dd$, the following operator can be defined
%
\begin{equation}
\label{scatt08.eq-unitarygroup} 
(\Ww_b\,\phi)(k)\;=\;
 \exp\left(
    \frac{1}{2}
     \int_0^b du\;
      \dive(\hX)(\theta_u(k))
     \right)\;
      \phi(\theta_b(k))\,,
\end{equation}

\noindent because the singularities of $\hX$ {are} not reached, due to the restriction on the support of $\phi$. The unitarity of $\Ww_b$ follows from the change of variables $k\mapsto \theta_b(k)$ and from the Jacobian formula
\begin{equation}
\label{scatt08.eq-detcalc} 
\det(\theta'_b(k))\;=\;
 \exp\left(
  \int_0^b du \;
   \dive(\hX)(\theta_u(k))
     \right)\,.
\end{equation}

\noindent This latter relation follows from integrating $\partial_b \ln \det( \theta'_b(k))=\dive(\hX)( \theta_b(k))$ with the initial condition $\det(\theta'_0)=1$. Furthermore, the group property $\theta_b\circ\theta_u=\theta_{b+u}$ immediately implies $\Ww_b\Ww_u=\Ww_{b+u}$. It can be checked, by a direct calculation, that $\|\Ww_b\phi\|=\|\phi\|$ for $\phi\in\Dd$. In addition, using the Lebesgue dominated convergence theorem, $\lim_{b\to0}\Ww_b\phi=\phi$ for $\phi\in\Dd$. It follows, from a $3\epsilon$ argument, that $\Ww_b$ can be extended as a one-parameter, strongly continuous group of unitary operators on $L^2(\TM^d)$. By Stone's theorem the generator $\hA=\frac{1}{\imath}\partial_b \Ww_b|_{b=0}$ is self-adjoint and $\Ww_b=\exp(\imath b \hA)$. Also \cite[Corollary 3.1.7]{BR} implies that $\Dd$ is a core for $\hA$ because $\Dd$ is left invariant under $\Ww_b$. The derivation of equation (\ref{scatt08.eq-unitarygroup}) leads to

\begin{equation}
\label{scatt08.eq-dilaction} 
\hA\,\phi \;=\;
 \frac{1}{\imath}
  \left(
    \hX(\phi)+\frac{1}{2}\,\dive(\hX)\,\phi
  \right)\;,
\end{equation}

\noindent where $\hX(\phi)=\langle \hX|\nabla\rangle\phi$ is the action of the vector field on the function $\phi\in \Dd$. Note that the multiplicative (zero order) operator $\frac{1}{2}\,\dive(\hX)$ is needed to make the r.h.s. of \eqref{scatt08.eq-dilaction} symmetric w.r.t. the scalar product in $L^2(\TM^d)$. The desired commutator property $\imath [A,H_0]=F(H_0)$ now follows directly from \eqref{scatt08.eq-dilaction} because $\imath [\hA,\hH_0]=\hX(\Ee)=F(\hH_0)$. This can be summarized as follows:

\begin{theo}
\label{theo-dilation} Let $\Ee$ be a Morse function with only one maximum and one local minimum and let $F$ be a smooth function vanishing linearly at the two extremal values $E_-$ and $E_+$ and nowhere else. Let $\Ww_b$ be defined by  {\rm \eqref{scatt08.eq-unitarygroup}} for $\phi\in\Dd$ and with $\hX$ and $\theta_b$ given by {\rm \eqref{scatt08.eq-Xvec}} and its flow. Then $\Ww_b$ is a strongly continuous one-parameter group of unitary operators on $L^2(\TM^d)$. Its generator $\hA=\frac{1}{\imath}\partial_b \Ww_b|_{b=0}$ is self-adjoint with core $\Dd$ and satisfies 
$$
\imath [\hA,\hH_0]\;=\;F(\hH_0)\;,
\qquad
\imath [\hA,f(\hH_0)]\;=\;\one\;.
$$ 
\end{theo}

A few comments conclude this section. The vector field $\hX$ defined by \eqref{scatt08.eq-Xvec} has singularities stemming from critical points of $\Ee$ with indefinite signature (where $F$ does not vanish). This leads to singularities in both the principal and subprincipal symbol of the differential operator $\hA$ as given in \eqref{scatt08.eq-dilaction}. As shown in Section~\ref{scatt08.sec-flow}, the singularity of the principal symbol is integrable in dimension $d\geq 2$ while the subprincipal symbol is integrable for $d\geq 3$. It has been shown above that this does not prevent \eqref{scatt08.eq-dilaction} from defining a self-adjoint operator. There is another similarity between $A=\Ff^*\hA\Ff$ and the usual dilation operator used for the Laplacian in $L^2(\RM^d)$. Let $X_j=\Ff^*\hX_j\Ff$ be the operator on $\ell^2(\ZM^d)$ associated with the $j$th component $\hX_j$ of $\hX$. Also let $Q=(Q_1,\ldots,Q_d)$ be the position operator defined by $Q_j\,\phi(n) =n_j\,\phi(n)$, for $n\in \ZM^d$ and $\phi$ decreasing sufficiently fast. Then the Fourier transform of the r.h.s. of \eqref{scatt08.eq-dilaction} leads to

\begin{equation}
\label{scatt08.eq-dil}
A\;=\;\frac{1}{2}\,
 \sum_{j=1}^d
  \left( X_j\,Q_j+Q_j\,X_j\right)\,.
\end{equation}

\noindent Comparing with the usual dilation operator on $\RM^d$, $X_j$ can be interpreted as the lattice analog of the $j$th component of the momentum operator.

\subsection{Change of variables and REF representation}
 \label{scatt08.ssect-change}

\noindent This section is devoted to the definition and the properties of the {\it rescaled energy and Fermi surface} (REF) representation. The proof of Theorem~\ref{theo-dilation} was mainly based on the change of variables $\theta_b:\TM^d\to\TM^d$ with Jacobian \eqref{scatt08.eq-detcalc}. It will be supplemented by the {\em coarea formula}  (see {\it e.g.} \cite{Sak} for a proof and note that $\criM$ is of zero measure). If $\nu_E$ denotes the Riemannian volume measure on $\Sigma_E$ (induced by the euclidean metric on $\TM^d$),

\begin{equation}
\label{scatt08.eq-varchange2}
\int_{\TM^d} dk\;\phi(k)\;=\;
 \int^{E_+}_{E_-} dE\;
  \int_{\Sigma_E}\nu_E(d\sigma)\;
   \frac{1}{|\nabla\Ee(\sigma)|}\;
    \phi(\sigma)\,.
\end{equation}

\noindent This holds for $\phi$ in the set $\Dd$. For the reference energy surface $\Sigma=\Sigma_{E_r}$, the measure is simply denoted by $\nu=\nu_{E_r}$. The coarea formula leads to the following:

\begin{lemma}
\label{scatt08.lem-varchange} Let $\phi\in\Dd$. Then its integral can be written in the following three equivalent ways:

\begin{eqnarray}
\int_{\TM^d} dk\;\phi(k)  & = & 
 \int_\RM db
  \int_{\Sigma}\nu(d\sigma)\;
   \Big|\det(\theta'_b|_{T_\sigma\Sigma})\Big|\;
    \Big|\hX(\theta_b(\sigma))\Big|\;
     \phi\left\{\theta_b(\sigma)\right\}\,,
      \label{scatt08.eq-varchange3a}\\
& = & 
\int_\RM db
 \int_{\Sigma}\nu(d\sigma)\;
  \exp\left(
       \int_0^b du \;\dive(\hX)(\theta_u(\sigma))
        \right)\;
   \Big|\hX(\sigma)\Big|\;
    \phi\left(\theta_b(\sigma)\right)\,,
     \label{scatt08.eq-varchange3} \\
& = & 
\int^{E_+}_{E_-}dE
 \int_{\Sigma}\nu(d\sigma)\;
   \frac{
          |\det(\theta'_{f(E)}|_{T_\sigma\Sigma})|
        }{|\nabla \Ee({\theta}_{f(E)}(\sigma))|}\;
     \phi\left(\theta_{f(E)}(\sigma)\right)\,,
     \label{scatt08.eq-varchange4} 
\end{eqnarray}
where $\theta'_b|_{T_\sigma\Sigma}$ denotes the derivative of $\theta_b$ restricted to the tangent space of $\Sigma$ at $\sigma$ {\rm (}so that this is a $(d-1)\times (d-1)$ matrix{\rm )}.
\end{lemma}

\noindent  {\bf Proof:} Starting from the coarea formula (\ref{scatt08.eq-varchange2}), the substitution $b=f(E)$ given in (\ref{scatt08.eq-energychange}) and the diffeomorphism of Proposition~\ref{scatt08.prop-Morsediffeo} will be used in the following change of variables:

\begin{eqnarray*}
\int_{\TM^d} dk\;\phi(k) & = & 
 \int_\RM db
  \int_{\Sigma_{f^{-1}(b)}}
   \nu_{f^{-1}(b)}(d\sigma)\;
    \frac{F(f^{-1}(b))}{|\nabla \Ee(\sigma)|}\;
     \phi(\sigma) \\ 
& = & 
 \int_\RM db\;
  \int_{\Sigma}\nu(d\sigma)\;
   \Big|\det(\theta'_b|_{T_\sigma\Sigma})\Big|\;
    \frac{F(\Ee(\theta_b(\sigma)))}{|\nabla \Ee(\theta_b(\sigma))|}\;
     \phi(\theta_b(\sigma)) \,.
\end{eqnarray*}

\noindent In the second equality, the identity $F(f^{-1}(b))=F(\Ee(\sigma))$ for $\sigma\in\Sigma_{f^{-1}(b)}$ was used. Replacing the definition of $\hX$ already shows \eqref{scatt08.eq-varchange3a} as well as \eqref{scatt08.eq-varchange4}. Next $\theta_b'$ can be decomposed as $\theta'_b|_{T_\sigma\TM^d}=\theta'_b|_{T_\sigma\Sigma}\oplus \theta'_b|_{(T_\sigma\Sigma)^\perp}$ implying 
\begin{equation}
\label{scatt08.eq-factorize}
|\det(\theta'_b|_{T_\sigma\TM^d})|\;=\;|\det(\theta'_b|_{T_\sigma\Sigma})|\,
|\theta'_b|_{(T_\sigma\Sigma)^\perp}|
\;.
\end{equation}
In order to compute $\theta'_b|_{(T_\sigma\Sigma)^\perp}$ it should be remarked that the derivative of the equation $\partial_b\theta_b=\hX\circ\theta_b$ is $\partial_b\theta'_b=\hX'\circ\theta_b\,\theta'_b$, leading to $\theta'_b(\hX(\sigma))=\hX(\theta_b(\sigma))$. As the one-dimensional space ${(T_\sigma\Sigma)^\perp}$ is spanned by $\hX(\sigma)$, it follows that 
$$
|\theta'_b|_{(T_\sigma\Sigma)^\perp}|
\;=\;
\Bigl|\theta'_b\Bigl( \frac{\hX(\sigma)}{|\hX(\sigma)|}\Bigr)\Bigr|
\;=\;
\frac{|\hX(\theta_b(\sigma))|}{|\hX(\sigma)|}
\;.
$$
Consequently

\begin{equation}
\label{scatt08.eq-Jacobians}
\Big|\det(\theta'_b|_{T_\sigma\Sigma})\Big|\;
 \Big|\hX(\theta_b(\sigma))\Big|\;=\;
  \exp\left(
    \int_0^b du \;\dive(\hX)( \theta_u(\sigma))
      \right)\;
   \Big|\hX(\sigma)\Big|\,.
\end{equation}

\noindent Replacing this in \eqref{scatt08.eq-varchange3a} proves \eqref{scatt08.eq-varchange3}.
\hfill $\Box$

\vspace{.2cm}

The following notation will be useful

\begin{equation}
\label{scatt08.eq-JacFac}
d_{b}(\sigma)\;=\;
 \Big|\det(\theta'_b|_{T_\sigma\Sigma})\Big|^{\frac{1}{2}}\;
  \Big|\hX(\theta_b(\sigma))\Big|^{\frac{1}{2}}\;=\;
   \exp\left(
      \frac{1}{2}\;\int_0^{b} du \;\dive(\hX)(\theta_u(\sigma))
       \right)\;
    \Big|\hX(\sigma)\Big|^{\frac{1}{2}}\,.
\end{equation}

\noindent From (\ref{scatt08.eq-varchange3}), it follows that the map
$\Uu$ defined on $\Dd$ by

\begin{equation}
\label{scatt08.eq-Udef}
(\Uu \phi)_{b}(\sigma)\;=\;
 d_{b}(\sigma)\;
 \phi(\theta_{b}(\sigma))\,,
\hspace{2cm}
  \phi\in \Dd\subset L^2(\TM^d)\,,
\end{equation}

\noindent extends to a unitary from $L^2(\TM^d)$ to $L^2(\RM)\otimes L^2(\Sigma,\nu)$. The variable $b$ is the rescaled energy difference w.r.t. the reference quasi-Fermi surface $\Sigma$. Expressing this in terms of $\Ww_b$ (see equation~(\ref{scatt08.eq-unitarygroup})), leads to $(\Uu \phi)_{b}(\sigma)=|\hX(\sigma)|^{\frac{1}{2}}(\Ww_{b}\phi)(\sigma)$. The inverse, acting on $\psi\in L^2(\RM) \otimes L^2(\Sigma,\nu)$, is given by

$$(\Uu^* \psi)(k)\;=\;
 d_{b}(\theta_{-b}(k))^{-1}\;
 \psi_{b}(\theta_{-b}(k))\,,
\hspace{2cm}
  b=f(\Ee(k))\,.
$$

\noindent Note that $\Uu$ is unitary. The expression $\widetilde{H}_0=\Uu\widehat{H}_0\Uu^*=\Uu\Ff H_0\Ff^*\Uu^*$ will be called the REF representation of $H_0$. Any operator in the REF representation will carry a tilde. The operator $(\widetilde{B}\psi)_{b}=b\psi_{b}$ is the {\em rescaled energy}. Its conjugate operator clearly is $\widetilde{A}$ with $(\widetilde{A}\psi)_{b}=\frac{1}{\imath}\partial_{b}\psi_{b}$. Both of these operators are unbounded and have the standard self-adjoint domains. The following result states that these notations are consistent with the above.

\begin{proposi}
\label{scatt08.prop-rep}
The following relations hold

$$
\Uu\,\hH_0\,\Uu^*\;=\;
 f^{-1}(\widetilde{B})\otimes {\bf 1}_{\Sigma}\,,
\hspace{2cm}
  \Uu\,f(\hH_0)\,\Uu^*\;=\;\widetilde{B}\,,
\hspace{2cm}
   \Uu\,\hA\,\Uu^*\;=\;\widetilde{A}\,.
$$
\end{proposi}

\noindent  {\bf Proof:} The only point to be checked is how the commutation relations of $H_0$ and $A$, as proved in Theorem~\ref{theo-dilation}, are implemented under $\Uu$. The first identity results from (\ref{scatt08.eq-Udef}) and

$$
f^{-1}(b)\phi(\theta_{b}(\sigma))\;=\;
 \Ee(\theta_{b}(\sigma))
  \phi(\theta_{b}(\sigma))\;=\;
   \Big(\hH_0\phi\Big)(\theta_{b}(\sigma))\,.
$$

\noindent The second formula is obtained from the first one through (unbounded) functional calculus. The third one follows from

$$
(\widetilde{A}\otimes{\bf 1}\,\Uu\phi)_{b}(\sigma)\;=\;
\frac{1}{\imath}\,
 \partial_{b}\, 
  (|\hX|^{\frac{1}{2}}\,
   e^{\imath b\hA}\phi)(\sigma)\;=\;
    ({\frac{1}{2}}\,
     e^{\imath b\hA}\,\hA\, \hA\, \phi)(\sigma)\;=\;
      (\Uu \hA\,\phi)_{b}(\sigma)\,,
$$

\noindent where $\Uu$ is expressed in terms of the unitary group $\Ww_b=e^{\imath b\hA}$ up to the factor $|\hX(\sigma)|^{\frac{1}{2}}$ which does not depend on $b$.
\hfill $\Box$

\vspace{.2cm}

It is worth comparing the previous construction to the usual one used in scattering theory on $\RM^d$, where $H_0=-\Delta$ is the Laplacian acting on $L^2(\RM^d)$. Then, the (unitary) Fourier transform $\Ff:L^2(\RM^d)\mapsto L^2(\RM^d)$ diagonalizes $H_0$, that is, $\Ff H_0\Ff^*$ is the operator of multiplication by $\Ee(k)=k^2$. This function has only one critical point at $k^*_-=0$ corresponding to the minimum of energy $E_-=0$.  The vector field $\hX$ is defined as in \eqref{scatt08.eq-Xvec}, now with $k\in\RM^d$. Let the reference energy be $E_r= 1$ so that the (quasi-) Fermi surface $\Sigma$ is the unit sphere $\SM^{d-1}$. Furthermore let $F(E)=2E$, which vanishes at the only critical value. Then $\hX(k)=k$ and $f(E)=\int^E_1\frac{de}{2e}=\frac{1}{2}\ln(E)$. The flow is $\theta_b(\sigma)=e^b\sigma$. As ${\rm div}(\hX)=d$, it follows that $d_b(\sigma)=e^{\frac{1}{2} db}$. Therefore the unitary transformation $\Uu:L^2(\RM^d)\to L^2(\RM)\otimes L^2(\SM^{d-1})$ to the REF representation is given by

$$
(\Uu\phi)_b(\sigma)=
 e^{\frac{1}{2}db}\,\phi(e^b\sigma)\,.
$$

\noindent This transformation is discussed and used, {\it e.g.}, by Jensen \cite{Jen} and also \cite{KR3}.

\vspace{.2cm}

 \subsection{EF representation}
 \label{scatt08.sec-EFrep}

\noindent Another natural useful representation is the {\it energy and Fermi surface} (EF) representation. A local version of this representation is used in the paper by Birman and Yafaev \cite{BY}. It is associated with the unitary map $\Vv:L^2(\TM^d)\to L^2([E_-,E_+])\otimes L^2(\Sigma,\nu)$ defined on $\Dd$ by

$$
(\Vv\phi)_E(\sigma)\;=\;
 \frac{
     |\det(\theta'_{f(E)}|_{T_\sigma\Sigma})|^{\frac{1}{2}}
      }{
     |\nabla \Ee({\theta}_{f(E)}(\sigma))|^{\frac{1}{2}}
      }\;
   \phi({\theta}_{f(E)}(\sigma))\,,
\hspace{2cm}
 \phi\in\Dd\,.
$$

\noindent The unitarity follows directly from \eqref{scatt08.eq-varchange4}. It is related to the unitary operator $\Uu$ as follows

\begin{equation}
\label{scatt08.eq-REFlinkEF}
(\Vv\phi)_E(\sigma)\;=\;
 \frac{1}{F(E)^{\frac{1}{2}}}\;
  \frac{1}{|\hX(\sigma)|^{\frac{1}{2}}}\;
    (\Uu\phi)_{f(E)}(\sigma)\,.
\end{equation}

\noindent The EF representation of an operator on $L^2(\TM^d)$ is then obtained by conjugation with $\Vv$. It will carry a circle instead of a tilde, such as $\cH_0=\Vv\widehat{H}_0\Vv^*$, $\cA=\Vv\widehat{A}\,\Vv^*$ and so on. Any operator that is a direct integral in the REF representation is also a direct integral in the EF representation. The first example of this type is the Hamiltonian $H_0$ itself: 

$$
(\cH_0\phi)_E(\sigma)\;=\;E\,\phi_E(\sigma)\,.
$$

\noindent More generally, given any fibered operator $\widetilde{O}=\int^{\oplus} db\,\widetilde{O}_b$ in the REF representation, its EF representation is given by $\cO=\int^{\oplus} dE\,\cO_E$ with $\cO_E=\widetilde{O}_{f(E)}$. Another example will be the scattering matrix below. The dilation operator in the EF representation can be easily deduced from \eqref{scatt08.eq-REFlinkEF}:

$$
(\cA\phi)_E(\sigma) \;=\; F(E)\,
  \frac{1}{\imath}\,\partial_E
   \phi_E(\sigma)+
    \frac{1}{2\imath}\;F'(E)\,\phi_E(\sigma)\,,
$$

\noindent where $\phi$ is in the domain of $\cA$, in particular, its derivative is square integrable and $\phi$ vanishes at the boundaries of $[E_-,E_+]$. 

\vspace{.2cm}

 \subsection{Boundary values of the free resolvent}
 \label{scatt08.ssect-LAP}

\noindent Let $\Lambda\subset\ZM^d$ be a finite set. Eventually, $\Lambda$ will be the support of the perturbation. Associated with $\Lambda$ is the subspace $\ell^2(\Lambda)=\CM^{|\Lambda|}$. Let $\piso^\ast:\CM^{|\Lambda|}\to \ell^2(\ZM^d)$ be the canonical injection obtained by extending elements of $\ell^2(\ZM^d)$ by zero outside $\Lambda$. It is a partial isometry such that $\piso^*\piso$ is the $|\Lambda|$-dimensional projection in $\ell^2(\ZM^d)$ onto the subspace of elements supported by $\Lambda$, while $\piso\,\piso^*=\one_{\CM^{|\Lambda|}}$. The finite volume Green matrix is defined by:

$$
G^\piso_0(z)\;=\;\piso \;(z-H_0)^{-1}\;\piso^*\;.
$$

\noindent This is a matrix of size $|\Lambda|\times |\Lambda|$. If $\Lambda=\{0\}$ it will be called the Green function. An important basic fact about the Green matrix is its Herglotz property, that is, $-\Im m\,G_0^\piso(z)=\imath(G_0^\piso(z)-G_0^\piso(z)^*)/2>0$ for $\Im m(z)>0$. This implies, in particular, that $G_0^\piso(z)$ is invertible for $\Im m(z)\neq 0$. The boundary values of $G_0^\piso(z)$ on the real axis will be analyzed in this section. Gieseker, Kn\"orrer and Trubowitz \cite{GKT} studied thoroughly the Fermi surfaces for dimensions $d\geq 2$ and for generic periodic  potentials. They showed that it is an algebraic variety and constructed a compactification. They also investigated the nature of the van Hove singularities, which, in two dimension produce a logarithmic divergence of the density of states, namely the diagonal elements of $\Im m\,G_0^\piso(E-\imath 0)$. For $d=2$ and the discrete Laplacian, these limit behaviors can also be read off the explicit formulas for the Green function given in \cite{Eco}, but for $d\geq 3$ only numerical results and toy models seem to be known.

\begin{proposi}
\label{scatt08.prop-Green3d} 
Let $d\geq 3$ and let $\Ee$ be analytic. The weak limits $G^\piso_0(E\pm\imath 0)=\lim_{\epsilon\downarrow 0} G^\piso_0(E\pm\imath \epsilon)$ exist. Furthermore:

\vspace{.1cm}

\noindent {\rm (i)} Away from the critical values of $\Ee$, the map $E\in\RM\mapsto G^\piso_0(E\pm\imath 0)$ is real analytic. At the critical 

points it is H\"older continuous.

\vspace{.1cm}

\noindent {\rm (ii)} $\Im m\,G^\piso_0(E-\imath 0)= - \Im m\,G^\piso_0(E+\imath 0)$ vanishes on $(-\infty,E_-]\cup[E_+,\infty)$. It is a positive matrix 
with 

nonzero diagonal entries on $(E_-,E_+)$.

\vspace{.1cm}

\noindent {\rm (iii)} The map $E\in\RM\mapsto \Re e\,G^\piso_0(E)$ is negative and decreasing on $(-\infty,E_-]$ and positive and 
decreasing 

on $[E_+,\infty)$. Furthermore, $G^\piso_0(\pm\infty)=0$.

\vspace{.1cm}

\noindent {\rm (iv)} For $E\in[E_-,E_+]$ close to $E_\pm$,
$$
\Im m\,G^\piso_0(E-\imath 0)
\;=\;
D_\pm\;|E-E_\pm|^{\frac{d}{2}-1}\;M^\piso_\pm
\;+\;
\Oo(|E-E_\pm|^{\frac{d}{2}})
\;,
$$

where $M^\piso_\pm=|v^\piso_\pm\rangle\langle v^\piso_\pm|$ is the projection on the vector $v^\piso_\pm=(|\Lambda|^{-\frac{1}{2}}\,e^{\imath n\cdot k^*_\pm})_{n\in\Lambda}\in\CM^{|\Lambda|}$ and

$$D_\pm \;=\; 
     \frac{2^{\frac{d}{2}-1}\pi\, |\Lambda|\,|\SM^{d-1}|}{(2\pi)^d}\;|\det(\Ee''(k_\pm^*))|^{\frac{1}{2}}\; .
$$

\vspace{.1cm}

\noindent {\rm (v)} There are matrices $N^\piso_\pm<0$ such that

$$
\Re e \,G^\piso_0(E)\;=\;G^\piso_0(E_\pm)\;+\;
\left\{
\begin{array}{cc}
\Oo(E-E_\pm)& d=3\;,
\\
D_\pm |E-E_\pm|\ln\left(\frac{1}{|E-E_\pm|}\right)\,M^\piso_\pm+ \Oo(E-E_\pm)
& d=4\;,
\\
(E-E_\pm)\,N^\piso_\pm+o(E-E_\pm)
& d\geq 5\;.
\end{array}
\right.
$$
\end{proposi}

\noindent  {\bf Proof:} The proofs given below are detailed extensions of the work of van Hove \cite{VH}. For $m,n\in\Lambda$, the matrix elements of $G^\piso_0(z)$ are given by

$$\langle m|G^\piso_0(z)|n\rangle \;=\;
   \langle m|(z-H_0)^{-1}|n\rangle \;=\;
    \int_{\TM^d} \frac{d^dk}{(2\pi)^d}\;
     \frac{e^{\imath (n-m)\cdot k}}{z- \Ee(k)}\,.
$$
\noindent {\bf (i) Outside the critical values: } By construction the matrix $G^\piso_0(z)$ is holomorphic for $z\notin \sigma(H_0)$. In particular, since the spectrum of $H_0$ is the interval $\sigma(H_0)=[E_-,E_+]$, it follows that the map $E\in\RM\setminus[E_-,E_+]\mapsto G_0^\piso(E)$ is real analytic and converges to zero at $\pm \infty$. Moreover, its derivative is negative. In particular, if the limit of this matrix exists at $E_\pm$, this limit is a negative matrix at $E_-$ and a positive matrix at $E_+$. Now, since $\Ee$ is analytic, it follows that it has only a finite number of critical points and it admits a holomorphic continuation in $(\TM+\imath \RM)^d$ in a small neighborhood of the form $B_\eta = \{k+\imath \kappa \in (\TM+\imath \RM)^d\,|\, \max_{1\leq i\leq d}{|\kappa_i|} <\eta \}$.  It follows that, for $\epsilon>0$ small enough, the manifold defined as the set $\TM_\epsilon^d = \{k+\imath \epsilon\nabla \Ee(k)\,|\, k\in\TM^d\}$ is entirely contained in $B_\eta$. Using the Cauchy formula, it follows that 

$$\langle m|G^\piso_0(z)|n\rangle \;=\;
    \int_{\TM_\epsilon^d} \frac{d^dk'}{(2\pi)^d}\;
     \frac{e^{\imath (n-m)\cdot k'}}{z- \Ee(k')}\,.
$$

\noindent Since $k'\in \TM_\epsilon^d$, it follows that $k'=k+\imath \epsilon\nabla\Ee(k)$ for some $k\in\TM^d$, so that, using a Taylor expansion,

$$\Im m\,\Ee(k') \;=\; \epsilon\,|\nabla\Ee(k)|^2\;+\; \Oo(\epsilon^2)\;.
$$

\noindent Consequently, if $E\in[E_-,E_+]\setminus \Ee(\cri)$ is not a critical value, there is $\rho >0$ such that, if $|z-E|<\rho$, the distance of $\dist\{z, \Ee(\TM_\epsilon^d)\}>0$ does not vanish. In particular, $G^\piso_0(z)$ extends as a holomorphic function of $z$ from $\Im m(z) <0$ to a neighborhood of $E$. In particular, the boundary value $G^\piso_0(E-\imath 0)$ is analytic in $E$ in $[E_-,E_+]\setminus \Ee(\cri)$. A similar argument applies to $G^\piso_0(E+\imath 0)$. 

\vspace{.1cm}

\noindent {\bf (ii) Partitioning:} For any $k^*\in\cri$, let $B_\delta(k^*)$ be the open ball centered at $k^*$ of radius $\delta>0$. Let also $\overline{B}_{\delta/2}(k^*)$ be the closed ball also centered at $k^*$ of radius $\delta/2$. Let $U_{\mbox{\rm\tiny reg}}$ be the open set obtained by removing from $\TM^d$ the union of the balls $\overline{B}_{\delta/2}(k^*)$, $k^*\in\cri$. It follows that the family $\{U_{\mbox{\rm\tiny reg}}\}\cup\{B_\delta(k^*)\,|\, k^*\in\cri\}$ is a finite open cover of $\TM^d$. Let then $\{\chi_{\mbox{\rm\tiny reg}}\}\cup \{\chi_{k^*}\,|\, k^*\in\cri\}$ be a smooth partition of unity associated with this open cover. The previous integral can be decomposed into a sum
\begin{equation}
\label{scatt08.eq-part}
\langle m|(z-H_0)^{-1}|n\rangle \;=\;
   G_{\mbox{\rm\tiny reg}}(z) + \sum_{k^*\in\cri} G_{k^*}(z)\,,
\hspace{1cm}
   G_{k^*}(z) \;=\; \int_{B_\delta(k^*)} \frac{d^dk}{(2\pi)^d}\;\chi_{k^*}(k)\;
     \frac{e^{\imath (n-m)\cdot k}}{z- \Ee(k)}\,.
\end{equation}
\noindent The contribution $G_{\mbox{\rm\tiny reg}}$ is regular because the integral vanishes around all critical points. Using the coarea formula and the results of Appendix~\ref{sec-Borel}, it follows that $G_{\mbox{\rm\tiny reg}}$ is holomorphic in the complement of the spectrum of $H_0$ and its boundary values are smooth everywhere on the real line.

\vspace{.1cm}

\noindent {\bf (iii) Non extremal critical points:} The boundary values of the $G_{k^*}$'s, however, may not be smooth because of the contribution of the critical point. Let $k^*$ be one of the critical points of signature $d=(d_+,d_-)$ with $d_\pm \neq 0$ and in the following $G_\ast= G_{k^*}$ will denote its contribution to the previous decomposition. If $\delta$ is small enough, the Morse lemma \cite{Nic} implies that there exists a neighborhood $U$ of $k^*$ containing $B_\delta(k_*)$ and a diffeomorphism $\varphi:B_\delta(0)\to U$ such that $\varphi(0)=k^*$ and $\Ee_\varphi=\Ee\circ \varphi$ is quadratic:
$$
\Ee_\varphi(k)
\;=\;
 E_\ast \,+\, \frac{1}{2}\,\sum_{i=1}^{d_+} k_i^2 \,-\,\frac{1}{2}\,\sum_{j=d_++1}^{d} k_j^2
 \;,
 $$
for $\|k\| < \delta$ and where $E_\ast=\Ee(k^*)$. This diffeomorphism has a Jacobian matrix $J= \varphi'(0)$ satisfying $J\mbox{\rm diag}(\one_{d_+},-\one_{d_-})J^*= \Ee''(k^*)^{-1}$. In particular, the Jacobi determinant of $\varphi$ stays close to $|\det(\Ee''(k^*))|^{-1/2}$ over the neighborhood $U$ and is a smooth function. It follows that the integral defining $G_\ast$ is given by

$$G_\ast(z) \;=\;
   \int_{\|k\|<\delta} \frac{d^dk}{(2\pi)^d}\;
    \bigl|\det(\varphi'(k))\bigr|\,\chi_{k^*} (\varphi(k))\;
     \frac{e^{\imath (n-m)\cdot \varphi(k)}}{z- \Ee_\varphi(k)}\,.
$$

\noindent It will be convenient to use the following polar variables

$$k_i \;=\; r_+ \omega_+\, \hspace{.5cm} \mbox{\rm if}\;\; 1\leq i\leq d_+\;,
  \hspace{2cm}
    k_j \;=\; r_- \omega_-\, \hspace{.5cm} \mbox{\rm if}\;\; d_+< j\leq d\;,
$$

\noindent were $r_\pm \geq 0$ are the radial variables and $\omega_\pm \in \SM^{d_\pm -1}$ the angular ones. It follows that

$$G_\ast(z) \;=\;
   \int_{r_+^2+r_-^2<\delta^2} 
    \frac{r_+^{d_+-1}dr_+\, r_-^{d_--1}dr_-}{(2\pi)^d}\;
     \frac{F(r_+,r_-)}{z- E_\ast - \frac{1}{2}(r_+^2-r_-^2)}\;,
$$

\noindent where $F$ is a smooth function with support inside the disk $r_+^2+r_-^2<\delta^2$ given by

$$F(r_+,r_-) \;=\;
   \int_{\SM^{d_+ -1}\times \SM^{d_- -1}} d\omega_+ \,d\omega_-\;
    \bigl|\det(\varphi'(k))\bigr|\,\chi_{k^*}  (\varphi(k))\;
     e^{\imath (n-m)\cdot \varphi(k)}\,.
$$

\noindent Equivalently $G_\ast$ can be expressed as 

$$G_\ast(z) \;=\;
   \int_{\RM} \frac{\rho(e) de}{z-E_\ast -e}\,,
$$

\noindent where $\rho$ is defined by

$$\rho(e) \;=\;
     \int_{r_+^2+r_-^2<\delta^2} 
    \frac{r_+^{d_+-1}dr_+\, r_-^{d_--1}dr_-}{(2\pi)^d}\;
     F(r_+,r_-)\; 
      \delta\left(
        \frac{r_+^2-r_-^2}{2}-e
            \right)\;.
$$

\noindent If $e >0$, the usual rule followed by the Dirac distribution $\delta$ leads to
\begin{equation}
\label{scatt08.eq-rho}
\rho(e) \;=\;
  \int_0^\delta \frac{dr}{(2\pi)^d}\;\,
   r^{d_--1} (e+r^2)^{(d_+-2)/2} \,\;
    F(\sqrt{r^2+e},r)
\end{equation}
\noindent For $e <0$, a similar formula holds by exchanging $d_+$ with $d_-$ and $F(r,r')$ with $F_s(r,r')=F(r',r)$. The previous expression shows that, if $d_-\geq 2$, the Lebesgue dominated convergence theorem implies that the limits $\rho(\pm0)$ exist and are equal. In particular, $\rho$ is continuous at $e=0$. Moreover, since $d\geq 3$, if $d_+=1$, then $d_-=d-1\geq 2$. Then $r^{d_--1} (e+r^2)^{(d_+-2)/2}= r^{d-2} (e+r^2)^{-1/2}\leq r^{d-5/2}$ showing that, again, $\rho$ is continuous at $e=0$. 

\vspace{.1cm}

Equation~(\ref{scatt08.eq-rho}) also shows that $\rho$ is differentiable for $e \neq 0$. Moreover, its derivative is given by the sum of two terms $\rho'_1+\rho'_2$ with

$$\rho'_1(e)\;=\; \frac{d_+-2}{2}
   \int_0^\delta \frac{dr}{(2\pi)^d}\;\,
   r^{d_--1} (e+r^2)^{d_+/2-2} \,\;
    F(\sqrt{r^2+e},r)\,,
$$
$$\rho'_2(e)\;=\; \frac{1}{2}
   \int_0^\delta \frac{dr}{(2\pi)^d}\;\,
   r^{d_--1} (e+r^2)^{d_+/2-3/2} \,\;
    \partial_1 F(\sqrt{r^2+e},r)\,.
$$

\noindent The same argument as before shows that, if $d\geq 3$, $\rho'_2$ admits a finite limit as $\pm e\downarrow 0$. However, these two limits may not be equal, if $F\neq F_s$. On the other hand, if $d\geq 5$, $\rho'_1$ also admits limits and the two limits coincide. For $d=3,4$, however, it follows that $d_+ <4$ so that $\rho'_1$ may diverge at $e\rightarrow 0$. Nevertheless, the integrand can be bounded by 

$$r^{d_--1} (e+r^2)^{d_+/2-2} \;\leq\; 
    e^{-\alpha}\, r^{d-5+2\alpha}\;,
$$

\noindent which is integrable if $\alpha >1/2$ for $d=3$ and $\alpha>0$ for $d=4$. Hence in both cases, there is $K>0$ such that

$$\left|\partial_e\rho\right| \;\leq\;
    \frac{K}{e^\alpha}
\hspace{1cm}\Longrightarrow \hspace{1cm}
    |\rho(e)|\;\leq\; \frac{K}{1-\alpha}\,e^{1-\alpha}\,,
$$

\noindent showing that $\rho$ is H\"older continuous at the critical points. Using the Plemelj-Privalov theorem (Lemma~\ref{scatt08.lem-hilb} of Appendix~\ref{sec-Borel}), it follows that the same is true for the boundary values of $G_\ast$.

\vspace{.1cm}

\noindent {\bf (iv) Near the extrema: }The behavior near the maximum or the minimum can be treated similarly so that it is enough to consider only the minimum at $k_-^*$. Again by the Morse lemma, there is a neighborhood $U$ of $k_-^*$ containing $B_\delta(k_-^*)$ and a diffeomorphism $\varphi:B_\delta(0)\to U$ with $\varphi(0)=k_-^*$ and such that $\Ee\circ \varphi(k) = E_-+ (1/2) \sum_{i=1}^d k_i^2$. Introducing the polar coordinates $r= \|k\|$ and $\omega\in\SM^{d-1}$ so that $k=r\omega$, the contribution $G_-(z)=G_{k_-^*}(z)$ is given by the integral

$$G_-(z) \;=\; 
   \int_0^\delta \frac{r^{d-1}dr}{(2\pi)^d} \;
    \frac{e^{\imath(n-m)\cdot k_-^*}\;F(r)}{z-E_- -\frac{1}{2}r^2}\,,
\hspace{1cm}
F(r) \;=\;
  \int_{\SM^{d-1}} d\omega \,
   |\varphi'(r\omega)|\, \chi_-(\varphi(r\omega))\, 
    e^{\imath (n-m)\cdot(\varphi(r\omega)-k_-^*)}\,.
$$

\noindent with $\chi_-$ a smooth function with support in $U(k_-^*)$ which is equal to $1$ on the ball $\|k-k_-^*\|\leq \delta/2$. In particular, $F$ is smooth and bounded in $0<r<\delta$, it vanishes in a neighborhood of $r=\delta$ and all its derivatives have a limit at $r=0$. Consequently, the integration domain can be extended to $[0,\infty)$ without change. At this point two remarks should be made:

\vspace{.1cm}

\noindent (1) $F(0)=|\det(\varphi'(0))|\, |\SM^{d-1}| > 0$ and the Morse lemma shows that $|\det(\varphi'(0))|=\det(\Ee''(k_-^*))^{1/2}$.

\vspace{.1cm}

\noindent (2) The expression $e^{\imath (n-m)\cdot k_-^*}$ is the matrix element $|\Lambda|\langle m| M_-^\piso |n\rangle$ of the projection matrix $M_-^\piso$. 

\vspace{.1cm}

\noindent The change of variable $e=r^2/2$ yields

\begin{equation}
\label{eq-G-}
G_-(z) \;=\; \frac{e^{\imath(n-m)\cdot k_-^*}}{(2\pi)^d}
   \int_0^\infty de\;\frac{(2e)^{\frac{d}{2}-1}F(\sqrt{2e})}{z-E_- -e}\;.
\end{equation}

\noindent Since $d\geq 3$, the function $e\in [0,\infty)\mapsto e^{\frac{d}{2}-1}F(\sqrt{2e})$ is continuous and vanishes at $e=0$ like $e^{d/2-1}$. Hence it can be continued as a H\"older continuous function on the entire real line with support in $[0,\frac{\delta^2}{2})$. Consequently, thanks to the Lemma~\ref{scatt08.lem-hilb}, $G_-(E\pm \imath 0)$ is also continuous w.r.t. $E$. In particular, it has a finite value at $E=E_-$. Since the other contributions to $G_0^\piso$ are regular near $E_-$, $G_0^\piso(E\pm\imath 0)$ is also a H\"older continuous function of $E$ near $E=E_-$. 

\vspace{.2cm}

All contributions in equation~(\ref{scatt08.eq-part}) other than $G_-$ being analytic near $E=E_-$, it follows that any singularity of $G_0^\piso(E\pm\imath 0)$ near $E=E_-$ is coming from $G_-$. In addition, the imaginary part of the other contributions to $G_0^\piso$ vanishes on the real axis at $E=E_-$ since $H_0$ is selfadjoint. Hence the only contribution to its imaginary part is coming from $G_-$. Thanks to the Lemma~\ref{scatt08.lem-hilb} it follows from \eqref{eq-G-} that this imaginary part is exactly 

$$\Im m \;G_0^\piso(E\pm\imath 0) \;=\;\mp\,\pi\,
   (2(E-E_-))^{\frac{d}{2}-1} 
    |\Lambda|\,M_-^\piso\;
     \frac{\det(\Ee''(k_-^*))^{\frac{1}{2}}\, |\SM^{d-1}|}{(2\pi)^d}\; 
      +\;\Oo \bigl((E-E_-)^{\frac{1}{2}}\bigr) \;.
$$

\noindent On the other hand, the real part can be estimated by considering the subdominant contribution of $G_-$ given by the difference

\begin{equation}
\label{scatt08.eq-gminusderiv}
G_-(E\pm\imath 0)- G_-(E_-) \;=\; 2(E-E_-)\;
   \frac{e^{\imath(n-m)\cdot k_-^*}}{(2\pi)^d}
   \int_0^\infty de\;\frac{(2e)^{\frac{d}{2}-2}F(\sqrt{2e})}{E-E_- -e \pm\imath 0 }\;.
\end{equation}

\noindent The same argument as before shows that the integral defines a continuous function of $E$ on the real line if $d\geq 5$. Consequently, $E\in\RM\mapsto G_0^\piso(E\pm\imath 0)$ is continuously differentiable in a  small neighborhood of $E=E_-$. Since the derivative is negative outside of the spectrum of $H_0$, it follows that the claim (v) of the Proposition~\ref{scatt08.prop-Green3d} holds for $d\geq 5$.

\vspace{.2cm}

For $d=4$, equation~(\ref{scatt08.eq-gminusderiv}) shows that Lemma~\ref{scatt08.lem-hilbplus} applies. For indeed, the function $e\in[0,\infty|\mapsto F(\sqrt{2e}) \in\CM$ is smooth because the Taylor expansion of $F(r)$ near the origin contains only even terms and $F(0)\neq 0$. Consequently 

$$G_-(E\pm\imath 0)- G_-(E_-) \;=\; 
   (E-E_-)\ln(|E-E_-|)\;
    \frac{e^{\imath(n-m)\cdot k_-^*}F(0)}{(2\pi)^{d-1}} \;+\; \Oo(|E-E_-|)\;.
$$

\noindent Since all other contributions to the real part of $G_0^\piso(E\pm\imath 0)$ are regular at $E=E_-$, it follows that 

$$\Re e\; G_0^\piso(E\pm\imath 0) \,=\,G_0^\piso(E_-) + 
   |\Lambda|M_-^\piso\;
   \frac{\det(\Ee''(k_-^*))^{1/2}\, |\SM^{d-1}|}{(2\pi)^{d-1}}\,
    (E-E_-)\ln(|E-E_-|)\,+\, \Oo(|E-E_-|)\,.
$$

\vspace{.1cm}

\noindent At last, for $d=3$, returning to the variable $r=\sqrt{2e}$, equation~(\ref{scatt08.eq-gminusderiv}) becomes

$$G_-(E\pm\imath 0)- G_-(E_-) \;=\; 2(E-E_-)\;
   \frac{e^{\imath(n-m)\cdot k_-^*}}{(2\pi)^3}
   \int_0^\infty dr\; \frac{F(r)}{E-E_- -\frac{r^2}{2} \pm\imath 0 }\;.
$$

\noindent The integral on the r.h.s. can be decomposed into two contributions

$$I(E) \;=\;
    \int_0^\infty dr\;\frac{F(r)}{E-E_- -\frac{r^2}{2} \pm\imath 0 }\;=\;
      I_1(E)\;+\;I_2(E)\;,
$$

\noindent with

\begin{equation}
\label{scatt08.eq-I12}
I_1(E)\;=\; 
   F(0)\, 
    \int_0^\infty dr\;\frac{1}{E-E_- -\frac{r^2}{2} \pm\imath 0 }\;,
\hspace{1cm}
   I_2(E)\;=\; 
    \int_0^\infty dr\;\frac{F(r)-F(0)}{E-E_- -\frac{r^2}{2} \pm\imath 0 }\;.
\end{equation}

\noindent The first part can be computed explicitly to give

$$I_1(E) \;=\; \pm \,\imath \;\frac{\pi\,F(0)}{\sqrt{2(E-E_-)}}\;,
\hspace{2cm}
  \mbox{\rm if}\;\; E>E_- \,.
$$

\noindent This part is singular and gives a nontrivial contribution to $G_0^\piso(E\pm\imath 0)$ of the form

$$G_0^\piso(E\pm\imath 0) \;=\;G_0^\piso(E_-)
   \,\pm \,\imath\; \pi\, \sqrt{2(E-E_-)}\;
    |\Lambda|\;M_-^\piso\;
   \frac{\det(\Ee''(k_-^*))^{\frac{1}{2}}\, |\SM^{d-1}|}{(2\pi)^d}\;
     +\;\hat{I}_2(E)\,,
$$

\noindent where $\hat{I}_2(E)$ comes from the contribution of $I_2$. Since the matrix $M_-^\piso$ is a projection, the singularity does not contribute to the real part of this expression. On the other hand, the integral $I_2$ can be treated by using two remarks: (a) $F(r)-F(0) = \Oo(r^2)$ at $r=0$, (b) for $0 \leq E-E_-\leq \frac{\delta^2}{4}$ and, for $r>\delta$, the integral converges to a smooth function of $E$. Hence the contribution of the integral coming from $r>\delta$ does not produce any singularity, while the contribution for $r\leq \delta$ is regular at $r=0$ leading to a contribution that is continuous at $E=E_-$ thanks to the Lemma~\ref{scatt08.lem-hilb}. This finishes the proof of Proposition~\ref{scatt08.prop-Green3d}.
\hfill $\Box$

\vspace{.3cm} 
 
 \subsection{Localized states in the REF representation}
 \label{sec-ONB}

\noindent The REF representation of the localized state at site $m\in\ZM^d$ is $\psi_m=\Uu\Ff\,|m\rangle$.  The states $(\psi_m)_{m\in\ZM^d}$ form an orthonormal basis in $L^2(\RM)\otimes L^2(\Sigma,\nu)$. More explicitly, they are given by

\begin{equation}
\label{scatt08.eq-ONbasis}
\psi_{m,b}(\sigma)\;=\;
 \frac{1}{(2\pi)^{\frac{d}{2}}}\;
  d_b(\sigma)\;
   e^{\imath m\cdot \theta_b(\sigma)}\,,
\end{equation}

\noindent for any $\sigma\in\Sigma$ avoiding $\criM$.
It will be convenient below to consider $\psi_{m,b}$ as a state in $L^2(\Sigma,\nu)$. These restricted localized states are not normalized, but their norm is independent of $m$:

$$
\|\psi_{m,b}\|^2_{L^2(\Sigma,\nu)}\;=\;
 \frac{1}{(2\pi)^d}\;
  \int_\Sigma \nu(d\sigma)\;
   |d_b(\sigma)|^2\,.
$$

\noindent This norm as well as scalar products between these states are linked to the resolvent.

\begin{lemma}
\label{scatt08.lem-DOS}
The following holds

$$\langle\psi_{n,b}|\psi_{m,b}\rangle_{L^2(\Sigma,\nu)}\;=\;
   \frac{F(f^{-1}(b))}{\pi}\;
    \langle n|
   \,\mp\Im m\Bigl((f^{-1}(b)\pm\imath 0-H_0)^{-1}\Bigr)\,
    |m\rangle\,.
$$
\end{lemma}

\noindent  {\bf Proof:} Thanks to the coarea formula and the Plemelj-Privalov theorem (see Lemma~\ref{scatt08.lem-hilb})

\begin{eqnarray*}
\langle n|\,\Im m\bigl((E\pm\imath 0-H_0)^{-1}\bigr)\,|m\rangle
& = &
\frac{1}{2\imath}\int^{E_+}_{E_-} de\;
\left(\frac{1}{E\pm\imath 0-e}-\frac{1}{E\mp\imath 0-e}\right)\;\int_{\Sigma_{e}}
\frac{\nu_e(d\sigma)}{(2\pi)^d}\;
\frac{e^{\imath(m-n)\cdot \sigma}}{|\nabla \Ee(\sigma)|}
\\
& = &
\mp \pi\;
\int_{\Sigma_{E}}
\frac{\nu_E(d\sigma)}{(2\pi)^d}\;
\frac{1}{|\nabla \Ee(\sigma)|}\;e^{\imath(m-n)\cdot \sigma}
\,.
\end{eqnarray*}

\noindent Always using $E=f^{-1}(b)$, the map $\theta_b:\Sigma\to\Sigma_E$ is a diffeomorphism. Thus the associated change of variables gives

\begin{eqnarray*}
\langle n|\,\Im m\bigl((E\pm\imath 0-H_0)^{-1}\bigr)\,|m\rangle
& = &
\mp \,\pi\;
\int_{\Sigma}
\frac{\nu(d\sigma)}{(2\pi)^d}\;
\bigl|\det(\theta'_b|_{T_\sigma\Sigma})\bigr|
\;\frac{1}{|\nabla \Ee(\theta_b(\sigma))|}
\;e^{\imath(m-n)\cdot \theta_b(\sigma)}
\\
& = &
\frac{\mp \,\pi}{F(f^{-1}(b))}\;
\int_{\Sigma}
\frac{\nu(d\sigma)}{(2\pi)^d}\;
\bigl|\det(\theta'_b|_{T_\sigma\Sigma})\bigr|
\;|\hX(\theta_b(\sigma))|
\;e^{\imath(m-n)\cdot \theta_b(\sigma)}\,.
\end{eqnarray*}

\noindent Now the formula follows from the definition of $\psi_{m,b}$ and \eqref{scatt08.eq-JacFac}.
\hfill $\Box$

\vspace{.2cm}

\begin{coro}
\label{scatt08.lem-partialiso} Let us introduce the operator $\Rr_b= \sum_{m\in\Lambda} |\psi_{m,b}\rangle\langle m|$ mapping $\ell^2(\Lambda)=\CM^{|\Lambda|}$ onto the subspace of $L^2(\Sigma,\nu)$ spanned by the $\left(\psi_{m,b}\right)_{m\in\Lambda}$. The range of its adjoint $\Rr_b^*= \sum_{m\in\Lambda} |m\rangle\langle \psi_{m,b}|$ is denoted by $\Ff_b\subset\CM^{|\Lambda|}$. Further let $\piso_b$ be a partial isometry from $\Ff_b$ onto the subspace of $L^2(\Sigma,\nu)$ spanned by the $\left(\psi_{m,b}\right)_{m\in\Lambda}$.

\vspace{.1cm}

\noindent {\rm (i)} The following holds
$$\Rr_b^\ast\Rr_b \;=\; 
   \frac{F(E)}{\pi}\;
    \Im m\,G_0^\piso(E-\imath 0)\,,
\hspace{2cm}
b=f(E)\,.
$$

\vspace{.1cm}

\noindent {\rm (ii)} If $P_b$ denotes the orthogonal projection in $\CM^{|\Lambda|}$ onto the subspace $\Ff_b$, then
$$\Rr_b \;=\; \piso_b \,
   \sqrt{
    \frac{F(E)}{\pi}
         }\;
    \left(
     \Im m\,G_0^\piso(E-\imath 0)
    \right)^{\frac{1}{2}}\;
      P_b\,,
\hspace{2cm}
b=f(E)\,.
$$

\vspace{.1cm}

\noindent {\rm (iii)} The map $b\in\RM\mapsto \Rr_b\in \Bb\left(\CM^{|\Lambda|},L^2(\Sigma,\nu)\right)$ is norm continuous.
\end{coro}

\noindent  {\bf Proof:} (i) is a re-phrasing of Lemma~\ref{scatt08.lem-DOS} and (ii)  just the usual polar decomposition. (iii) Since $\Rr_b$ has finite rank, the norm continuity follows form the strong continuity. In turns the strong continuity follows from the continuity of the inner products $\langle\psi_{n,b}|\psi_{m,b}\rangle_{L^2(\Sigma,\nu)}$.  The latter property follows from Lemma~\ref{scatt08.lem-DOS} and from the continuity of $F$, $f^{-1}$ and the imaginary part of the Green function (see Proposition~\ref{scatt08.prop-Green3d}), with respect to $E$ or to $b$. 
\hfill $\Box$

\vspace{.2cm}

It is worth remarking that $P_b=\Pi_b^*\Pi_b$ and $\Pi_b=\Pi_b P_b$. Furthermore, $\Im m\,G_0^\piso(f^{-1}(b)\pm\imath 0)$ commutes with $P_b$. The next lemma is a technical result which will be needed to deal with threshold singularities in dimension $d=3$.

\begin{lemma}
\label{scatt08.lem-limitstate}
Let $d\geq 2$ and let the extrema of $\Ee$ be isotropic in the sense that $\Ee''(k^*_\pm)$ is a multiple of the identity matrix. Then

\begin{equation}
\label{scatt08.eq-limitstate}
\lim_{b\to\pm\infty}\;
 \frac{1}{\|\psi_{m,b}\|_{L^2(\Sigma,\nu)}}\;
  \psi_{m,b} \;=\;
   e^{\imath m\cdot k^*_\pm}\;
    \psi_\pm\;,
\end{equation}
where $\psi_\pm\in L^2(\Sigma,\nu)$ are normalized states given by

$$
\psi_\pm(\sigma) \;=\;
 C\;\exp\left(\frac{1}{2}\;
  \int_0^{\infty} du \;
   \bigl(
    \dive(\hX)(\theta_u(\sigma))\pm d
   \bigr)\right)\;
    |\hX(\sigma)|^{\frac{1}{2}}\,,
     \qquad C>0\,.
$$
\end{lemma}

\noindent  {\bf Proof:} Let us first argue that $\psi_\pm$ are well-defined and normalizable. The isotropy hypothesis and \eqref{scatt08.eq-divXvicinity2} imply $\mbox{\rm div}(\hX)(k^*_\pm+k) =\mp\,d+\Oo(|k|)$. But $\theta_u(\sigma)$ converges to $k^*_\pm$ as $u\to\pm\infty$ at an exponential rate. Therefore the integral in the exponential exists and hence $\psi_\pm$ are well-defined. Thanks to the definition \eqref{scatt08.eq-ONbasis} of $\psi_{m,b}$ and to equation~(\ref{scatt08.eq-JacFac}) the conclusion of the lemma follows.
\hfill $\Box$

\begin{rem}
\label{scatt08.rem-niso}
{\em Without the isotropy assumption that $\Ee''(k_\pm^*)$ is a multiple of the identity, the l.h.s. of \eqref{scatt08.eq-limitstate} does not converge to a state in $L^2(\Sigma,\nu)$. 
}
\hfill $\Box$
\end{rem}

Lemma~\ref{scatt08.lem-DOS} also implies that the states $\psi_{m,b}$ are in general not orthogonal in $L^2(\Sigma,\nu)$, not even linearly independent as show the next results.

\begin{lemma}
\label{scatt08.lem-typrank}
Using the notation of the {\rm Corollary~\ref{scatt08.lem-partialiso}}, one has:

\vspace{.1cm}

\noindent {\rm (i)} $\Ff_b= \Ran\bigl(\Im m\,G_0^\piso(E-\imath 0)\bigr)$ whenever $E_-<E<E_+$ and $ b=f(E)$.

\vspace{.1cm}

\noindent {\rm (ii)}  In any interval of $\RM$ not containing a critical value $f(\Ee(\cri))$, there is a discrete subset without 

accumulation points outside of which the dimension of $\Ff_b$ is constant. 
\end{lemma}

\noindent  {\bf Proof:}  The first result follows directly from Lemma~\ref{scatt08.lem-DOS}. Since $F$ does not vanish on $(E_-,E_+)$, the image is also the image of $\Im m\,G_0^\piso(E-\imath 0)$. In particular,

$$\dim (\Ff_b) \;=\; \rank\bigl(\Im m\,G_0^\piso(E-\imath 0)\,\bigr)\;,
\hspace{2cm}
b=f(E)\,.
$$
As $E\mapsto \Im m\,G_0^\piso(E-\imath 0)$ is real-analytic away from the critical values $\Ee(\cri)$, the statement (ii) follows from analytic perturbation theory.
\hfill $\Box$

\vspace{.2cm}

The next question concerns whether the rank is indeed changing as a function of $E$. In addition, it is important to have examples leading to a non-maximal typical rank because this will allow produce embedded eigenvalues later on. The following result will help to construct such examples.

\begin{lemma}
\label{scatt08.lem-kernel}
With the hypothesis of {\rm Lemma~\ref{scatt08.lem-typrank}}, the orthogonal complement $\Ff_b^\perp$ in $\CM^{|\Lambda|}$ is given by

$$\Ff_b^\perp \;=\; \Ker (\Im m\,G^\piso_0(E-\imath 0))\,,
\hspace{2cm}
   b=f(E)\,.
$$

\noindent In addition, a vector $v= (v_m)_{m\in\Lambda}$ belongs to $\Ff_b^\perp$ {\rm (}with $b=f(E)${\rm )} if and only if its Fourier transform $\hat{v}(k)= \sum_{m\in\Lambda} v_m\,e^{\imath m\cdot k}$ is vanishing identically on the energy surface $\Sigma_E$.
\end{lemma}

\noindent  {\bf Proof:} The first relation comes directly from the first result of Lemma~\ref{scatt08.lem-typrank}. As in the proof of Lemma~\ref{scatt08.lem-DOS},

$$\langle v | \Im m\,G^\piso_0(E-\imath 0)|v\rangle\;=\;
   \pi\; \int_{\Sigma_{E}}
    \frac{\nu_E(d\sigma)}{(2\pi)^d}\;
     \frac{|\hat{v}(\sigma)|^2}{|\nabla \Ee(\sigma)|}\;.
$$

\noindent In particular, since $\Im m\,G^\piso_0(E-\imath 0) \geq 0$, it follows that $v\in\Ker \bigl(\Im m\,G^\piso_0(E-\imath 0)\bigr)$ if and only if $\hat{v}$ vanishes $\nu_E$-almost everywhere on $\Sigma_E$. But since $\hat{v}$ is a trigonometric polynomial, it has to vanish everywhere on $\Sigma_E$. 
\hfill $\Box$

\begin{proposi}
\label{scatt08.prop-fperp}
If $E\in [E_-,E_+]$ is not a critical energy, a vector $v=(v_m)_{m\in\Lambda}$ belongs to $\Ff_b^\perp$, with $b=f(E)$, if and only if there is $w\in\ell^2(\ZM^d)$ such that $\Pi^* v=(E\id -H_0)w$. Then $w$ admits a Fourier transform having the same degree of regularity as $\Ee$. More precisely, if $\Ee$ is of class $C^r$ with $r\in\NM\cup\{\infty\}\cup\{\omega\}$, so is the Fourier transform of $w$. Moreover, if $\Ee$ is a trigonometric polynomial, then $w$ has a finite support and the previous claim applies to  any $E\in (E_-,E_+)$.
\end{proposi}

\noindent  {\bf Proof:} A direct calculation shows indeed that, if $\Pi^* v=(E\id -H_0)w$ with some $w\in\ell^2(\ZM^d)$, then $\langle v|\Im m\,G_E^\piso(E-\imath 0)|v\rangle =0$. Conversely, if $v\in \Ff_b^\perp$, then $\hat{v}$ is a trigonometric polynomial vanishing on $\Sigma_E$ thanks to the Lemma~\ref{scatt08.lem-kernel}. Therefore, since $E$ is not critical, the energy surface $\Sigma_E$ is a smooth manifold and $\hat{w}(k)= \hat{v}(k)/(E-\Ee(k))$ is analytic in a neighborhood of this surface. Since it is analytic outside as well, the result follows. The Fourier coefficients of $\hat{w}$ defines a vector $w\in\ell^2(\ZM^d)$, which decay exponentially fast at infinity  by analyticity. In addition, if $\Ee$ is a polynomial, it follows that $\hat{w}$ is a polynomial as well by Hilbert's Nullstellensatz, which applies even if $E$ is critical. Therefore $w$ has finite support.
\hfill $\Box$ 

\vspace{.2cm}

The last proposition suggests examples of situations for which the kernel is trivial or not.

\begin{exam}
\label{scatt08.exam-1pt}
{\em If $\Lambda$ is reduced to one point, then $\rho^\piso(E)$ being positive for $E_-<E<E_+$ by Proposition~\ref{scatt08.prop-Green3d}, the kernel is trivial.
}
\hfill $\Box$
\end{exam}

\begin{exam}
\label{scatt08.exam-Ffb}
{\em Let $H_0$ be the discrete Laplacian in dimension $d=2$ with band function $\Ee(k_1,k_2)=\cos(k_1)+\cos(k_2)$. If $\Lambda=\{(0,0),(-1,0),(0,-1), (1,0),(0,1),(1,1),(1,2),(2,1)\}$ and $\hat{w}(k)=e^{\imath k_1}-1$, then neither $H_0w$ nor $w$ are supported in $\Lambda$. Nevertheless $(\frac{1}{2}+H_0){w}$ is supported by $\Lambda$. Hence for the energy $E=-\frac{1}{2}=f(b)$ the set $\Ff_E^\perp$ is nonempty.
}
\hfill $\Box$
\end{exam}

\begin{exam}
\label{scatt08.exam-Ffb2}
{\em Again $H_0$ is the discrete Laplacian in dimension $d=2$. Here we choose the set $\Lambda=\{(0,0),(-1,0),(0,-1),(1,1),(1,2),(2,1)\}$, then $\hat{w}(k)=e^{\imath k_1}e^{\imath k_2}-1$ is supported in $\Lambda$ and so is $\Ee(k)\hat{w}(k)$. Therefore for any $E=f^{-1}(b)$ the vector $\hat{v}=(E-\Ee)\hat{w}$ is the Fourier transform of a compactly supported vector. 
}
\hfill $\Box$
\end{exam}

If $\Ee$ is a trigonometric polynomial, more can be said. In such a case there is a finite set $S\subset\ZM^d$ such that $\Ee(k)= \sum_{s\in \ZM^d} \Ee_s e^{\imath s\cdot k}$ with $\Ee_s\neq 0\iff s\in S$. This set $S$ is called the support of $H_0$.  Since $H_0$ is self-adjoint, $\Ee$ is real-valued so that $\Ee_{-s}=\overline{\Ee_s}$ for all $s\in S$. In particular, $S$ is invariant under the parity map $s\mapsto -s$. The $S$-interior of a finite set will now be defined as the set of its points that cannot jump to the outside using hopping terms from $H_0$:

\begin{defini}
\label{scatt08.def-Sint}
Let $\Lambda$ and $S$ be subsets of $\ZM^d$. Then the $S$-interior $\Lambda^S$ of $\Lambda$ is the set of $x\in \Lambda$ such that $x+S\subset \Lambda$.
\end{defini}

\begin{exam}
\label{scatt08.exam-int}
{\em If $\Ee$ is a trigonometric polynomial with support $S$ and if $\Lambda$ is a finite set with nonempty $S$-interior, then any $w$ supported by $\Lambda^S$ satisfies $(E-H_0)w(x) =0$ outside $\Lambda$. Hence the dimension of $\Ff_E^\perp$ is at least equal to the number of points in the $S$-interior of $\Lambda$.
}
\hfill $\Box$
\end{exam}

\begin{proposi}
\label{scatt08.th-Ffbpoly}
Let $\Lambda\subset\ZM^d$ be finite and let $\Ee$ be a polynomial. Then, there is a finite subset $\vs=\vs(\Lambda)$ in the spectrum of $H_0$, such that $\dim(\Ff_b^\perp)$ is constant for $E$ outside of $\vs$, if $f(b)=E$. 
\end{proposi}

The proof of this proposition requires several steps that are described in the next four subsections. 

\subsubsection{Prime vectors and convexity}
\label{scatt08.sssect-conv}

\begin{lemma}[Prime vectors]
\label{scatt08.lem-prime}
A vector $a=(a_1,\ldots,a_d)\in\ZM^d$ is called prime if it satisfies one of the following equivalent definitions:

\vspace{.1cm} 

\noindent {\rm (i)} Any $\lambda >0$ such that $\lambda a\in \ZM^d$ must satisfy $\lambda \geq 1$.

\vspace{.1cm} 

\noindent {\rm (ii)} The greatest common divisor of the coordinates of $a$ is equal to $1$.

\vspace{.1cm} 

\noindent {\rm (iii)} The map $\phi_a: x\in\ZM^d\mapsto a\cdot x\in \ZM$ is onto.
\end{lemma}

\noindent {\bf Proof of the equivalence: } (i)$\Rightarrow$(ii) Let $p\geq 1$ be the greatest common divisor of the coordinates of $a$. It follows that $a/p\in\ZM^d$. Therefore $1/p \geq 1$ implying that $p=1$.

\vspace{.1cm}

\noindent (ii)$\Rightarrow$(iii) Let the greatest common divisor of the coordinates of $a$ be equal to one. The map $\phi_a$ is a group homomorphism. In particular, its image is a subgroup of $\ZM$. Therefore there is an integer $p\geq 1$ such that this image coincides with $p\ZM$. The coordinates of $a$ are given by $a_i=\phi_a(e_i)$, where $\{e_1,\ldots,e_d\}$ denotes the standard basis of $\ZM^d$. Thus, there are integers $b_i$ such that $a_i=pb_i$ for all $i$'s. Since the greatest common divisor of the coordinates is $1$, it follows that $p=1$.

\vspace{.1cm}

\noindent (iii)$\Rightarrow$(i) Let $a$ be such that $\phi_a$ is onto. Then there is $y\in\ZM^d$ such that $\phi_a(y)=1$. Let $\lambda >0$ satisfiy $\lambda a\in \ZM^d$. It follows that $\lambda \in \QM$ and that $\phi_{\lambda a}(x) = \lambda \phi_a(x)\in \ZM$. In particular, $\phi_{\lambda a}(y)=\lambda \in \ZM$, showing that $\lambda \geq 1$. Hence (i) holds.
\hfill $\Box$

\begin{defini}
\label{scatt08.def-halfplane}
A {\rm (}positive{\rm )} half-plane in $\ZM^d$ is a set of the form $\hs_{a,m}^+= 
\{x\in\ZM^d\,|\, a\cdot x \geq m\}$ where $a$ is a prime vector in $\ZM^d$ and $m\in\ZM$. The associated {\rm (}oriented{\rm )} affine hyperplane $\hs_{a,m}$ is defined similarly as $\hs_{a,m}=\{x\in\ZM^d\,|\, a\cdot x =m\}$.
\end{defini}

\begin{defini}
\label{scatt08.def-polytope}
Let $\Lambda\subset \ZM^d$. 

\vspace{.1cm}

\noindent {\rm (i)} A prime vector $a$ is a $\Lambda$-direction if there is $m\in\ZM^d$ such that $\Lambda\subset \hs_{a,m}^+$. 

\vspace{.1cm}

\noindent {\rm (ii)} An oriented affine hyperplane $\hs_{a,m}$ {\rm (}resp. half-plane $\hs_{a,m}^+${\rm )} is called a contact hyperplane {\rm (}resp. 

a contact half-plane{\rm )} for $\Lambda$ whenever $\Lambda\subset \hs_{a,m}^+$ and $\Lambda\cap\hs_{a,m}\neq \emptyset$. 

\vspace{.1cm}

\noindent {\rm (iii)} The convex hull of $\Lambda$, denoted by $\Conv(\Lambda)$, is the intersection of all its contact half-planea. If 

no prime vector is a $\Lambda$-direction, then $\Conv(\Lambda)=\ZM^d$.  

\vspace{.1cm}

\noindent {\rm (iv)} $\Lambda$ is convex whenever it coincides with its convex hull. 

\vspace{.1cm}

\noindent {\rm (v)} A finite convex set is called a polytope.
\end{defini}

It is easy to check that a polytope has a finite number of contact hyperplanes.

\begin{lemma}
\label{scatt08.lem-sldz}
Let $a$ be a prime vector in $\ZM^d$. Then there is a matrix $A\in \SL(d,\ZM)$, such that $\phi_a\circ A(n,y)=n$ for all $(n,y)\in \ZM\times \ZM^{d-1}$. 
\end{lemma}

\noindent  {\bf Proof:} Thanks to the Lemma~\ref{scatt08.lem-prime}, there is a vector $b_1\in\ZM^d$ such that $\phi_a(b_1)=1$. On the other hand, $\Ker(\phi_a)$ is a subgroup of $\ZM^d$ and therefore it is free. In particular, it admits a basis $\{b_2,\ldots,b_d\}$ (minimal set of generators). Clearly the vectors $b_j$ can be chosen to be prime. Then let $A$ be the $d\times d$ matrix with column given by the family $\{b_1,\ldots,b_d\}$. By construction, $A$ has integer coefficients. Moreover, $b_1 =Ae_1$ and therefore $\phi_a(Ae_1)=1$. In addition, $\phi_a(Ae_j)=\phi_a(b_j)=0$ whenever $j\geq 2$. Consequently, if $y\in \ZM^d$ is such that $Ay=0$, then $y_1=0$ and $\sum_{j=2}^d y_jb_j=0$. Since $\Ker(\phi_a)$ is free, this implies that $y_j=0$ for all $j$'s. Hence $A$ is one-to-one. On the other hand, if $x\in\ZM^d$, then $x-\phi_a(x) b_1\in\Ker({\phi_a})$. Therefore there are $y_2,\ldots,y_d\in \ZM$ such that $x-\phi_a(x) b_1 = \sum_{j=2}^d y_jb_j$. Setting $y_1=\phi_a(x)$ and $y=(y_1,\ldots,y_d)$ it follows that $x=Ay$. Hence $A$ is also onto. Therefore the matrix $A$ is invertible and its inverse has also integer entries. In particular, changing the sign of one of the $b_j$'s if necessary, $\det(A)= 1$.
\hfill $\Box$

  \subsubsection{The $A$-Fourier transform}
  \label{scatt08.sssect-AFourier}

\noindent Let $s\in\ZM^d$ and let $A\in \SL(d,\ZM)$. Then let $T_s$ and $U_A$ be the operators acting on $\ell^2(\ZM^d)$ defined by 

$$\left(T_s\psi\right)(x) \;=\;
   \psi(x-s)\,,
\hspace{1cm}
  \left(U_A\psi\right)(x) \;=\;
   \psi(A^{-1}x)\,,
\hspace{1cm}
    \psi\in \ell^2(\ZM^d)\,.
$$

\noindent Then both $T_s$ and $U_A$ are unitary operators. Moreover, if $s,t\in \ZM^d$, one has $T_{s+t}=T_sT_t$ and $T_0=\id$. In a similar way, if $A,B\in \SL(d,\ZM)$ then $U_AU_B= U_{AB}$ and $U_{\rm id}=\one$. In particular, $(U_A)^{-1}=U_{A^{-1}}= U_A^\ast$. In addition, $U_AT_sU_A^{-1}= T_{As}$ leading to

$$H_0\;=\; \sum_{s\in S} \Ee_s\,T_s
\hspace{1cm}\Longrightarrow \hspace{1cm}
   U_AH_0U_A^{-1}\;=\; 
    \sum_{s\in AS} \Ee_{A^{-1}s}\,T_s\,,
$$

\noindent Thus $U_A$ changes the support of $H_0$ from $S$ into $AS$. Let now $a$ be a prime vector and let $A\in\SL(d,\ZM)$ be chosen to satisfy $\phi_a(Ae_1)=1$. The partial Fourier transform $\Ff_A$ is the unitary transformation from $\ell^2(\ZM^d)$ into $\ell^2(\ZM)\otimes L^2(\TM^{d-1})$ defined by 

$$(\Ff_A\psi)_n(p) \;=\;
    \sum_{y\in\ZM^{d-1}} \psi\big(A(n,y)\big)\,e^{\imath p\cdot y}\,,
\hspace{2cm}
    \psi\in \ell^2(\ZM^d)\;,\;\;p\in\TM^{d-1}\;.
$$

\noindent It follows that

$$(\Ff_A H_0\psi)_n(p) \;=\;
   \sum_{r\in \phi_a(S)} \Ee_r^A(p) 
     (\Ff_A\psi)_{n-r}(p)\,,
\hspace{2cm}
 \Ee_r^A(p)\;=\; 
   \sum_{t\in\ZM^{d-1},\, (r,t)\in A^{-1}S}
    \Ee_{A(r,t)} e^{\imath p\cdot t}\,.
$$

\noindent Since $S$ is finite, each of the $\Ee_r^A$'s is a trigonometric polynomial. Moreover, $S$ is invariant under the reflection $s\mapsto -s$, so that $\phi_a(S)\subset [-r_a,r_a]$ if $r_a(S)= \max{\phi_a(S)}$ and both $\pm r_a\in \phi_a(S)$. Consequently, since $\phi_a(A(r,t))=r$, it follows that $\Ee_r^A\neq 0$ if and only if $r\in\phi_a(S)$. In particular, $\Ee_{\pm r_a}^A\neq 0$.

  \subsubsection{The case of convex $\Lambda$}
  \label{scatt08.sssect-Lconv}

\noindent From the definition of the $S$-interior, it follows immediately that the $S$-interior of a convex set $\Lambda$ coincides with the intersection $\Lambda^S= \bigcap_{(a,m)}\hs_{a,m+r_a(S)}^+$ over the pairs $(a,m)$ such that $\hs_{a,m}^+$ is an oriented contact half-space for $\Lambda$. This leads to the following result.

\begin{proposi}
\label{scatt08.prop-Ffb}
Let  $\Lambda\subset \ZM^d$ be a polytope and let $S$ be the support of $H_0$. Then the equation $(E-H_0)w(x) =0$ for $x\notin \Lambda$ admits a solution $w$ with finite support if and only if $w$ is supported by the $S$-interior of $\Lambda$.
\end{proposi}

\noindent {\bf Proof:} Let $a\in\ZM^d$ be a prime vector and $m\in\ZM$ be such that $\hs_{a,m}$ is an oriented contact hyperplane of $\Lambda$. Let also $A\in \SL(d,\ZM)$ be chosen such that $\phi_a(Ae_1)=1$. Then for $w\in\ell^2(\ZM^d)$ let $w_n(p)$ denote the partial Fourier transform $(\Ff_Aw)_n(p)$. It follows that, since $\phi_a(S)\subset [-r_a(S),r_a(S)]$, $w$ satisfies $(E-H_0)w(x)=0$ for $\phi_a(x)< m$, namely

\begin{equation}
\label{scatt08.eq-Fba}
\sum_{|r|\leq r_a(S)} \left(E\,\delta_{r,0}-\Ee_r^a(p)\right) 
     w_{n-r}(p)\;=\;0\,,
\hspace{2cm}
     \forall\; n<m\,.
\end{equation}

\noindent In the following $r_a(S)$ will be denoted by $r_a$. In particular, $\Ee_{-r_a}\neq 0$. Since $w$ is finitely supported, there is $N\in\NM$ such that $w_n=0$ for $n<-N$. In addition, each component $w_n(p)$ is a trigonometric polynomial in $p$. Writing the equation~(\ref{scatt08.eq-Fba}) for $n=-N-r_a$ leads to

$$\Ee_{-r_a}(p) w_{-N}(p)\;=\;0\,
\hspace{1cm}\Longrightarrow\hspace{1cm}
    w_{-N}\;=\;0\,.
$$

\noindent Proceeding to write the equation~(\ref{scatt08.eq-Fba}) for $n=-N-r_a+l$ for $l=1,\ldots, m-1+N+r_a$, gives, by the same argument, $w_{n}=0$ for $n<m+r_a$. Hence the support of $w$ is contained in the half-plane $\hs_{a,m+r_a}^+$. Since this is true for any contact hyperplane $\hs_{a,m}$, the support of $w$ is contained in the $S$-interior of $\Lambda$. Conversely, if $w$ is supported by $\Lambda^S$, it follows that $(E-H_0)w$ is supported by $\Lambda$.
\hfill $\Box$

  \subsubsection{Conclusion of the proof of Proposition~\ref{scatt08.th-Ffbpoly}}
  \label{scatt08.sssect-Theopoly}

\noindent {\bf Proof of Proposition~\ref{scatt08.th-Ffbpoly}:} Since $\Lambda\subset \Conv(\Lambda)$, it follows that the equation $(E-H_0)w(x)=0$ is satisfied for $x\notin \Conv(\Lambda)$. Thanks to the Proposition~\ref{scatt08.prop-Ffb}, it follows that $w$ is supported in the $S$-interior of $\Conv(\Lambda)$. Let $R$ be the orthogonal projection on the subspace $\ell^2(\Conv(\Lambda)^S)$ and let $P$ the orthogonal projection on $\ell^2(\Lambda)$. Then both $P$ and $R$ are finite dimensional. In addition, the previous equation is satisfied if and only if $w\in \Ker((\id -P)(E-H_0)R)=\Ker(R(E-H_0)(\id -P)(E-H_0)R)$. The matrix $A(E)= R(E-H_0)(\id -P)(E-H_0)R$ is finite dimensional, acts on $\ell^2(\Conv(\Lambda)^S)$ and is a polynomial in the variable $E\in [E_-,E_+]$. Therefore
its kernel has a constant dimension away from a finite set by analytic perturbation theory. 
\hfill $\Box$

\vspace{.3cm}

\section{Scattering by a finite range perturbation}
\label{scatt08.sect-finiteRange}

\noindent This section is dedicated to the scattering of a lattice particle by a finite range perturbation. In  the first part, the Green matrix of the perturbed Hamiltonian will be investigated, in the second part, various formulas will be derived for the wave operators, the scattering matrix and the time delay operator. At last, the Levinson theorem will be proved.

\vspace{.1cm}

Let $\Lambda\subset\ZM^d$ be a finite subset, with $|\Lambda|$ points. Let $\piso:\ell^2(\ZM^d)\to\CM^{|\Lambda|}$ be the corresponding partial isometry (see the introduction of Section~\ref{scatt08.ssect-LAP}). The perturbation will be a finite rank selfadjoint operator $V:\ell^2(\ZM^d)\to\ell^2(\ZM^d)$, supported on $\Lambda$, namely $V=\piso^*\piso V\piso^*\piso$. Hence $V$ is encoded in the $|\Lambda|\times|\Lambda|$ matrix  $V^\piso=\piso V\piso^*$. For convenience, $V^\piso$ will be assumed to be invertible (this hypothesis can be dropped if the kernel is eliminated like in Section~\ref{sec-ONB}). The perturbed Hamiltonian describing the scattering is then $H=H_0+V$. A typical example for a local perturbation is a potential with support $\Lambda$, namely $V=\sum_{n\in\Lambda}v_n\,|n\rangle\langle n|$ with $v_n\in\RM\backslash\{0\}$.

\vspace{.2cm}

 \subsection{Green function}
 \label{scatt08.ssect-Greenspec}

Let $G^\piso(z)=\piso (z-H)^{-1}\piso^*$ be the Green matrix of the perturbed Hamiltonian. As for the unperturbed case, it is also a Herglotz matrix which is invertible for $\Im m(z)\neq 0$. The following formulas are well-known.

\begin{lemma}
\label{scatt08.lem-Greenperturb}
For $z\in\CM\setminus\RM$,

\begin{equation}
\label{scatt08.eq-Greenperturb1}
G^\piso(z)\;=\;
 \bigl(G_0^\piso(z)^{-1}-V^\piso\bigr)^{-1}\;=\;
  \bigl(\one-G_0^\piso(z)V^\piso\bigr)^{-1}G_0^\piso(z)\,,
\end{equation}

\noindent Let the $T$-matrix be defined by

\begin{equation}
\label{scatt08.eq-Tmatrix}
T(z)\;=\;\piso^*\,T^\piso(z)\,\piso\,,
\hspace{2cm}
 T^\piso(z)\;=\;
  \bigl(\one-V^\piso G_0^\piso(z)\bigr)^{-1}V^\piso\,.
\end{equation}

\noindent Then

\begin{equation}
\label{scatt08.eq-Greenperturb2}
\frac{1}{z-H}\;=\;
 \frac{1}{z-H_0}+
  \frac{1}{z-H_0}\,T(z)\,\frac{1}{z-H_0}\,,
\end{equation}
\end{lemma}

\noindent  {\bf Proof:} The resolvent identity yields

$$\frac{1}{z-H_0}\;=\;
   \frac{1}{z-H}-\frac{1}{z-H_0}\,V\,\frac{1}{z-H}\;=\;
    \left({\bf 1}-\frac{1}{z-H_0}\,V\right)\,\frac{1}{z-H}\,.
$$

\noindent Applying $\piso$ and $\piso^*$ from the left and right respectively gives

$$G_0^\piso(z)\;=\;
   \bigl({\bf 1}-G_0^\piso(z)V^\piso\bigr)G^\piso(z)\,.
$$

\noindent Now $G_0^\piso(z)$ is Herglotz and thus invertible since $z\notin \RM$. Hence, $G_0^\piso(z)^{-1}-V^\piso$ is also Herglotz and invertible, leading to the invertibility of ${\bf 1}-G_0^\piso(z)V^\piso= G_0^\piso(z)(G_0^\piso(z)^{-1}-V^\piso)$. To prove \eqref{scatt08.eq-Greenperturb2}, the resolvent identity gives the factor ${\mathbf 1}-\frac{1}{z-H_0}\,V$. This operator is invertible, because it is a finite rank perturbation of ${\mathbf 1}$ and any element in its kernel is an eigenvector of $H=H_0+V$ with eigenvalue $z\notin\RM$, namely the kernel is trivial. Using the identity $({\mathbf 1}-A)^{-1} A =({\mathbf 1}-A)^{-1}(A-{\mathbf 1}+{\mathbf 1})=  ({\mathbf 1}-A)^{-1} -{\mathbf 1}$, the inverse can be written as

$$ \left({\bf 1}-\frac{1}{z-H_0}\,V\right)^{-1}\;=\;
   {\bf 1}+
    \frac{1}{z-H_0}\,
     \piso^*\,V^\piso\,\piso\,
      \left({\bf 1}-\frac{1}{z-H_0}\,V\right)^{-1}\,.
$$

\noindent Since 

$$\piso \left({\bf 1}-\frac{1}{z-H_0}\,V\right)\;=\;
   \left({\bf 1}-G^\piso_0(z)\,V^\piso\right)\piso\;,
$$

\noindent it follows that

$$\piso\,\left({\bf 1}-\frac{1}{z-H_0}\,V\right)^{-1}\;=\;
   \left({\bf 1}-G^\piso_0(z)\,V^\piso\right)^{-1}\,\piso\,,
$$

\noindent When combined with the resolvent identity this completes the proof.
\hfill $\Box$

\vspace{.2cm}

 \subsection{Spectral analysis}
 \label{scatt08.ssect-spec}
 
Because the perturbation has finite range, the essential spectrum of $H$ is given by the essential spectrum of $H_0$. However, $H$ may have some discrete spectrum, which, since $H$ is selfadjoint, is given by the simple poles of the resolvent on the real axis. Thanks to Proposition~\ref{scatt08.prop-Green3d}, it follows from equation~\eqref{scatt08.eq-Greenperturb1} that the only way to get a polar singularity in the Green matrix of $H$ is for ${\bf 1}-G^\piso_0(z)\,V^\piso=\left((V^\piso)^{-1}-G^\piso_0(z)\right)V^\piso$ to have a nontrivial kernel (recall that we restrict ourselves to the case of invertible $V^\piso$). This can be analyzed using the  determinant of ${\bf 1}-G^\piso_0(z)\,V^\piso$ which is also called the {\em perturbation determinant} \cite{Yaf}. Furthermore, if $E$ is an eigenvalue of $H$,
\begin{equation}
\label{eq-embedcount}
\mbox{\rm multiplicity of}\;
E\;=\;
 \dim\Ker\left(
 (V^\piso)^{-1}-G^\piso_0(E\pm\imath 0)
\right)\,.
\end{equation}
\noindent If $E\not\in[E_-,E_+]$, it is called an {\em isolated eigenvalue} while, if $E\in(E_-,E_+)$, it is called an {\em embedded eigenvalue}. For $E=E_\pm$, a non-trivial kernel of $(V^\piso)^{-1}-G^\piso_0(E_\pm)$ leads to a {\em threshold singularity} which will be dealt with below. With any $E\in\RM$ is associated the subspace of $\CM^{|\Lambda|}$

$$\Ss_E\;=\;
   \Ker\bigl(\,(V^\piso)^{-1}-\Re e\;G^\piso_0(E)\,\bigr)\,.
$$

\noindent Then the multiplicity of the eigenvalue $E\in\RM\setminus [E_-,E_+]$ of $H$ is also equal to $\dim(\Ss_E)$. The embedded eigenvalues are characterized in the next result (where the space $\Ff_E$ is the space $\Ff_b$ for $E=f(b)$ used in Corollary~\ref{scatt08.lem-partialiso}). 

\begin{proposi}
\label{scatt08.prop-embedded}
Let $V$ have finite range. Then $E\in(E_-,E_+)$ is an embedded eigenvalue if and only if $\Ff^\perp_E\cap\Ss_E$ is non-trivial and the dimension of this intersection is equal to the multiplicity of $E$. If $\Ee$ is analytic, the associated eigenvectors are decaying exponentially fast at infinity. If $\Ee$ is a trigonometric polynomial, the  the eigenvectors have a finite support.
\end{proposi}

\begin{rem}
\label{scatt08.rem-KV}
{\em The last statement, namely that the eigenvectors have compact support, was proved in a slightly different context in \cite{KV}. 
}
\hfill $\Box$
\end{rem}

\noindent  {\bf Proof:}  Let $E=f^{-1}(b)\in(E_-,E_+)$ be an embedded eigenvalue and $v\in\CM^{|\Lambda|}$ be the associated vector in the kernel of $(V^\piso)^{-1}-G^\piso_0(E\pm\imath 0)$. Because $\Im m(G^\piso_0(E-\imath 0))\geq 0$ and $V^\piso$ is self-adjoint, this is equivalent to having  $\Im m\,G^\piso_0(E-\imath 0)v=0$ and $(V^\piso)^{-1}v=\Re e \,G^\piso_0(E)v$, or alternatively $v\in\Ff^\perp_E\cap\Ss_E$. As shown in Lemma~\ref{scatt08.lem-kernel} and in Proposition~\ref{scatt08.prop-fperp}, for any vector $v\in\Ff^\perp_E\cap\Ss_E$ there is $w\in\ell^2(\ZM^d)$ such that $v=(E\id-H_0)w$. Moreover $w$ decays exponentially fast at infinity and has finite support if $\Ee$ is a trigonometric polynomial.
\hfill $\Box$

\begin{exam}
\label{scatt08.exam-gpim1}
{\em Let $\Lambda$ be such that $\Ff_E^\perp\neq \{0\}$ (see Examples~\ref{scatt08.exam-Ffb} to \ref{scatt08.exam-int}). Chosing $V^\Pi=(\Re e\,G^\piso_0(E))^{-1}$ whenever the latter is invertible leads to a Hamiltonian $H$ with an embedded eigenvalue of multiplicity $\dim(\Ff^\perp_E)$. 
}
\hfill $\Box$
\end{exam}
\begin{exam}
\label{scatt08.exam-barrier}
{\em Let us present another way to construct Hamiltonians with embedded eigenvalues, again by perturbing a periodic $H_0$ with a polynomial energy band $\Ee$. Let $P=\Pi^*\Pi$ and $Q=\one-\Pi^*\Pi$ and then set $V= -PH_0Q-QH_0P$. Clearly $V$ has finite range. Now $H=H_0+V$ splits into a direct sum of $PH_0P$ and $QH_0Q$. The former has finite rank and admits a spectrum of eigenvalues with eigenvectors supported in $\Lambda$. By the minimax principle, it follows that all these eigenvalues belong to $[E_-,E_+]$. On the other hand, $QH_0Q$ is a finite rank perturbation of $H_0$, so it has the same essential spectrum. Hence the eigenvalues of $PH_0P$ are embedded indeed.  
}
\hfill $\Box$
\end{exam}

It is natural to address the question of whether embedded eigenvalues exist if $V$ is a multiplication operator with support in $\Lambda$. The following result gives a negative answer.

\begin{proposi}
\label{scatt08.prop-noembed}
Let $H_0$ be a trigonometric polynomial and  $V=\sum_{n\in\Lambda}v_n\,|n\rangle\langle n|$ is a potential with entries $v_n\not = 0$ for $n\in\Lambda$. Then $H_0+V$ has no embedded eigenvalues.
\end{proposi}

\noindent {\bf Proof:} Clearly $V^\piso$ is invertible in $\CM^{|\Lambda|}$. Thanks to Propositions~\ref{scatt08.prop-fperp} and \ref{scatt08.prop-embedded}, if $E$ is an embedded eigen\-value, then there is a vector $v\in\CM^{|\Lambda|}$ such that $(V^\piso)^{-1}v- G_0^\piso(E)) v=0$ and $\Pi^*v=(E-H_0)w$ for some $w\in\ell^2(\ZM^d)$. It then follows from Proposition~\ref{scatt08.prop-Ffb}  that $w$ is supported in the $S$-interior $\Lambda^S$ of $\Lambda$, if $S$ is the support of $H_0$. If $\Lambda^S$ is empty, then $v=0$ and there is no embedded eigenvalue. Now let $\Lambda^S$ be non-empty. Because $w$ is supported by $\Lambda^S$, one has $G_0^\piso(E)\Pi(E\,\one-H_0)w=\Pi w$ so that replacing $v=\Pi (E\,\one-H_0)w$ shows $\Pi(E\,\one-H_0)w=\Pi Vw$ and thus $(E\,\one-H_0)w=Vw$. Now the r.h.s is supported in $\Lambda^S$ since $V$ is a multiplication operator. Applying again Proposition~\ref{scatt08.prop-Ffb} shows that $w$ is supported in the $S$-interior $(\Lambda^S)^S$ of $\Lambda^S$. Now this procedure can be iterated. As the iterated $S$-interior of $\Lambda$ is empty after a finite number of steps, one concludes that $w=0$.
\hfill $\Box$

\vspace{.2cm}

Let us next investigate threshold singularities in more detail, now considering only the case of dimension $d\geq 3$. They appear at either one of the band edges $E_\pm$ whenever $(V^\piso)^{-1}-G^\piso_0(E_\pm\pm\imath 0)$ has a non-trivial kernel, which is equivalent to $\Ss_{E_\pm}$ being non-trivial. Hence $\dim(\Ss_{E_\pm})$ is the multiplicity of the threshold singularity at $E_\pm$. Again it is easy to produce such singularities by an adequate choice of $V$. Such a singularity can either be a threshold eigenvalue or a threshold resonance (the latter is also called a half-bound state) depending upon whether the equation $H\psi=E_\pm\psi$ has a square integrable solution $\psi$ or not \cite{New,KJ}. In order to analyze the threshold singularities let the following space be defined

\begin{equation}
\label{scatt08.eq-Fborthpm}
\Tt_\pm \;=\;
 \left\{
  \,v\in\CM^{|\Lambda|} \,|\,
  \hat{v}\,\mbox{ has a zero of order at least  } 5-d 
   \mbox{ at } k_\pm^*\,
 \right\}\,.
\end{equation}

\noindent The definition of $\Tt_\pm$ is somewhat similar to $\Ff_b^\perp$ as given in Proposition~\ref{scatt08.prop-fperp}. However, $\Tt_\pm$ is likely to be larger than the limit of $\Ff_b^\perp$ as $b\to\pm\infty$ because it tests zeros only at a single point. In addition,  $\Tt_\pm$ coincides with $\CM^{|\Lambda|}$ for $d\geq 5$.

\begin{proposi}
\label{scatt08.prop-halfbounded}
Let $d\geq 3$ and $V$ be of finite range.  Then the multiplicity of $E_\pm$ as threshold eigenvalue of $H$ is equal to $\dim\bigl(\Ss_{E_\pm}\cap\Tt_{\pm}\bigr)$. Moreover, the multiplicity of $E_\pm$ as threshold resonance of $H$ is equal to $\dim\bigl(\Ss_{E_\pm}\bigr)-\dim\bigl(\Ss_{E_\pm}\cap\Tt_{\pm}\bigr)$. In particular, for $d\geq 5$ all threshold singularities lead to threshold eigenvalues. 
\end{proposi}

\noindent  {\bf Proof:} Let $v\in\Ss_{E_\pm}$. Then $V^\piso G^\piso_0(E_\pm)v=v$ and $V(E_\pm-H_0)^{-1}\piso^*v=\piso^*v$. Hence there is a threshold eigenvector whenever the equation $\piso^*v=(E_\pm-H_0)\psi$ admits a solution $\psi\in\ell^2(\ZM^d)$. If so, then $H\psi=E_\pm \psi$ indeed. Using the Fourier transform, the equation leads to the following solution $\hat{\psi}(k)=\hat{v}(k)/(E_\pm-\Ee(k))$. The denominator satisfies $E_\pm-\Ee(k)=\langle (k-k_\pm^*)|\Ee''(k_\pm^*)|(k-k_\pm^*)\rangle+\Oo((k-k_\pm^*)^3)$. This quadratic singularity is integrable in dimension $d\geq 3$ so that indeed $\psi\in\ell^\infty(\ZM^d)$ is well-defined. If $d\geq 5$, then the singularity is also square integrable so that $\psi\in \ell^2(\ZM^d)$ is an eigenvector. In dimension $d=4$, $\hat{\psi}$ is square integrable only if $\hat{v}$ has a zero at $k_\pm^*$. In dimension $d=3$ the zero has to be of order $2$. Combining this leads to the definition \eqref{scatt08.eq-Fborthpm} and to the conclusion above.
\hfill $\Box$

\vspace{.3cm}

 \subsection{The wave operator as an integral operator}
 \label{scatt08.ssect-waveint}

\noindent The potential being finite rank, the Kato-Rosenblum theorem for trace class scattering theory \cite{RS,Yaf} implies that the wave operators

$$\Omega_\pm \;=\;
   \mbox{\rm s-}\!\!\!\lim_{t\rightarrow \pm \infty}\;
     e^{\imath Ht}\;e^{-\imath H_0t}\,,
$$

\noindent exist and are complete, that is, $\Ran(\Omega_+)=\Ran(\Omega_-)=P_{\mbox{\tiny\rm ac}}(H)$ where $P_{\mbox{\tiny\rm ac}}(H)$ is the projection on the absolutely continuous subspace of $H$. Then the wave operators are partial isometries satisfying

\begin{equation}
\label{scatt08.eq-Omegaid}
\Omega_\pm^\ast \, \Omega_\pm \;= \;\id\;, \qquad
   \Omega_\pm \, \Omega_\pm^\ast \;=\; P_{\mbox{\tiny\rm ac}}(H)\;=\;
   \id\,-\,P_{\mbox{\tiny\rm pp}}(H)\,,
\end{equation}

\noindent where $P_{\mbox{\tiny\rm pp}}(H)$ is the projection on the pure-point spectrum of $H$ and the last equality holds because there is no singular continuous spectrum. In addition, if one sets $\Omega(t)=e^{\imath H t}e^{-\imath H_0 t}$, then $\Omega(t)e^{\imath H_0 s}=e^{\imath H s}\Omega(t-s)$ for all $s\in\RM$. Passing to the limit $t\rightarrow \pm\infty$ yields the intertwining relation

$$
\Omega_{\pm}\,g(H_0)\;=\;
 g(H)\,\Omega_\pm\,,
\hspace{2cm}
  g\in C_0(\RM)\;.
$$

\noindent Birman's invariance principle \cite{RS,Yaf} can now be expressed as follows. The function $f:(E_-,E_+)\to\RM$ defined in \eqref{scatt08.eq-Fchoiceconclusion} is smooth and has positive derivative. It is therefore admissible for the invariance principle so that

\begin{equation}
\label{scatt08.eq-invarprin}
\Omega_\pm\;=\;
 \mbox{\rm s-}\!\!\!\lim_{t\rightarrow \pm \infty}\;
  P_{\mbox{\tiny\rm ac}}(H)\;
   e^{\imath f(H)t}\;
    e^{-\imath f(H_0)t}\;.
\end{equation}

\noindent However, $H$ may have some spectrum outside $(E_-,E_+)$ and possibly eigenvalues at $E_\pm$ so that $f(H)$ may not be well defined. However, using the completeness of the wave operators, $P_{\mbox{\tiny\rm ac}}(H)$ can be inserted discounting all eigenvalues. Since $H$ has the same essential spectrum as $H_0$, this eliminates all ambiguity in the definition. It will allow to derive an explicit formula for $\widehat{\Omega}_\pm=\Ff\Omega_\pm\Ff^*$ which will serve as a tool to calculate the wave operator and the scattering operator.

\vspace{.2cm}

\begin{proposi}
\label{scatt08.prop-waveop}
The following formula holds

$$
\bigl((\widehat{\Omega}_\pm-\one)\phi\bigr)(k)\;=\;
 \lim_{\epsilon\downarrow 0}\;
  \int_{\TM^d}\frac{dk'}{(2\pi)^d}\,
   \sum_{n,m\in\Lambda} \,
    \langle n|
     \,T(\Ee(k')\mp\imath \epsilon)\,
      |m\rangle\;
       \frac{e^{\imath(k\cdot n-k'\cdot m)}}
             {\Ee(k')\mp\imath \epsilon-\Ee(k)}\;
         \phi(k')\,.
$$
\end{proposi}

\noindent  {\bf Proof:} It follows from DuHamel's formula and a Tauberian lemma \cite{RS} that

$$
\Omega_\pm \;=\;
 \id \,\pm\,\imath\,
  \mbox{\rm s-}\!\!\lim_{t\rightarrow \infty}\;
   \int_0^t ds\;     e^{\pm\imath Hs}\;V\;e^{\mp\imath H_0s}\;=\;
 \id \,\pm\,\imath\,
  \mbox{\rm s-}\!\lim_{\epsilon\downarrow 0}\;
   \int_0^\infty ds\;  e^{-\epsilon s}\;   
     e^{\pm\imath Hs}\;V\;e^{\mp\imath H_0s}\;.
$$

\noindent Hence

$$\left((\widehat{\Omega}_\pm-{\bf 1})\phi\right)(k) \;=\; 
   \pm \,\imath\lim_{\epsilon\downarrow 0}\;
    \int_0^\infty ds\;  e^{-\epsilon s}\;
     \left(
       \Ff\, e^{\pm\imath Hs}\;V\;e^{\mp\imath H_0s}\,\Ff^*\,\phi
     \right) (k)\,.
$$

\noindent In the following, the notation $V_{l,m}=\langle l|V|m\rangle$ will be used. In addition, 

$$\langle m|\,e^{\mp\imath H_0s}\,\Ff^*\,|\phi\rangle \;=\;
   \int_{\TM^d}
    \frac{dk'}{(2\pi)^d}\;
     e^{-\imath k'\cdot m}\;
     \,e^{\mp\imath \Ee(k')s}\,\phi(k')
$$

\noindent Consequently the previous formula leads to

$$\left((\widehat{\Omega}_\pm-{\bf 1})\phi\right)(k) \;=\; 
   \pm \imath 
    \sum_{l,m\in\Lambda} \,V_{l,m}\;
     \lim_{\epsilon\downarrow 0}\;
      \int_0^\infty ds\;  e^{-\epsilon s}\;
       \left(
        \Ff\, e^{\pm\imath Hs}|l\rangle
       \right) (k)
        \int_{\TM^d}
    \frac{dk'}{(2\pi)^d}\;
     e^{-\imath k'\cdot m}\;
     \,e^{\mp\imath \Ee(k')s}\,\phi(k')\,.
$$

\noindent The integral over $s$ can be performed to give

$$\left((\widehat{\Omega}_\pm-{\bf 1})\phi\right)(k) \;=\;
   \lim_{\epsilon\downarrow 0}\;
    \int_{\TM^d}\frac{dk'}{(2\pi)^d}\;
     \sum_{l,m\in\Lambda} V_{l,m}\,e^{-\imath k'\cdot m}\;
      \left(
        \Ff\;\frac{1}{\Ee(k')\mp\imath \epsilon-H}\;|l\rangle
      \right)(k)\; \phi(k')\,.
$$

\noindent In the previous expression, it becomes possible to compute the part in the parenthesis.  For indeed, using the resolvent identity as in Lemma~\ref{scatt08.lem-Greenperturb} yields

$$\frac{1}{z\id-H}\;\piso^* \;=\; 
   \frac{1}{z\id -H_0}\; \piso^*\;
    \frac{1}{\id- V^\piso G_0^\piso(z)}\,,
$$

\noindent and remarking that $l\in\Lambda$. Hence, passing to the Fourier space leads to

$$\left(
   \Ff\;\frac{1}{z-H}\;|l\rangle
  \right) (k) \;=\;
    \sum_{n\in\Lambda}\;
     \frac{1}{z-\Ee(k)}\,e^{\imath k\cdot n}\,
      \langle n|
       \left(
         \id -V^\piso G_0^\piso(z)
       \right)^{-1}\,|l\rangle\,.
$$

\noindent Replacing this in the above expression for $\widehat{\Omega}_\pm-{\bf 1}$ completes the proof.
\hfill $\Box$

\vspace{.3cm}

 \subsection{The wave operator in the REF representation}
 \label{scatt08.ssect-waveREF}

\noindent In this section, the REF representation will be used to calculate the wave operator $\widetilde{\Omega}_\pm=\Uu\widehat{\Omega}_\pm\Uu^*$ in dimension $d\geq 3$. It is an operator on  $L^2(\RM)\otimes L^2(\Sigma,\nu)$.  From Proposition~\ref{scatt08.prop-waveop}, the definition \eqref{scatt08.eq-Udef}, the change of variables formula \eqref{scatt08.eq-varchange3} and the definition \eqref{scatt08.eq-ONbasis} of the states $\psi_{m,b}$, it follows that

$$
((\widetilde{\Omega}_\pm-{\bf 1})\phi)_b \,=\,
 \lim_{\epsilon\downarrow 0}\;
  \int db'\;
   \sum_{n,m\in\Lambda} 
    |\psi_{n,b}\rangle\;
     \frac{
       \langle n|\,T(f^{-1}(b')\mp\imath \epsilon)\,|m\rangle
          }{f^{-1}(b')\mp\imath \epsilon-f^{-1}(b)}\;
        \langle \psi_{m,b'}|\phi_{b'}\rangle\,,
$$

\noindent where $\langle \psi_{m,b'}|\phi_{b'}\rangle$ stands for the inner product in the Hilbert space $L^2(\Sigma,\nu)$ and the integral of $b'$ carries over $\RM$. Thanks to Corollary~\ref{scatt08.lem-partialiso}, the sums over $n$ and $m$ can be computed to give

\begin{equation}
\label{eq-Omegaformula}
((\widetilde{\Omega}_\pm-{\bf 1})\phi)_b\,=\,
   \lim_{\epsilon\downarrow 0}
    \int \frac{db'}{\pi}\,
     \frac{F(f^{-1}(b))^{\frac{1}{2}}\,F(f^{-1}(b'))^{\frac{1}{2}}}
          {f^{-1}(b')\mp\imath \epsilon-f^{-1}(b)}
     \,\Pi_b^*\,
      \bigl|\Im m\,G^\piso_0(f^{-1}(b))\bigr|^{\frac{1}{2}}
       \bigl(e^{\frac{b'}{2}}+e^{-\frac{b'}{2}}\bigr)\,
        (\widetilde{O}_{\pm}\phi)_{b'}\,,
\end{equation}

\noindent where $\widetilde{O}_{\pm}=\int\! db\,\widetilde{O}_{\pm,b}$ with

\begin{equation}
\label{scatt08.eq-Odef}
\widetilde{O}_{\pm,b} \;=\;
 \lim_{\epsilon\downarrow 0}\;
  \frac{1}{e^{\frac{b}{2}}+e^{-\frac{b}{2}}}\;
   T^\piso(f^{-1}(b)\mp\imath \epsilon)\; 
   \bigl|\Im m\,G^\piso_0(f^{-1}(b))\bigr|^{\frac{1}{2}}\;
    \Pi_{b}\;.
\end{equation}

\noindent It is part of the proof of the following result to show that the limit in \eqref{scatt08.eq-Odef} exists and that the expression \eqref{scatt08.eq-Tmatrix} for the $T$-matrix can be replaced to give

\begin{equation}
\label{scatt08.eq-Odef2}
\widetilde{O}_{\pm,b}\;=\;
 \frac{1}{e^{\frac{b}{2}}+e^{-\frac{b}{2}}}\;
  \Bigl(\,(V^\piso)^{-1}-\Re e \,G^\piso_0(f^{-1}(b))\,
   \mp\,\imath\,
    \bigl|\Im m\,G^\piso_0(f^{-1}(b))\bigr|\Bigr)^{-1}\; 
     \bigl|\Im m\,G^\piso_0(f^{-1}(b))\bigr|^{\frac{1}{2}}
      \;\Pi_{b}\,.
\end{equation}

\begin{theo}
\label{scatt08.theo-waveop3d}
Let $d\geq 3$ and let $V$ have finite support. In addition, $F$ is chosen as in {\rm \eqref{scatt08.eq-Fchoice}} and the following will be assumed:

\vspace{.1cm}

\noindent {\rm (i)} If $d=3$, the threshold singularities have multiplicity at most $1$ and any vector $w_\pm$ in the kernel 

of $(V^\piso)^{-1}-G^\piso_0(E_\pm)$ has a Fourier transform satisfying $\hat{w}_\pm(k^*_\pm)\not = 0$.

\vspace{.1cm}

\noindent {\rm (ii)} If $d=4$, there are no threshold singularities.  

\vspace{.1cm}

\noindent {\rm (iii)} The embedded eigenvalues lie neither on the critical values of $\Ee$ nor in the set $\vs$ described in 

{\rm Proposition~\ref{scatt08.th-Ffbpoly}}. The corresponding zeros of $b\in\RM\mapsto (V^\piso)^{-1}-G^\piso_0(f^{-1}(b))$  are of first order in the 

real part.

\vspace{.1cm}

\noindent Then the operators $\widetilde{O}_{\pm,b}$ are well-defined, continuous in $b$ and uniformly bounded. The wave operators are given by

\begin{equation}
\label{scatt08.eq-Omegarep3d2}
\widetilde{\Omega}_\pm \;=\;
 {\bf 1}+
  \sum_{\kappa=\pm 1}
   \imath\;\Pi_{\widetilde{B}}^*\;
    \bigl|\Im m\,G^\piso_0(f^{-1}({\widetilde{B}}))\bigr|^{\frac{1}{2}}\,
     e^{\kappa\frac{\widetilde{B}}{2}}\;
     \left(
       \pm\,{\bf 1}\,+\,
       \frac{e^{\pi\widetilde{A}}-\kappa\,\imath}
            {e^{\pi\widetilde{A}}+\kappa\,\imath}
     \right)
\,\widetilde{O}_{\pm}\,.
\end{equation}
\end{theo}

Formula \eqref{scatt08.eq-Omegarep3d2} shows that the wave operator can be calculated in terms of $\widetilde{O}_{\pm}$ and the dilation operator $\widetilde{A}$. It is similar to those obtained by Kellendonk and Richard for continuous scattering systems \cite{KR2,KR3,KR4}, however, we stress that we also allow for embedded eigenvalues, a fact that is closely linked to proving that the inverse in \eqref{scatt08.eq-Odef2} exists. In fact, if the kernels of $(V^\piso)^{-1}-\Re e\,G^\piso_0(E)$ and of $\Im m\,G^\piso_0(E)$ have a non-trivial intersection, the associated pole in \eqref{scatt08.eq-Odef2} is attained on the orthogonal complement of the range $\Ff_b$ of $\bigl|\Im m\,G^\piso_0(f^{-1}(b))\bigr|^{\frac{1}{2}}$ and thus one can prove that the inverse on $\Ff_b$ exists (in the sense of Lemma~\ref{scatt08.lem-kerA+iB}). Energies at which this happens are exactly the embedded eigenvalues by Proposition~\ref{scatt08.prop-embedded}. The intersection of the kernels is supposed to be a regular singular point and this allows to argue for the continuity of $\widetilde{O}_{\pm,b}$. Hypothesis (iii) holds generically for $V^\piso$ (within the non-generic situations of embedded eigenvalues).  Another comment is that the condition (i) imposed for $d=3$ implies that the threshold singularity is a threshold resonance. This is because, by Proposition~\ref{scatt08.prop-halfbounded}, $\hat{w}_\pm(k^*_\pm)\not =0$ implies $\Ss_{E_\pm}\cap\Tt_\pm=\{0\}$. Again this is the generic behavior in dimension $d=3$. It is possible to treat threshold singularities for $d=4$, but this is technically more involved and not carried out here.

\vspace{.2cm}

\noindent {\bf Proof} of Theorem~\ref{scatt08.theo-waveop3d}: Let us first suppose that $\widetilde{O}_{\pm}$ are well-defined and bounded with fibers $\widetilde{O}_{\pm,b}$ depending continuously on $b$, and then show how \eqref{scatt08.eq-Omegarep3d2} follows from \eqref{eq-Omegaformula}.  Thanks to the formulas \eqref{scatt08.eq-Fchoiceconclusion}, $\Ee(\theta_{b}(\sigma))=f^{-1}(b)=E_r+\Delta\tanh(b)$ and $F(f^{-1}(b))=\Delta\cosh^{-2}(b)$, a bit of algebra now leads to
$$((\widetilde{\Omega}_\pm-{\bf 1})\phi)_b \;=\;
   \Pi^*_b\;
    \bigl|\Im m\;G^\piso_0(f^{-1}(b))\bigr|^{\frac{1}{2}}\,
     \int \frac{db'}{\pi}\;
      \frac{1}{\sinh(b'-b)\mp\imath 0} \;
       \bigl(e^{\frac{b'}{2}}+e^{-\frac{b'}{2}}\bigr)\;
        (\widetilde{O}_{\pm}\phi)_{b'}\,.
$$
\noindent In the previous formula, $\widetilde{O}_{\pm}\phi$ is a vector in  the Hilbert space $L^2(\RM)\otimes \CM^{|\Lambda|}$. As previously let $\widetilde{A}=-\imath\partial_b$ be the generator of the translation group in $L^2(\RM)\otimes L^2(\Sigma,\nu)$ as well as $L^2(\RM)\otimes \CM^{|\Lambda|}$. Changing the integration variable $b'$ to $u=b'-b$ leads to  $(\widetilde{O}_{\pm}\phi)_{u+b}=(e^{\imath \widetilde{A}u}\widetilde{O}_{\pm}\phi)_{b}$. Hence
$$
\Big((\widetilde{\Omega}_\pm-{\bf 1})\phi\Big)_b\;=\;
 \sum_{\kappa=\pm 1}\;
  \Pi_b\;
   \bigl|\Im m\;G^\piso_0(f^{-1}(b))\bigr|^{\frac{1}{2}}\,
    e^{\kappa\frac{b}{2}}
     \int\frac{du}{\pi}\;
       \frac{1}{\sinh(u)\mp\imath 0} \;
        e^{\kappa\frac{u}{2}}
         \left(e^{\imath \widetilde{A}u}\widetilde{O}_{\pm}\phi
         \right)_{b}\,.
$$
\noindent Now \eqref{scatt08.eq-Omegarep3d2} is obtained from the following identity:
$$\int\frac{du}{\imath\pi}\;
 \frac{1}{\sinh\bigl(u\bigr)\mp\imath 0}\;
   e^{\kappa\frac{u}{2}}\;
    e^{\imath \widetilde{A}u}\;=\;
     \pm\one\;+\;
      \frac{
        e^{\pi\widetilde{A}}-\kappa\,\imath
           }{
       e^{\pi\widetilde{A}}+\kappa\,\imath
            }\;.
$$

It remains to show the above mentioned properties of the operators $\widetilde{O}_{\pm,b}$ defined in \eqref{scatt08.eq-Odef2}. We first check that for every $b\in\RM$ they are well-defined and continuous in $b$. This analytical issue is tied to embedded eigenvalues. In fact, away from them there would be nothing to prove because the inverse in \eqref{scatt08.eq-Odef2} exists by \eqref{eq-embedcount} and the fact that $A+\imath B$ is invertible for any invertible operator $A=A^*$ and non-negative operator $B\geq 0$ (see the proof of Lemma~\ref{scatt08.lem-kerA+iB}). Now focussing on embedded eigenvalues, let us set
$$
A_b\;=\;
(V^\piso)^{-1}-\Re e \,G^\piso_0(f^{-1}(b))
\;,
\qquad
B_b\;=\;
\bigl|\Im m\,G^\piso_0(f^{-1}(b))\bigr|
\;.
$$
These matrices $A_b$ and $B_b$ have nothing to do with the dilation operator and the rescaled energy operator, and only appear again in the following lines and Appendix~\ref{app-inverses}. Then $A_b$ is self-adjoint and $B_b$ is non-negative, and by Proposition~\ref{scatt08.prop-Green3d}(i) both are real analytic in $b$ as long as $f^{-1}(b)$ is not a critical value of $\Ee$.  Now the properties of hypothesis (iii) of Theorem~\ref{scatt08.theo-waveop3d} guarantee that Lemma~\ref{lem-linindep} of Appendix~\ref{app-inverses} can be applied because, in particular, the zeros of $b\in\RM\mapsto A_b+\imath B_b$ are of first order in $A_b$. This lemma implies that $\widetilde{O}_{\pm,b}$ is even analytic for $b=f^{-1}(E)$ away from the set $\Ee(\cri)$ of critical energies and away from the exceptional set $\vs$ of Proposition~\ref{scatt08.th-Ffbpoly}. Continuity at the latter points follows again from Proposition~\ref{scatt08.prop-Green3d} because there are no embedded eigenvalues there. In particular, let us also note that the dimension of $\Pi_b$ changes at points $b$ corresponding to energies in $\vs$, but in \eqref{scatt08.eq-Odef2} this does not lead to discontinuities due to the factor $|\Im m\,G^\piso_0(f^{-1}(b))|^{\frac{1}{2}}$ directly following $\Pi_b$. 

\vspace{.1cm}

Now we check that the operator $\widetilde{O}_{\pm}$ seen as a linear map from $L^2(\RM)\otimes L^2(\Sigma,\nu)$ to $L^2(\RM)\otimes \CM^{|\Lambda|}$ is actually bounded. This means that we have to analyze the limits $b\to\pm\infty$ of $\widetilde{O}_{\pm}$, which depend on the behavior at the thresholds. Here the factor  $e^{\frac{b}{2}}+e^{-\frac{b}{2}}$ introduced in \eqref{scatt08.eq-Odef} will turn out to be crucial. We start by expanding the inverse in \eqref{scatt08.eq-Odef2} around the band edge $E_\pm$ using items (iv) and (v) of Proposition~\ref{scatt08.prop-Green3d}:
\begin{equation}
\label{scatt08.eq-expandinv}
(V^\piso)^{-1}-G^\piso_0(f^{-1}(b)\mp\imath 0)\;=\;
 (V^\piso)^{-1}-G^\piso_0(E_\pm)
  \;\mp\imath\;
   D_\pm M^\piso_\pm\,e^{-|b|(d-2)}+
    \Oo(e^{-|b|d},|b|e^{-2|b|})\,.
\end{equation}
\noindent If there is no threshold singularity, then by definition $(V^\piso)^{-1}-G^\piso_0(E_\pm)$ is invertible and thus the inverse in \eqref{scatt08.eq-Odef2} remains bounded as $b\to\pm\infty$ and the other factors lead to $\lim_{b\to\pm\infty}\widetilde{O}_{\pm,b}=0$. If there is a threshold singularity of multiplicity $\dim(\Ss_{E_\pm})=1$ in $d=3$, the operator $(V^\piso)^{-1}-G^\piso_0(E_\pm)$ has a kernel of dimension $1$ spanned by some vector $w_\pm$. If, in addition, $w_\pm$ is not orthogonal to the one-dimensional projection $M^\piso_\pm$, namely if $\langle v^\piso_\pm |w_\pm\rangle=|\Lambda|^{-\frac{1}{2}}\hat{w}(k^*_\pm)\not =0$, then $\bigl((V^\piso)^{-1}-G^\piso_0(f^{-1}(b)\mp\imath 0)\bigr)^{-1}=\Oo(e^{|b|})$. As $\bigl|\Im m\,G^\piso_0(f^{-1}(b))\bigr|^{\frac{1}{2}}=\Oo(e^{-\frac{|b|}{2}})$ (because $d=3$) the added prefactor $(e^{\frac{b}{2}}+e^{-\frac{b}{2}})^{-1}$ assures that $\widetilde{O}_{\pm,b}$ remains bounded as $b\to\pm\infty$. (Its convergence is analyzed in Proposition~\ref{scatt08.prop-R} below.)

\vspace{.1cm}

For $d\geq 5$, the term of order $e^{-|b|(d-2)}$ in \eqref{scatt08.eq-expandinv} is dominated by the terms of order $e^{-2|b|}$. Hence, \eqref{scatt08.eq-expandinv} becomes rather
$$(V^\piso)^{-1}-G^\piso_0(f^{-1}(b)\mp\imath 0)\;=\;
   (V^\piso)^{-1}-G^\piso_0(E_\pm) -N^\piso_\pm\,e^{-2|b|}+
    \Oo(e^{-3|b|})\,,
$$
\noindent which follows also from Proposition~\ref{scatt08.prop-Green3d}. In this case, $N^\piso_\pm$ is invertible so that for an arbitrary threshold singularity $\bigl((V^\piso)^{-1}-G^\piso_0(f^{-1}(b)\mp\imath 0)\bigr)^{-1}=\Oo(e^{2|b|})$, even in the most singular case where $(V^\piso)^{-1}=G^\piso_0(E_\pm)$. This is compensated in \eqref{scatt08.eq-Odef2} by the other factor  $(e^{\frac{b}{2}}+e^{-\frac{b}{2}})^{-1}\bigl|\Im m\,G^\piso_0(f^{-1}(b))\bigr|^{\frac{1}{2}}=\Oo(e^{-\frac{|b|}{2}(d-1)})$. Consequently $\widetilde{O}_{\pm}$ is bounded for $d\geq 5$.  This concludes the proof.
\hfill $\Box$

\vspace{.2cm}

 \subsection{The scattering operator}
 \label{scatt08.ssect-scat}

\noindent Whenever the wave operators are complete, the scattering operator is defined by:

$$
S\;=\;\Omega_+^* \Omega_-\,.
$$

\noindent It is unitary and satisfies $[S,H_0]=0$. Hence, in the REF representation, $[\widetilde{S},\widetilde{B}]=0$ and thus $\widetilde{S}=\int db\,\widetilde{S}_b$ with unitary operators $\widetilde{S}_b$ on $L^2(\Sigma,\nu)$. The intertwining relation and the invariance principle \eqref{scatt08.eq-invarprin} imply that for any admissible function $f$ with $f'>0$, one has

$$\mbox{\rm s-}\!\!\!\lim_{t\to\pm\infty}\;
   e^{\imath t f(H_0)}\,
    \Omega_\pm\, 
     e^{-\imath t f(H_0)}\;=\;
   \Omega_\pm^*\Omega_\pm \;=\;{\bf 1}\,,
\hspace{2cm}
  \mbox{\rm s-}\!\!\!\lim_{t\to\mp\infty}\;
   e^{\imath t f(H_0)}\,
    \Omega_\pm \,
     e^{-\imath t f(H_0)}\;=\;
      \Omega_\mp^*\Omega_\pm\,.
$$

\noindent The second expression is either $S$ or $S^*$. Let now $f$ be chosen as in \eqref{scatt08.eq-Fchoiceconclusion}. In the REF representation, Proposition~\ref{scatt08.prop-rep} then leads to

\begin{equation}
\label{scatt08.eq-Blimits}
\mbox{\rm s-}\!\!\!\lim_{t\to\pm\infty}\;
 e^{\imath t \widetilde{B}}\,
  \widetilde{\Omega}_\pm\, 
   e^{-\imath t \widetilde{B}} \;=\;
    {\bf 1}\,,
\hspace{1cm}
\mbox{\rm s-}\!\!\!\lim_{t\to\infty}\;
 e^{\imath t \widetilde{B}}\,
 \widetilde{\Omega}_-\,
  e^{-\imath t \widetilde{B}} \;=\;
   \widetilde{S}\,,
\hspace{1cm}
\mbox{\rm s-}\!\!\!\lim_{t\to-\infty}\;
 e^{\imath t \widetilde{B}}\,
  \widetilde{\Omega}_+\, 
   e^{-\imath t \widetilde{B}}\;=\;
    \widetilde{S}^*\,.
\end{equation}

\noindent Using the explicit formula for $\widetilde{\Omega}_-$ given in Theorem~\ref{scatt08.theo-waveop3d} now leads to an explicit expression for the on-shell scattering matrix. The structure of such formulas (in particular, the EF representation of the formula \eqref{scatt08.eq-Sformula} in the proof below) is well-known and has appeared in various guises (see \cite{New,Yaf} for a list of references).

\begin{theo}
\label{scatt08.theo-Sin3d}
Let the assumptions of {\rm Theorem~\ref{scatt08.theo-waveop3d}} hold. Then the on-shell scattering matrix $\widetilde{S}_b$ is a unitary operator on $L^2(\Sigma,\nu)$ depending continuously on $b$ and given by

$$\widetilde{S}_b\;=\;
 (\one-\piso_b^*\piso_b)+
  \piso_b^* (C_b-\imath)(C_{b}+\imath)^{-1}\piso_b\,,
$$

\noindent where the selfajoint $L\times L$ matrix $C_b:P_b\,\CM^{|\Lambda|}\to P_b\,\CM^{|\Lambda|}$ is defined by

$$C_b\;=\; P_b
  \bigl|\Im m\, G^\piso_0(f^{-1}(b))\bigr|^{-\frac{1}{2}}
   \Bigl((V^\piso)^{-1}-\Re e\,G^\piso_0(f^{-1}(b))\Bigr)
    \bigl|\Im m \,G^\piso_0(f^{-1}(b))\bigr|^{-\frac{1}{2}}\,
     P_b\,.
$$
\end{theo}

\noindent {\bf Proof:}  For any function $g$ the following formula holds $e^{\imath t \widetilde{B}}g(\widetilde{A}) e^{-\imath t \widetilde{B}}=g(\widetilde{A}-t)$. The limits $t\to\pm\infty$ can be taken whenever $g$ has limits at infinity. The function appearing in \eqref{scatt08.eq-Omegarep3d2} is of that type. The middle formula in equation~(\ref{scatt08.eq-Blimits}) and the expression of $\widetilde{\Omega}_-$ given in Theorem~\ref{scatt08.theo-waveop3d} leads to the calculation of $\widetilde{S}$, namely

\begin{equation}
\label{scatt08.eq-Sformula0}
\widetilde{S}_b\;=\;
 \one+
  \sum_{\kappa=\pm 1}
   \imath\;\Pi_{b}^*\;
    \bigl|\Im m\,G^\piso_0(f^{-1}(b))\bigr|^{\frac{1}{2}}\,
      e^{\kappa\frac{b}{2}}\;
      (-2)\;\widetilde{O}_{\pm,b}\,,
\end{equation}

\noindent Because Theorem~\ref{scatt08.theo-waveop3d} states that $\widetilde{O}_{\pm,b}$ is continuous in $b$, this formula already shows that $\widetilde{S}_b$ is continuous in $b$. Using equation~\eqref{scatt08.eq-Odef2}, it now follows that
$$\widetilde{S}_b\;=\;
   \one-2\,\imath\;
    \Pi_{b}^*\;
     \bigl|\Im m\,G^\piso_0(f^{-1}(b))\bigr|^{\frac{1}{2}}\,
      \Bigl((V^\piso)^{-1}-G^\piso_0(f^{-1}(b)+ \imath 0)\Bigr)^{-1}\,
       \bigl|\Im m\,G^\piso_0(f^{-1}(b))\bigr|^{\frac{1}{2}}\;
        \Pi_{b}\,.
$$

\noindent After simplification, one gets
\begin{equation}
\label{scatt08.eq-Sformula}
\widetilde{S}_b\;=\;
   \one-2\,\imath\;
     \piso^*_b(C_{b}+\imath)^{-1}\piso_b\,.
\end{equation}
\noindent This allows to prove the claim.
\hfill $\Box$

\vspace{.2cm}

Similar formulas hold for the  EF-representation of the scattering matrix. The comments made in Section~\ref{scatt08.sec-EFrep} and the results of Theorem~\ref{scatt08.theo-Sin3d} lead to (with $C_E=C_b$ for $b=f(E)$), 

$$\cS_E\;=\;
 \widetilde{S}_{f(E)}\;=\;
  (\one-\piso_{E}^*\piso_{E})+
   \piso_{E}^* (C_E-\imath)(C_E+\imath)^{-1}\piso_{E}\,.
$$

\noindent It is now possible to get results on the asymptotics of the scattering matrix.

\begin{proposi}
\label{scatt08.prop-Slimits}
Let the assumptions of {\rm Theorem~\ref{scatt08.theo-waveop3d}} hold. 

\vspace{.1cm}

\noindent {\rm (i)} If there are no threshold singularities or if $d\geq 5$, then $\lim_{b\to\pm\infty}\;\widetilde{S}_b=\one$. 

\vspace{.1cm}

\noindent {\rm (ii)} If for $d=3$ there is a threshold singularity of multiplicity $1$ at $E_\pm$ and the extremum of $\Ee$ at $E_\pm$ 

is isotropic in the sense that $\Ee''(k_\pm^*)$ is a multiple of the identity, then 

$$
\lim_{b\to\pm\infty}\;\widetilde{S}_b\;=\;
 \one -2\,|\psi_\pm\rangle\,\langle\psi_\pm|\,,
$$

where $\psi_\pm\in L^2(\Sigma,\nu)$ are the states given in {\rm Lemma~\ref{scatt08.lem-limitstate}}. 
\end{proposi}

\noindent {\bf Proof.} (i) If there are no threshold singularities, then $\lim_{b\to\pm\infty}\widetilde{O}_{\pm,b}=0$ as was shown in Section~\ref{scatt08.ssect-waveREF}. As $\bigl|\Im m\,G^\piso_0(f^{-1}(b))\bigr|^{\frac{1}{2}}\,e^{\kappa\frac{b}{2}}$ is bounded, it follows from \eqref{scatt08.eq-Sformula0} that $\lim_{b\to\pm\infty}\;\widetilde{S}_b=\one$. For $d\geq 5$, it has been shown that $\widetilde{O}_{\pm,b}$ is uniformly bounded even in the presence of threshold singularities. As the factor $\bigl|\Im m\,G^\piso_0(f^{-1}(b))\bigr|^{\frac{1}{2}}\,e^{\kappa\frac{b}{2}}$ vanishes in the limits $b\to\pm\infty$, the same conclusion holds thanks to equation~(\ref{scatt08.eq-Sformula0}).

\vspace{.2cm}

\noindent (ii) For $d=3$, it is assumed that there is a threshold singularity of multiplicity $1$ and the extrema are isotropic. Starting from \eqref{scatt08.eq-Sformula} the inverse of $C_b+\imath$ can be expanded using Proposition~\ref{scatt08.prop-Green3d} to give
\begin{eqnarray*}
(C_b+\imath)^{-1}  & = &
 D_\pm\, e^{-|b|}\,M^\piso_\pm
  \Bigl(
      (V^\piso)^{-1}-
       \Re e\,G^\piso(E_\pm)+
        \imath\,D_\pm\, 
         e^{-|b|}\,M^\piso_\pm
  \Bigr)^{-1}\, M^\piso_\pm +
   \Oo(e^{-\frac{|b|}{2}})\\
& =&
-\,\imath\,M^\piso_\pm +
\Oo(e^{-\frac{|b|}{2}})\,,
\end{eqnarray*}
\noindent where, in the second equality, Lemma~\ref{scatt08.lem-limit} stated below is used. It can indeed be applied thanks to the hypothesis stated in Theorem~\ref{scatt08.theo-waveop3d}. On the other hand $\Rr_b v^\piso_\pm=\psi_{0,b}+\Oo(e^{\mp b})$ so that it follows that $\Pi^*_b M^\piso_\pm\Pi_b= \|\psi_{0,b}\|^{-2}\,|\psi_{0,b}\rangle\langle\psi_{0,b}|+\Oo(e^{\mp b})$. Hence Lemma~\ref{scatt08.lem-limitstate} allows to conclude the proof.
\hfill $\Box$

\begin{lemma}
\label{scatt08.lem-limit}
Let $P\in \mbox{\rm Mat}(n\times n,\CM)$ be a one-dimensional orthogonal projection and $A=A^*\in \mbox{\rm Mat}(n\times n,\CM)$ have a one-dimensional kernel not lying in the kernel of $P$. Then 
$$
\lambda P(A+\imath\lambda P)^{-1}P\;=\;-\,\imath \,P+\Oo(\lambda)
\;.
$$
\end{lemma}

\noindent  {\bf Proof:} This follows from a short calculation using Cramer's rule.
\hfill $\Box$

\vspace{.3cm}

 \subsection{The contributions of the threshold singularities}
 \label{scatt08.sec-threshold}

\noindent Just as the scattering operator is obtained as a rescaled energy boost of the wave operator in \eqref{scatt08.eq-Blimits}, it is natural to study the dilation operator boost of the wave operator. As the scattering operator is obtained as boost of $\widetilde{\Omega}_-$, it is sufficient to consider

$$
\widetilde{R}_\pm \;=\;
 \mbox{\rm s-}\!\!\!\lim_{t\to\pm\infty}\;
   e^{\imath t \widetilde{A}}\,
    \widetilde{\Omega}_-\, 
     e^{-\imath t \widetilde{A}}\,.
$$

\noindent It ought to be remarked that the limits in $\widetilde{R}_\pm$ approach the critical values $E_\pm$ respectively. This is because the identity $e^{\imath t \widetilde{A}}g(\widetilde{B}) e^{-\imath t \widetilde{A}}=g(\widetilde{B}+t)$ holds for any function $g$. From the definition it follows that $[\widetilde{R}_\pm,\widetilde{A}]=0$ so that
$\widetilde{R}_\pm=\int da\,\widetilde{R}_{\pm,a}$ with operator fibers
$\widetilde{R}_{\pm,a}$ acting on $L^2(\Sigma,\nu)$. Since the operator $ \widetilde{A}$ has continuous spectrum, it follows that $\lim_{t\rightarrow \pm\infty} e^{\imath t \widetilde{A}}K e^{-\imath t \widetilde{A}}=0$ for any compact operator $K$. In particular, since $\Omega_-$ is unitary modulo  a compact operator, $\widetilde{R}_\pm$ is unitary. Consequently each $\widetilde{R}_{\pm,a}$ is unitary. The following operators are associated to $\widetilde{R}_\pm$ in a similar manner as the time delay is associated to the scattering operator:,

$$
\widetilde{T}_\pm\;=\;
 \pm\;\frac{1}{\imath}\;
  (\widetilde{R}_\pm)^{-1}\,
   [\widetilde{B},\widetilde{R}_\pm]\,.
$$

\begin{proposi}
\label{scatt08.prop-R}
Let the assumptions of {\rm Theorem~\ref{scatt08.theo-waveop3d}} hold. 

\vspace{.1cm}

\noindent {\rm (i)} If there are no threshold singularities or if $d\geq 5$, then $\widetilde{R}_\pm=\one$ and $\widetilde{T}_\pm=0$.

\vspace{.1cm}

\noindent {\rm (ii)} If $d=3$, if there is a threshold resonance at $E_\pm$ of multiplicity $1$ and if the extrema of $\Ee$ are 

isotropic, then

$$\widetilde{R}_{\pm,a}\;=\;
   \Bigl(\one\,-\,|\psi_\pm\rangle\langle\psi_\pm|\Bigr)\;+\;
    \frac{e^{\pi{a}}\mp\,\imath}{e^{\pi{a}}\pm\,\imath}\;
     |\psi_\pm\rangle\langle\psi_\pm|\,,
\hspace{2cm} 
\TR({T}_\pm)\;=\;\frac{1}{2}\,.
$$
\end{proposi}

\noindent  {\bf Proof:} Statement (i) is a consequence of the proof of Theorem~\ref{scatt08.theo-waveop3d} and the arguments below, so let us focus on the case $d=3$. Equation~(\ref{scatt08.eq-Omegarep3d2}) gives an explicit expression of the wave operator. Each term in the sum over $\kappa =\pm 1$ in its r.h.s. is the product of three terms that can be treated separately under the boost action. The first factor is

\begin{equation}
\label{scatt08.eq-firstfac}
\lim_{b\to\pm\infty}
   \Pi_{b}^*\;
    \bigl|\Im m\,G^\piso_0(f^{-1}(b))\bigr|^{\frac{1}{2}}\,
     e^{\kappa\frac{b}{2}}\;=\;
      (D_\pm)^\frac{1}{2}\;
       \delta_{\kappa,\pm}\;
        |\psi_\pm\rangle\langle v^\piso_\pm|\,,
\end{equation}

\noindent where $\delta_{\kappa,\pm 1}$ is the Kronecker delta. Details for the proof of the second equality are similar to the proof of Proposition~\ref{scatt08.prop-Slimits}. Taking into acount the Ket $\langle v^\piso_\pm|$ in \eqref{scatt08.eq-firstfac} permits to treat the last factor using Lemma~\ref{scatt08.lem-limit}. This gives
$$\lim_{t\to\pm\infty}\;
   e^{\imath t \widetilde{A}}\,
   \langle v^\piso_\pm|\,
    \widetilde{O}_{\pm}\, 
     e^{-\imath t \widetilde{A}}\;=\;
  \lim_{b\to\pm\infty}
   \langle v^\piso_\pm|
    \Bigl(\,(V^\piso)^{-1}-G^\piso_0(f^{-1}(b)+\imath 0)\,\Bigr)^{-1}\; 
     \frac{
        \bigl|\Im m\,G^\piso_0(f^{-1}(b))\bigr|^{\frac{1}{2}}
          }{
        e^{\frac{b}{2}}+e^{-\frac{b}{2}}
           }\;\Pi_{b}\,,
$$

\noindent The limit on the r.h.s. is given by $\imath (D_\pm)^{-\frac{1}{2}} \langle \psi_\pm|$. Replacing these factors leads to the formula for $R_{\pm,a}$. Using the integral over $a$, the trace of $T_\pm$ is given, up to the sign, by the rotation number of the map $a\in\RM\mapsto (e^{\pi a}\mp\,\imath)/(e^{\pi a}\pm\,\imath)\in\SM^1$. The latter is equal to $\pm\,\frac{1}{2}$. The sign in the definition of $T_\pm$ compensate the previous sign leading to the final result.
\hfill $\Box$

\vspace{.3cm}

 \subsection{The time delay operator}
 \label{scatt08.ssect-timedelay}

\noindent The time delay operator $T$ is the derivative of the scattering matrix w.r.t. the energy (the notation $T$ should not be confused the $T$-matrix). More formally, it is defined by $T=\frac{1}{\imath}\,S^{-1}[A,S]$ whenever $S$ is differentiable w.r.t. to the dilation $A$. In the REF it becomes

$$\widetilde{T}\;=\;
   \int db\;\widetilde{T}_b\,,
\hspace{2cm}
     \widetilde{T}_b\;=\;
      \frac{1}{\imath}\,
       (\widetilde{S}_b)^{-1}
        \partial_b\widetilde{S}_b\,,
$$

\noindent while in the EF representations it is given by

$$\cT\;=\; 
   \int^{E_+}_{E_-}dE\;\cT_E\,,
\hspace{2cm}
   \cT_E\;=\;
    \frac{1}{\imath}\;
     (\cS_E)^{-1}
      \partial_E\cS_E\,.
$$

\noindent The {\em total time delay} is the trace of $T$. The formula given in the following result is sometimes called the {\em spectral property} of the time-delay \cite{TO,New}

\begin{theo}
\label{scatt08.theo-Tin3d}
Let the assumptions of {\rm Theorem~\ref{scatt08.theo-waveop3d}}. In addition, suppose that $\Ff_E=\CM^{|\Lambda|}$ for almost all $E$ and that there are no threshold eigenvalues. Then, for almost all $E\in [E_-,E_+]$,

\begin{equation}
\label{scatt08.eq-Tin3d}
\TR_{L^2(\Sigma,\nu)}(\cT_E)\;=\;
 \lim_{\epsilon\downarrow 0}\;
  2\;\Im m \;\TR_{\ell^2(\ZM^d)}
   \left(
     \frac{1}{E-\imath \epsilon-H} \;-\;
     \frac{1}{E-\imath \epsilon-H_0}
   \right)\,.
\end{equation}
\end{theo}
\begin{rem}
\label{scatt08.rem-Tin3da}
{\em The condition $\Ff_E=\CM^{|\Lambda|}$ for almost all $E$, implies that $H$ has no embedded eigenvalues (see Proposition~\ref{scatt08.prop-embedded}).  If $H$ has embedded eigenvalues at energy $E$ with multiplicity $n(E)$, then the r.h.s. of equation (\ref{scatt08.eq-Tin3d}) must be modified by subtracting $n(E)/\epsilon$ to the trace to compensate for the singularity occurring at this energy. One should also be able to deal with a threshold singularity by subtracting the adequate contribution. However, no details are provided here.
}
\hfill $\Box$
\end{rem}
\begin{rem}
\label{scatt08.rem-Tin3db}
{\em The previous formula for the total time delay is well-known for potential scattering when $H_0$ is the Laplacian in $\RM^d$ (see \cite{CN} for a list of references). It can be proved by a direct calculation in the REF representation (following the lines of \cite{TO}) or by a computation inspired by the Birman-Krein formula \cite{Yaf} for the scattering phase, an approach used below.
}
\hfill $\Box$
\end{rem}

\noindent {\bf Proof of Theorem~\ref{scatt08.theo-Tin3d}: }The following notation will be used $\Pi_E=\piso_{f(E)}$, $C_E=C_{f(E)}$ etc. From equation~\eqref{scatt08.eq-Sformula}, it follows that
$$
\frac{1}{2\imath}\,\partial_E\cS_E\;=\;
 -\partial_E\Pi^*_E(C_E+\imath)^{-1}\Pi_E +
   \Pi^*_E(C_E+\imath)^{-1}\partial_E C_E(C_E+\imath)^{-1}\Pi_E-
    \Pi^*_E(C_E+\imath)^{-1}\partial_E\Pi_E\,.
$$
\noindent The equation $\Pi_E\Pi_E^*=\one_{\CM^{|\Lambda|}}$ implies $\partial_E\Pi_E\Pi_E^*=-\Pi_E\partial_E\Pi_E^*$. Hence,
$$\TR_{L^2(\Sigma,\nu)}(\cT_E)\;=\;
 \frac{1}{\imath}\;
  \TR_{\Ff_E}
   \left(
     \Pi_E(\cS_E)^*\Pi_E^*\,
      \Pi_E\partial_E \cS_E \Pi_E^*
   \right)\;=\;
    2\,\TR_{\CM^{|\Lambda|}}
     \left(
       (C_E^2+1)^{-1}\partial_E C_E
     \right)\,.
$$
\noindent This can be rewritten as
$$\TR_{L^2(\Sigma,\nu)}(\cT_E)\;=\;
   \frac{1}{\imath}\;
    \partial_E\,
     \ln\,\det
      \left(
       \frac{C_E-\imath}{C_E+\imath}
      \right)\,.
$$
\noindent Therefore dividing out the imaginary part of the Green function appearing in the definition of $C_E$ (see Theorem~\ref{scatt08.theo-Sin3d}) gives
\begin{eqnarray*}
\TR_{L^2(\Sigma,\nu)}(\cT_E)
& = &
\frac{1}{\imath}\;
\partial_E\,\ln\,\mbox{det}\Bigl(
P_E\bigl((V^\piso)^{-1}-G^\piso_0(E-\imath 0)\bigr)P_E
\bigl(P_E((V^\piso)^{-1}-G^\piso_0(E+\imath 0))P_E\bigr)^{-1}\Bigr)
\\
& = & 2\;\Im m\;\partial_E\,\ln\,\mbox{det}
\bigl(P_E((V^\piso)^{-1}-G^\piso_0(E-\imath 0))P_E\bigr)
\,.
\end{eqnarray*}
%
\noindent On the other hand, using \eqref{scatt08.eq-Greenperturb2} and the cyclicity of the trace, leads to
$$
\TR_{\ell^2(\ZM^d)}
 \bigl((z-H)^{-1}-(z-H_0)^{-1}\bigr)\;=\;
  \TR_{\CM^{|\Lambda|}}
   \Bigl(
    \piso(z-H_0)^{-2}\piso^*
     \bigl((V^\piso)^{-1}-G^\piso_0(z)\bigr)^{-1}
   \Bigr)\,.
$$
\noindent Since $\partial_zG^\piso_0(z)=-\piso(z-H_0)^{-2}\piso^*$, it follows that
$$
\TR_{\ell^2(\ZM^d)}
 \bigl((z-H)^{-1}-(z-H_0)^{-1}\bigr)\;=\;
  \partial_z\;
   \TR_{\CM^{|\Lambda|}}
    \Bigl( \ln((V^\piso)^{-1}-G^\piso_0(z)\bigr)\Bigr)\,.
$$
\noindent This leads to the identity.
\hfill $\Box$

\vspace{.3cm}

 \subsection{A Levinson-type theorem}
 \label{scatt08.ssect-Levinson}

\begin{theo}
\label{scatt08.theo-Levinson3d}
Let the assumptions of {\rm Theorem~\ref{scatt08.theo-waveop3d}} hold. Further let $N=\TR({P}_{\mbox{\rm\tiny pp}})$ be the number of bound states of $H$, including embedded eigenvalues and threshold eigenvalues. Then

\begin{equation}
\label{scatt08.eq-Levinson}
\frac{1}{2\pi}\;
 \TR_{\ell^2(\ZM^d)}(T)\;+\;N\;=\;
  \left\{
   \begin{array}{cc}
    -\,\frac{1}{2}\,\dim(\Ss_{E_+})-
        \frac{1}{2}\,\dim(\Ss_{E_-})
     & \mbox{\rm if }d=3\,,\\
     0 & \mbox{\rm if }d\geq 4\,,
\end{array}
\right.
\end{equation}

\noindent where $\dim(\Ss_{E_\pm})\in\{0,1\}$ is the multiplicity of the threshold resonance.
\end{theo}

Two proofs will be provided. The first one will require stronger hypothesis to show how the argument principle combined with the spectral property of the time delay can be used as in most standard references \cite{RS,New}. It might be possible to lift these hypothesis with more technical effort (see Section~\ref{sec-example} where this is done for a point interaction). The second proof requires only the stated hypothesis. It is based on the approach proposed by Kellendonk and Richard \cite{KR06} and uses on the index theorem for \CsS. 

\vspace{.2cm}

\noindent {\bf First proof} of Theorem~\ref{scatt08.theo-Levinson3d}: It will be assumed that there are no embedded eigenvalues and no threshold singularities and that $\Ff_E=\CM^{|\Lambda|}$ for almost all $E$ (these assumptions hold for the case of a single site impurity). The number $N$ of eigenvalues is obtained by counting the poles of the resolvent using the Cauchy formula and a contour integration. The contour is given by two circles, one large counterclockwise oriented circle $\gamma$ around the spectrum of $H$ and a second small clockwise oriented circle $\Gamma$ around the spectrum of $H_0$ (but not touching it). Then

\begin{equation}
\label{scatt08.eq-argprin}
N\;=\;
 \oint_{\Gamma+\gamma}
  \frac{dz}{2\pi\imath}\;
   \TR_{\ell^2(\ZM^d)}
    \bigl((z-H)^{-1}-(z-H_0)^{-1}\bigr)\,.
\end{equation}

\noindent The resolvent identity implies that the contribution of $\gamma$ vanishes in the limit where its radius goes to infinity. Then let $\Gamma$ converge to the concatenation of the two intervals $[E_-+\imath 0,E_++\imath 0]$ and $[E_--\imath 0,E_+-\imath 0]$. Since it has been assumed that there is no threshold singularity, the regularity of the Green function at the band edges implies that the small circle connecting these contours near the band edges have a vanishing contribution in the contour integral. Thus

$$N\;=\;
 \int_{E_-}^{E_+}
  \frac{dE}{2\pi\imath}\;
   \TR_{\ell^2(\ZM^d)}
    \left(
     \frac{1}{E+\imath 0-H}-
     \frac{1}{E+\imath 0-H_0}-
     \frac{1}{E-\imath 0-H}+
     \frac{1}{E-\imath 0-H_0}
    \right)\,.
$$

\noindent The formula for the total time delay, proved in Theorem~\ref{scatt08.theo-Tin3d}, gives

$$N\;=\; -\,
   \frac{1}{2\pi}\;
     \int^{E_+}_{E_-}dE\;\TR_{L^2(\Sigma,\nu)}(\cT_E)\;.
$$

\noindent The r.h.s. is nothing but the trace of $T$ expressed in the EF representation, leading to the result.
\hfill $\Box$

\vspace{.2cm}

\noindent {\bf Second proof} of Theorem~\ref{scatt08.theo-Levinson3d}: As a preamble let us construct an extension the algebra $C_0(\RM)$ of the continuous functions on $\RM$ vanishing at $+\infty$ and $-\infty$. Then set $C_{\infty}(\RM)=C(\overline{\RM})$ where $\overline{\RM}= \{-\infty\}\cup \RM\cup\{+\infty\}$ which are the continuous functions having limits at $+\infty$ and $-\infty$. The evaluation map $\mbox{ev}$ is the $\ast$-homomorphism defined by $\mbox{ev}:g\in C(\overline{\RM})\mapsto \mbox{ev}(g)= (g(+\infty),g(-\infty))\in \CM^2=\CM\oplus \CM$. The kernel of this map is precisely $C_0(\RM)$ leading to a short exact sequence $0\to C_0(\RM)\hookrightarrow C_\infty(\RM)\overset{\mbox{\rm\tiny ev}}{\to}\CM\oplus \CM\to 0$. Next let us consider a two-dimensional version of this extension. Let $C_\infty(\RM^2)$ be those continuous functions on $\RM^2$ having a continuous limit function built from the limits of $f$ in the directions of $\RM^2$. These directions can described by $\SM^1$ so that one obtains an exact sequence is $0\to C_0(\RM^2)\hookrightarrow C_\infty(\RM^2)\overset{\mbox{\rm\tiny ev}}{\to}C(\SM^1)\to 0$.  On the other hand the limit points at infinity can also be seen as a square the corners of which being given by points of coordinates $(\pm\infty,\pm\infty)$. Equivalently, $C_\infty(\RM^2)$ can be seen as the subalgebra of the $C^\ast$-tensor product $C_\infty(\RM)\otimes C_\infty(\RM)$ generated by functions that are continuous at the four corners of the square at infinity (let us point out that there is a unique $C^\ast$-norm on the tensor product since these algebras are abelian). The exact sequence relevant for the proof of Levinson's theorem is a non-commutative analog of this, where the two commuting coordinate functions in $C_\infty(\RM)\otimes C_\infty(\RM)$ are replaced by the operators $\widetilde{A}$ and $\widetilde{B}$ obeying the canonical commutation relations, and then everything is tensorized with the $C^\ast$-algebra of compact operators.

\vspace{.2cm}

Let $\js$ be the \Cs generated by operators of the form $f(\widetilde{A})\otimes K$ and $g(\widetilde{B})\otimes K$ with $f,g\in C_0(\RM)$ and $K$ in the set $\Kk$ of compact operators on $L^2(\Sigma,\nu)$. Let $\es$ denote the extension of $\js$ obtained by allowing $f$ and $g$ to have nonzero finite limits at $\pm\infty$. Evaluation at infinity of $\es$ gives the algebra $\as$ which is the subalgebra of $\bigl(C_\infty(\widetilde{A})\oplus C_\infty(\widetilde{B})\oplus C_\infty(\widetilde{A})\oplus C_\infty(\widetilde{B})\bigr)\otimes\Kk$ of operators having coinciding limits at each of the four corners. This leads to the following short exact sequence (see \cite{KR06}):

$$0\,\to\,\js\,
    \hookrightarrow\,\es\,
     \overset{\mbox{\rm\tiny ev}}{\to}\,\as\,\to\,0\,,
$$

\noindent It induces a six-term exact sequence in $K$-theory \cite{Bla98}, leading to a canonical index map $\mbox{\rm Ind}: K_1(\as)\to K_0(\js)$. As it turns out, using the unitary operators $\Uu\Ff$ where $\Uu$ is defined in equation (\ref{scatt08.eq-Udef}) and $\Ff$ is the Fourier transform, the elements of $\js$ are mapped into compact operators on $\ell^2(\ZM^d)$. Therefore $K_0(\js)=\ZM$ \cite{Bla98}. Hence the index is an integer. At this level of generality, the index is obtained as follows. Given a unitary element $S$ in $\as$, it is lifted to an element $\Omega\in\es$ such that $\mbox{ev}(\Omega)=S$. Then, $\mbox{ev}(\Omega\Omega^\ast-\id)=0=\mbox{ev}(\Omega^\ast\Omega-\id)$. This means that, if $\Omega$ can be chosen to be a partial isometry, then both $\Omega\Omega^\ast-\id$ and $\Omega^\ast\Omega-\id$ are compact projections. Therefore the index of $S$ is defined by $\mbox{\rm Ind}(S)= \TR(\Omega\Omega^\ast- \Omega^\ast\Omega)$ which is indeed an integer. At a more computational level, if there is a derivation $\partial$ acting in $\as$ and if there is a faithful $\partial$-invariant trace on $\as$, the index of $S$ is computed as $\mbox{\rm Ind}(S)= \TR(S^{-1}\partial S)/(2\pi)$ \cite{Bla98}. 

\vspace{.2cm}

In the present situation, combining the $S$-matrix, the operators $R_\pm$ and $\id$, one for each side of the square at infinity, make up a unitary element $\widetilde{S}_{\mbox{\rm\tiny tot}}=\widetilde{S}\oplus\widetilde{R}_+\oplus{\bf 1}\oplus\widetilde{R}_-$ of $\as$. In addition, in both cases, thanks to equation~(\ref{scatt08.eq-Blimits}) and to the definition of $R_\pm$ (see Section~\ref{scatt08.sec-threshold}), this unitary operator can be seen as $\mbox{ev}(\Omega_-)$ where $\Omega_-$ is one of the wave operators. If it can be proved that $\Omega_-\in\es$, then Levinson's theorem follows immediately  from the following two remarks: 

\vspace{.1cm}

\noindent (i) $\Omega_-^\ast\Omega_-=\id$ and $\id -\Omega_-\Omega_-^\ast$ is the projection onto the pure point spectrum of $H$. 

\vspace{.1cm}

\noindent (ii) The algebra $\as$ admits a trace $\TR_{\as}$ which consists of the trace on $\Kk$ followed by the integrals over $a$ and $b$ respectively along each side of the square at infinity. In addition, the derivatives $\partial_a$ and $\partial_b$ define a derivation $\partial_{\as}$ acting on $\as$ which leaves $\TR_{\as}$ invariant. These derivatives are also given by the commutators with $\widetilde{A}$ or $\widetilde{B}$ respectively. Therefore, the index of $\widetilde{S}_{\mbox{\rm\tiny tot}}$ is given by

$$\mbox{\rm Ind}(\widetilde{S}_{\mbox{\rm\tiny tot}})\;=\;
   \TR \left(\Omega\Omega^\ast- \Omega^\ast\Omega\right) \;=\;
    \frac{1}{2\pi} \;\TR_{\as}
     \left((\widetilde{S}_{\mbox{\rm\tiny tot}})^{-1}\partial_{\as} \widetilde{S}_{\mbox{\rm\tiny tot}}\right)\,,
$$

\noindent which is exactly equation~(\ref{scatt08.eq-Levinson}) when written in the language used previously. 

\vspace{.2cm}

The main remaining point is to prove that the wave operators are elements of the \Cs $\es$. In some sense, most of the technical preparations above have been dedicated to proving precisely this point. The formula in Theorem~\ref{scatt08.theo-waveop3d} shows that  $\widetilde{\Omega}_\pm$ is given by a concatenation of three operators: first the operators $\widetilde{O}_\pm=\int db\;\widetilde{O}_{\pm,b}$ which by Theorem~\ref{scatt08.theo-waveop3d} has fibers $\widetilde{O}_{\pm,b}$ depending continuously on $b$ with limits at $b=\pm\infty$ (due to Proposition~\ref{scatt08.prop-R}), then a smooth function of $\widetilde{A}$ which also has limits at infinity, and finally the imaginary part of the Green matrix also having these properties. One notes that the limits in the four corners coincide due to Propositions~\ref{scatt08.prop-Slimits} and \ref{scatt08.prop-R}.  For indeed 

$$
\lim_{b\to \pm \infty} \widetilde{S}_b \;=\;
  \one-2\,|\psi_\pm\rangle \langle \psi_\pm|\,,
\quad 
    \lim_{b\to \pm \infty} \one_b\;=\; \one\,,
\quad
\lim_{a\to -\infty} \widetilde{R}_{\pm,a} \;=\;
  \one-2\,|\psi_\pm\rangle \langle \psi_\pm|\,,
\quad   
    \lim_{a\to +\infty}\widetilde{R}_{\pm,a} \;=\; \one\,.
$$

\noindent In conclusion, $\widetilde{\Omega}_-\in\es$. 
\hfill $\Box$

\vspace{.2cm}

 \subsection{Example of a point interaction}
 \label{sec-example}

Here we discuss the example of the perturbation $V=\lambda\,|0\rangle\langle 0|$, $\lambda\in\RM$, localized on one site. Hence $\Pi=|0\rangle\langle 0|$, $\Lambda=\{0\}$ and,  one has $L=|\Lambda|=1$ even if $\Ee$ is a polynomial. Furthermore $G^\Pi_0(z)=G_0(z)$ is a number (and not a matrix of larger size). The behavior of the real and imaginary part of $G_0(E-\imath 0)$ can directly be read off Proposition~\ref{scatt08.prop-Green3d}. Note that, in particular, the imaginary part does not vanish on $(E_-,E_+)$. Let us introduce the critical coupling constants $\lambda_{\pm}=1/G_0(E_\pm)$. Note that $\lambda_{-}<0$ and $\lambda_{+}>0$. Because the perturbation determinant is $1-\lambda G_0(z)$, the operator $H=H_0+V$ has an eigenvalue smaller than $E_-$ if and only if $\lambda<\lambda_{-}$, and an eigenvalue larger than $E_+$ if and only if $\lambda>\lambda_{+}$. If $\lambda=\lambda_{\pm}$ there is a threshold singularity at $E_\pm$. In dimension $d=3$ and $d=4$ this singularity is a threshold resonance, whereas for $d\geq 5$ it is a threshold eigenvalue. There is never an embedded eigenvalue (also not for polynomial $\Ee$). 

\vspace{.2cm}

The scattering matrix given by \eqref{scatt08.eq-Sformula} differs from the identity only on a one-dimensional subspace:
$$
\cS_E
\;=\;(\one-\piso_{E}^*\piso_{E})\,+\,
\frac{\lambda^{-1}-G_0(E-\imath 0)}{\lambda^{-1}-G_0(E+\imath 0)}
\;\piso_{E}^* \piso_{E}
\;.
$$
Hence
$$
\TR(\cT_E)\;=\;
\frac{1}{\imath}\;
\partial_E\;\ln
\left(
\frac{\lambda^{-1}-G_0(E-\imath 0)}{\lambda^{-1}-G_0(E+\imath 0)}
\right)
\;.
$$
From this one readily deduces the winding number $\TR(T)$ if $\lambda\not =\lambda_\pm$. For the exceptional cases of threshold singularities, we need more precise information about the Green function as given in Proposition~\ref{scatt08.prop-Green3d}. For $d=3$ and $d\geq 5$ one has 
$$
G_0(E-\imath 0)\;=\;
G_0(E_\pm)+N_\pm (E_\pm-E)+\imath D_\pm|E-E_\pm|^{\frac{d-2}{2}}\,+\,o(E-E_\pm)
\;.
$$
The constants satisfy $D_\pm>0$, and, for $d\geq 5$, one also has $N_\pm<0$. Hence, if $\varphi(E)$ denotes the phase of $\lambda_{\pm}^{-1}-G_0(E-\imath 0)$, one has $\varphi(E_+)=0$ and $\varphi(E_-)=\pi$ with one loop in the clockwise orientation for $d\geq 5$, while in $d=3$ one has $\varphi(E_+)-\varphi(E_-)=\frac{\pi}{2}$ for $\lambda=\lambda_{\pm}$. Hence the total scattering phase is
$$
\TR(T)\;=\;
\int^{E_+}_{E_-}dE\;\,\TR(\cT_E)
\;=\;
2\pi\;
\left\{
\begin{array}{cc}
0 & \mbox{ for }\lambda\in(\lambda_{-},\lambda_{+})\;,
\\
-\,1 & \mbox{ for }\lambda<\lambda_{-}\mbox{ or }\lambda>\lambda_{+}\;,
\\
-\,1 & \mbox{ for }\lambda=\lambda_{\pm}\mbox{ and } d\geq 5\;,
\\
-\,\frac{1}{2} & \mbox{ for }\lambda=\lambda_{\pm}\mbox{ and } d=3\;.
\end{array}
\right.
$$
This fits with Levinson's theorem. In particular, in the threshold case for $d\geq 5$ there is a threshold eigenvalue, while for $d=3$ there is a threshold resonance (no bound state) but the correction on the r.h.s. of  \eqref{scatt08.eq-Levinson} is $-\frac{1}{2}$ because $\dim(\Ss_{E_\pm})=1$. 

\vspace{.2cm}

It is also instructive to complete the argument principle proof in the situation $\lambda\in[\lambda_-,\lambda_+]$. We start from \eqref{scatt08.eq-argprin} and remove $\gamma$ as above. Replacing the resolvent identity \eqref{scatt08.eq-Greenperturb2}, then gives
$$
0\;=\;
\int_\Gamma\frac{dz}{2\pi\imath}\;
\frac{\langle 0|(z-H_0)^{-2}|0\rangle}{\lambda^{-1}-G_0(z)}
\;=\;
\int_{G_0(\Gamma)}\frac{dG}{2\pi\imath}\;
\frac{1}{\lambda^{-1}-G}
\;,
$$
where in the second equality we used $\langle 0|(z-H_0)^{-2}|0\rangle=-\partial_z G_0(z)$. Now let us analyze the path $G_0(\Gamma)$ using the results of Proposition~\ref{scatt08.prop-Green3d}. It is always a closed curve going through the real axis twice exactly at $G_0(E_-)$ and $G_0(E_+)$. It is positively oriented and the limit curve is approached from the inside (as $\epsilon\downarrow 0$). The dimension $d$ now leads to the following crucial differences as to how the real axis is crossed. For $d=3$ it crosses transversally, as for an circle, while for $d\geq 4$, it is a spike pointing inward. Hence, if $\lambda=\lambda_\pm$, the singularity leads to a contribution equal to $\frac{1}{2}$ for $d=3$ and equal to $1$ for $d\geq 4$. In particular, this shows that $\TR(T)=-1$ for $\lambda=\lambda_\pm$ also for $d=4$, a case that was not covered by Theorem~\ref{scatt08.theo-Levinson3d}.

\appendix

\section{Boundary values of the Borel transform}
\label{sec-Borel}

\noindent For the convenience of the reader, this section is a reminder of properties of the Borel transform of a function defined on the real line. The first result is often called the Plemelj-Privalov theorem \cite{Mus}.

\begin{lemma}
\label{scatt08.lem-hilb}
Let $\rho:\RM\to \CM$ be a H\"older continuous function of exponent $\alpha \in (0,1]$ with compact support. Then, for any $\beta $ such that $0<\beta<\alpha$, its Borel transform

\begin{equation}
\label{eq-Borel}
G_\rho(z) \;=\;
   \int_\RM \frac{\rho(e)\,de}{z-e} 
\end{equation}

\noindent is holomorphic in $\CM\setminus\supp(\rho)$ and its boundary value on the real axis is H\"older continuous with exponent $\beta$. If $\rho$ is real-valued, then

$$
G_\rho(E\pm\imath 0) \;=\;\mp\,\imath\,\pi\, \pi\, \rho(E)\;+\;
   \dashint_\RM \frac{\rho(e)\,de}{E-e} 
$$
where $\dashint$ denotes the Cauchy principal value.
\end{lemma}

\noindent  {\bf Proof:} The holomorphy of $G_\rho$ outside $\supp(\rho)$ is a standard result that will not be proved here. By decomposing $\rho$ into its real and imaginary part, if necessary, there is no loss of generality in assuming that $\rho$ is real-valued. For $\epsilon >0$ and $E\in\RM$, $G_\rho(E+\imath\epsilon)$ is given by

$$G_\rho(E\pm\imath\epsilon) \;=\; 
   \int_\RM de\;\frac{\rho(e)(E-e\mp\imath\epsilon)}{(E-e)^2+\epsilon^2} \;=\;
    R_\rho(E,\epsilon)\;\mp\;\imath\, I_\rho(E,\epsilon)\,,
$$

\noindent where 

\begin{equation}
\label{scatt08.eq-ImRe}
I_\rho(E,\epsilon)\;=\; 
   \int_\RM de\;\frac{\rho(e)\epsilon}{(E-e)^2+\epsilon^2}\;,
\hspace{2cm}
   R_\rho(E,\epsilon)\;=\; 
   \int_\RM de\;\frac{\rho(e)(E-e)}{(E-e)^2+\epsilon^2}\;.
\end{equation}

\noindent The first term admits $\pi \rho(E)$ as a limit as $\epsilon \downarrow 0$. This is because, using the change of variables $e=E+\epsilon x$ and the Lebesgue dominated convergence theorem, gives

\begin{equation}
\label{scatt08.eq-Imf}
\lim_{\epsilon \downarrow 0}\;I_\rho(E,\epsilon)\;=\; 
   \lim_{\epsilon \downarrow 0}\;\int_\RM \frac{dx}{x^2+1}\; \rho(E+\epsilon x) 
    \;=\; \rho(E) \int_\RM \frac{dx}{x^2+1} \;=\;\pi \,\rho(E)\,.
\end{equation}

\noindent  Similarly, using the change of variable $u=e-E$ and the symmetry $u\mapsto -u$, one obtains

\begin{equation}
\label{scatt08.eq-Ref}
R_\rho(E,\epsilon)\;=\; 
   \int_0^{\infty} \frac{u\,du}{u^2+\epsilon^2}\; 
    \bigl(\rho(E-u)-\rho(E+u)\bigr)\,.
\end{equation}

\noindent For all $a>0$, the part of the integral corresponding to $0<a\leq u$ is also H\"older continuous of exponent $\alpha$ w.r.t. $E$, thanks to Lebesgue's dominated convergence theorem. In particular, if $E$ is not in the support of $\rho$, the integral over $u$ never reaches $u=0$ so that $R_\rho(E)$ is H\"older continuous outside the support of $\rho$. However, if $\rho$ is real valued, $R_\rho$ is the restriction to the complement of the support of $\rho$ (in the real line) of the real part of an holomorphic function and is therefore analytic. On the other hand, since $\rho$ is H\"older continuous of exponent $\alpha$ and with compact support, it follows that there is a constant $K>0$ for which  $|\rho(E+\delta\pm u) - \rho(E\pm u)|\leq K\delta^{\alpha}$ uniformly w.r.t. $E$ and $u$. In particular, $|\rho(E-u) - \rho(E+u)|\leq K(2u)^{\alpha}$ and 

$$|\rho(E+\delta +u) - \rho(E+ u)-\rho(E+\delta -u) + \rho(E- u)|
   \;\leq \;2K\min\{\delta^\alpha, (2u)^\alpha\} 
    \;\leq \;2^{1+\alpha-\beta}K \delta^{\beta}u^{\alpha-\beta}\,,
$$

\noindent for any $0<\beta<\alpha$. Using this estimate inside the part of the integral for which $u\in[0,1]$ and thanks to the dominated convergence theorem, it follows that $\lim_{\epsilon\downarrow 0}R_\rho(E,\epsilon)$ exists and is H\"older continuous of exponent $\beta$ for $E\in\supp(\rho)$. The last formula also follows from the above.
\hfill $\Box$

\begin{coro}
\label{scatt08.cor-hilb2}
Let $\rho:\RM\to \CM$ be $k$-times differentiable with $k$-th derivative H\"older continuous of exponent $\alpha \in (0,1]$ with compact support. Then, for any $\beta $ such that $0<\beta<\alpha$, its Borel transform $G_\rho(z)$ defined by {\rm \eqref{eq-Borel}} has a $k$-times differentiable boundary value on $\supp(\rho)$ for which the $k$th derivative is H\"older continuous with exponent $\beta$.
\end{coro}

\noindent  {\bf Proof:} Equations~(\ref{scatt08.eq-Imf}) and (\ref{scatt08.eq-Ref}) show that $\partial_E I_\rho(e,\epsilon)= I_{\rho'}(E,\epsilon)$ and similarly for $R_\rho$. Using Lemma~\ref{scatt08.lem-hilb} gives the result.
\hfill $\Box$

\begin{lemma}
\label{scatt08.lem-hilbplus}
Let $\rho$ be a real valued $C^1$-function on the real line with compact support such that $\rho(0)\neq 0$. Let $G_\rho^+$ denote the partial Borel transform defined by

$$G_\rho^+(z) \;=\;
   \int_0^{\infty} \frac{\rho(e)\,de}{z-e} \,.
$$

\noindent Then one has the following.

\vspace{.1cm}

\noindent {\rm (i)} $G_\rho^+$ is holomorphic outside of $[0,\infty)\cap\supp(\rho)$.

\vspace{.1cm}

\noindent {\rm (ii)} The boundary value $I_\rho(E)=\Im m\,G_\rho^+(E\mp\imath 0)$ of its imaginary part is $\pm\pi \rho(E)$ whenever $E>0$, 

it is $\pm\frac{\pi}{2} \rho(0)$ at $E=0$ and it vanishes for $E<0$.

\vspace{.1cm}

\noindent {\rm  (iii)} The boundary values $R_\rho(E)=\Re e\,G_\rho^+(E\pm\imath 0)$ is continuously differentiable for $E\neq 0$ and 

satisfies 

$$R_\rho(E) \;=\;
   \rho(0) \ln(|E|)\; + \;\Oo(1)\,,
   \qquad \mbox{as }E\to 0\;.
$$
\end{lemma}

\noindent  {\bf Proof:} The proof will use the results given in the proof of Lemma~\ref{scatt08.lem-hilb}. The first claim (i) can be proved in the same way indeed. 

\vspace{.1cm}

(ii) Thanks to equation~(\ref{scatt08.eq-ImRe}) and with the change of variable $e=E+\epsilon x$, the result of equation (\ref{scatt08.eq-Imf}) is still valid as long as $E>0$. This is because the interval of integration for the variable $x$ is $[-E/\epsilon, \infty)$ in the present case. Then, as $\epsilon \downarrow 0$, the dominated convergence theorem gives the same result since this interval of integration eventually becomes $\RM$. The same argument applied to $E<0$ gives and interval of integration $[|E|/\epsilon, +\infty)$ which shrink to the empty set as $\epsilon\downarrow 0$. At last, if $E=0$ the interval of integration is $[0,\infty)$, and the result is $\frac{\pi}{2}$ instead of $\pi$. 

\vspace{.1cm}

(iii) The argument given  in the proof of Lemma~\ref{scatt08.lem-hilb} applies also for $E\neq 0$, so that $R_\rho$ is continuously differentiable as well. Near $E=0$, however, the integral defining $R_\rho$ diverges. This can be seen as follows, using equation (\ref{scatt08.eq-Imf}) (note the change of sign)

$$R_\rho(E,\epsilon)\;=\; 
   \int_0^\infty de\; \frac{\rho(e)(E-e)}{(E-e)^2+\epsilon^2} \;=\;
    -\,\frac{1}{2}\,
     \int_0^\infty \rho(e)\; d\left(\ln((E-e)^2+\epsilon^2)\right)\;.
$$

\noindent Since $\rho$ has compact support, integrating by parts yields

$$R_\rho(E,\epsilon) \;=\; 
   \frac{\rho(0)}{2}\, \ln(E^2+\epsilon^2)
    \,-\,\frac{1}{2}
     \int_0^\infty de\;\rho'(e) 
     \ln\bigl((E-e)^2+\epsilon^2\bigr)\;.
$$

\noindent Note that the second integral converges as $\epsilon\downarrow 0$ to 

$$-\int_0^\infty de\;\rho'(e) \,\ln(|E-e|)\;.
$$

\noindent which is continuous with respect to $E$ near $E=0$. This can be seen by changing $e$ into $e-E$ and using the fact that the derivative $\rho'$ is continuous as well as the dominated convergence theorem. 
\hfill $\Box$

\section{Technicalities on certain operator inverses}
\label{app-inverses}

\begin{lemma}
\label{scatt08.lem-kerA+iB}
Let $A=A^*$ and $B\geq 0$ be bounded operators on a Hilbert space. Then $\Ker (A+\imath B)= \Ker(A)\cap\Ker(B)=\Ker(A-\imath B)$. In addition, the restriction  of the operator $K=(A+\imath B)^{-1}B$ to $\left(\Ker(A)\cap\Ker(B)\right)^\perp$ is well-defined. In particular, if $v\in (\Ker(A)\cap\Ker(B))^\perp$, then $w=(A+\imath B)^{-1}v$ exists as the unique solution $w\in(\Ker(A)\cap\Ker(B))^\perp$ of $(A+\imath B)w=v$.
\end{lemma}

\noindent  {\bf Proof:} If $w\in \Ker(A)\cap\Ker(B)$, then clearly $(A+\imath B)w=0$ so that $w\in \Ker(A+\imath B)$. Conversely, if $w\in \Ker(A+\imath B)$, then $0=\langle w|(A+\imath B)w\rangle =\langle w|Aw\rangle +\imath \langle w|Bw\rangle$. Since $A$ and $B$ are selfadjoint, it follows that $\langle w|Aw\rangle=\langle w|Bw\rangle=0$. Since $B$ is positive, $\|Bw\|^2=\langle w|B^2w\rangle \leq \|B\|\langle w|Bw\rangle=0$. Hence $w\in\Ker(B)$. This also implies $Aw=0$ so that $w\in \Ker(A)\cap \Ker(B)$. The same argument shows that $\Ker(A-\imath B)=\Ker(A)\cap \Ker(B)$. Let $P$ be the orthogonal projection onto $\Ker(A)\cap \Ker(B)$. Then it commutes with both $A$ and $B$. Therefore, both $(A+\imath B)$ and $B$ restrict to $\Ran(\id-P)$. Moreover, $\Ran (A+\imath B)= \Ker(A-\imath B)^\perp=\Ran(\id-P)$. Therefore the equation $(A+\imath B)w = v$ admits a solution, which is unique modulo $\Ker(A+\imath B)$ which is unique if, in addition, $Pw=0$, namely if $w\in\left(\Ker(A)\cap\Ker(B)\right)^\perp$.
\hfill $\Box$

\begin{lemma}
\label{lem-linindep}
For an interval $I\subset\RM$, let $b\in I\mapsto A_b\in\mbox{\rm Mat}(N\times N,\CM) $
and $b\in I\mapsto B_b\in\mbox{\rm Mat}(N\times N,\CM) $ be real analytic functions satisfying $A_b=A_b^*$ and $B_b\geq 0$. Let $P_b$ denote the projection on $\mbox{\rm Ran}(B_b)$. Then $\dim(P_b)$ is constant except for $b$ in a discrete set $\vs \subset I$.  We suppose that $\mbox{\rm Ker}(A_b)\cap \mbox{\rm Ker}(B_b)$ is non-trivial only for a discrete set of $b$'s not lying in $\vs $ and that those zeros are regular singular points for $A_b$, that is, these zeros of $b\in I\mapsto A_b$ are of first order. Then $b\in I\mapsto K_b=(A_b+ \imath B_b)^{-1}(B_b)^{\frac{1}{2}}$ is well-defined, continuous and real analytic except in those points of $\vs $ which are not zeros of $A_b$.
\end{lemma}

\noindent  {\bf Proof:}  First of all, by analytic perturbation theory the eigenvalues of $B_b$ vary analytically with $b$ and therefore $\dim(P_b)$ indeed only differs from the almost sure value $L$ on a discrete set $\vs $. On $\vs $, the dimension of $P_b$ is then smaller. By the proof of Lemma~\ref{scatt08.lem-kerA+iB}, if $\mbox{\rm Ker}(A_b)\cap \mbox{\rm Ker}(B_b)$ is trivial, the inverse of $A_b+ \imath B_b$ exists and, moreover, it is real analytic in $b$. Thus we may restrict our attention to one point, say $b=0$, at which $\mbox{\rm Ker}(A_0)\cap \mbox{\rm Ker}(B_0)$ is non-vanishing. Then $K_b$ is real analytic in $b\in I\backslash\{0\}$. By Riemann's theorem on removable singularities it is sufficient to show that $K_b$ remains bounded in a neighborhood of $b=0$, because then it is already analytic at $b=0$.  Now $\dim(\mbox{\rm Ker}(B_b))=N-L$ is constant for $b\not =0$ (possibly after having made $I$ smaller). Again by analytic perturbation theory there exists a real analytic function $b\in I\mapsto U_b$ of unitaries such that $U_b B_b U_b^*=\binom{0 \;\;0}{0\;\,\eta_b}$ where $\eta_b> 0$ is a diagonal matrix of size $L$. It is clearly sufficient to show that $b\in I\mapsto U_bK_bU_b^*=(U_bA_bU_b^*+ \imath U_bB_bU_b^*)^{-1}U_b(B_b)^{\frac{1}{2}}U_b^*$ is real analytic, which is equivalent to supposing that $B_b$ is diagonal. Next let us introduce the notations
$$
U_bA_bU_b^*
\;=\;
\begin{pmatrix}
\alpha_b & \beta_b \\
\beta_b^* & \gamma_b
\end{pmatrix}
\;,
$$
where $\alpha_b$ and $\gamma_b$ are self-adjoint square matrices of size $(N-L)\times (N-L)$ and $L\times L$ respectively. Because $\mbox{\rm Ker}(A_0)\cap \mbox{\rm Ker}(B_0)$ is non-trivial, it follows that $\mbox{\rm Ker}(\alpha_0)\cap \mbox{\rm Ker}(\beta_0^*)$ is non-trivial as well.  Moreover, $\beta_b^*$ is of size $L\times (N-L)$ and $\binom{0 \;\;0}{\beta_b^*\;0}=U_bP_bA_b(1-P_b)U_b^*$. Furthermore
$$
U_bK_bU_b^*
\;=\;
\begin{pmatrix}
\alpha_b & \beta_b \\
\beta_b^* & \gamma_b+\imath\eta_b
\end{pmatrix}^{-1}
\begin{pmatrix}
0 & 0 \\
0 & (\eta_b)^{\frac{1}{2}}
\end{pmatrix}
\;,
$$
where the appearing inverse only exists in the sense of Lemma~\ref{scatt08.lem-kerA+iB}. In order to calculate the inverse, let us introduce the Schur complement by
\begin{eqnarray}
\sigma_b
& = &
\alpha_b-\beta_b(\gamma_b+\imath\eta_b)^{-1}\beta_b^*
\nonumber
\\
& = & 
\alpha_b-\beta_b(\gamma_b\eta_b^{-1}\gamma_b+\eta_b)^{-1}\gamma_b\eta_b^{-1}\beta_b^*
+\imath \;\beta_b(\gamma_b\eta_b^{-1}\gamma_b+\eta_b)^{-1}\beta_b^*
\label{eq-sigma}
\end{eqnarray} 
which is an $(N-L)\times (N-L)$ matrix. The second formula shows also that $\Im m(\sigma_b)$ is non-negative. For $b\not =0$, one has $\mbox{\rm Ker}(\alpha_b)\cap \mbox{\rm Ker}(\beta_b^*)=\{0\}$ so that $\sigma_b$ is invertible because then the imaginary part is positive. Now the Schur complement formula gives
$$
\begin{pmatrix}
\alpha_b & \beta_b \\
\beta_b^* & \gamma_b+\imath\eta_b
\end{pmatrix}^{-1}
\;=\;
\begin{pmatrix}
\sigma_b^{-1} & -\sigma_b^{-1}\beta_b(\gamma_b+\imath\eta_b)^{-1} \\
-(\gamma_b+\imath\eta_b)^{-1}\beta_b^*\sigma_b^{-1} & (\gamma_b+\imath\eta_b)^{-1} +(\gamma_b+\imath\eta_b)^{-1} \beta_b^*\sigma_b^{-1}\beta_b(\gamma_b+\imath\eta_b)^{-1}
\end{pmatrix}
\;,
$$
so that
$$
U_bK_bU_b^*
\;=\;
\begin{pmatrix}
0 & -\sigma_b^{-1}\beta_b(\gamma_b+\imath\eta_b)^{-1}(\eta_b)^{\frac{1}{2}} \\
0 & (\gamma_b+\imath\eta_b)^{-1}(\eta_b)^{\frac{1}{2}} +(\gamma_b+\imath\eta_b)^{-1} \beta_b^*\sigma_b^{-1}\beta_b(\gamma_b+\imath\eta_b)^{-1}(\eta_b)^{\frac{1}{2}}
\end{pmatrix}
\;.
$$
This shows that  $U_bK_bU_b^*$ remains bounded in a neighborhood of $b=0$ if $\sigma_b^{-1}\beta_b$ remains bounded, or equivalently that $F_b=\sigma_b^{-1}\beta_b(\gamma_b\eta_b^{-1}\gamma_b+\eta_b)^{-\frac{1}{2}}$ remains bounded. Setting
$$
C_b
\;=\;
\alpha_b-\beta_b(\gamma_b\eta_b^{-1}\gamma_b+\eta_b)^{-1}\gamma_b\eta_b^{-1}\beta_b^*
\;,
\qquad
D_b\;=\;
\beta_b(\gamma_b\eta_b^{-1}\gamma_b+\eta_b)^{-\frac{1}{2}}
\;,
$$
this can be written as
$$
F_b\;=\;(C_b+ \imath D_b D_b^*)^{-1}D_b
\;,
$$
due to the formula \eqref{eq-sigma}. Now the zero of $A_b$ at $b=0$ is of first order by assumption, therefore the zeros of $\alpha_b$ and $\beta_b^*$ are of first order as well. It follows that the zero of $C_b$ is of first order as well. Hence the following lemma is applicable to deduce that $F_b$ is bouned.  
\hfill $\Box$

\vspace{.2cm}

\begin{lemma}
\label{lem-linindep2}
For an interval $I\subset\RM$, let $b\in I\mapsto C_b\in\mbox{\rm Mat}(N\times N,\CM) $
and $b\in I\mapsto D_b\in\mbox{\rm Mat}(N\times L,\CM) $ be real analytic functions satisfying $C_b=C_b^*$. We suppose that both are  invertible except on a discrete set of points and that, in case $\mbox{\rm Ker}(C_b)\cap \mbox{\rm Ker}(D_b^*)$ is non-trivial for some such $b$, the zero of $C_b$ is of first order. Then $b\in I\mapsto F_b=(C_b+ \imath D_b D_b^*)^{-1}D_b$ is well-defined and real analytic as well.
\end{lemma}

\noindent  {\bf Proof:} Again we focus on one point $b=0$ for which $\mbox{\rm Ker}(C_0)\cap \mbox{\rm Ker}(D_0^*)$ is one-dimensional. Then, also as above, we choose a basis such that $C_b$ is diagonal. In this basis, let us introduce the following notations for the matrix entries:
$$
C_b
\;=\;
\begin{pmatrix}
b \kappa_b & 0 \\ 0 & \delta_b
\end{pmatrix}
\;,
\qquad
D_b^*
\;=\;
\begin{pmatrix}
b \eta_b &  \beta_b \\ b\gamma_b & \alpha_b
\end{pmatrix}
\;.
$$
Here $\kappa_b$ and $\delta_b$ are diagonal invertible matrices (for all $b$) and the size of $\kappa_b$ is the multiplicity of the zero of $C_0$. Then
$$
D_bD_b^*
\;=\;
\begin{pmatrix}
b^2(\eta_b^*\eta_b+\gamma_b^*\gamma_b) &  b(\beta_b+\gamma_b^*\alpha_b) \\
b(\beta_b^*+\alpha_b^*\gamma_b) & \beta_b^*\beta_b+\alpha_b^*\alpha_b
\end{pmatrix}
\;.
$$
Thus we introduce the Schur complement for $(C_b+\imath D_bD_b^*)^{-1}$:
$$
\sigma_b
\;=\;
b\kappa_b+\imath\,b^2(\eta_b^*\eta_b+\gamma_b^*\gamma_b)-b^2\,
(\eta_b^*\beta_b+\gamma_b^*\alpha_b)
\bigl(
\delta_b+\imath\,(\beta_b^*\beta_b+\alpha_b^*\alpha_b)
\bigr)^{-1}
(\beta_b^*\eta_b+\alpha_b^*\gamma_b)
\;.
$$
Note that the appearing is well-defined also at $b=0$ because $\delta_0$ is invertible and the other contribution is imaginary. We also note that $\sigma_b=b\kappa_b+\Oo(b^2)$. Now we obtain
\begin{eqnarray*}
(D_b+\imath D_bD_b^*)^{-1}D_b
& = &
\begin{pmatrix}
b\kappa_b+\imath\,b^2(\eta_b^*\eta_b+\gamma_b^*\gamma_b) & \imath\, b(\eta_b^*\beta_b+\gamma_b^*\alpha_b) \\
\imath\,b(\beta_b^*\eta_b+\alpha_b^*\gamma_b) & \delta_b+\imath\,\beta_b^*\beta_b+\alpha_b^*\alpha_b
\end{pmatrix}^{-1}
\;
\begin{pmatrix}
b\eta_b & b\gamma_b^* \\
\beta_b^* & \alpha_b^*
\end{pmatrix}
\;.
\end{eqnarray*}
The upper left corner of the inverse is $(\sigma_b)^{-1}=\Oo(b^{-1})$ and thus singular, but all the other entries remain bounded for $b\to 0$ because $(\sigma_b)^{-1}$ is always multiplied by a factor $b$. The singularity of the upper left corner, however, is multiplied by $b$ from the second factor. In conclusion, $(C_b+\imath D_bD_b^*)^{-1}D_b$ remains bounded and therefore the singularity is removable.
\hfill $\Box$

\vspace{.5cm}

\end{document}